\documentclass[smallextended]{svjour3}       
\smartqed

\usepackage[ruled,vlined]{algorithm2e}

\SetAlFnt{\small} 
\SetAlCapFnt{\small}
\SetAlCapNameFnt{\small}
\SetAlCapHSkip{0pt}
\IncMargin{-\parindent}

\usepackage{graphicx}
\usepackage{lmodern}           	
\usepackage{amsmath} 
\usepackage{amssymb}

\usepackage{amsthm}
\usepackage[export]{adjustbox}

\usepackage{array}

  \newtheorem{dummy}{***}
  \newtheorem{thm}[dummy]{Theorem}

  \newtheorem{lem}[dummy]{Lemma}
  \newtheorem{defin}[dummy]{Definition}

\usepackage{tikz}
\usetikzlibrary{decorations.pathreplacing}
\usetikzlibrary{shapes}
\usetikzlibrary{arrows}
\usetikzlibrary{trees}
\usetikzlibrary{patterns}

\newlength{\hatchspread}
\newlength{\hatchthickness}
\newlength{\hatchshift}
\newcommand{\hatchcolor}{}
\tikzset{hatchspread/.code={\setlength{\hatchspread}{#1}},
         hatchthickness/.code={\setlength{\hatchthickness}{#1}},
         hatchshift/.code={\setlength{\hatchshift}{#1}},
         hatchcolor/.code={\renewcommand{\hatchcolor}{#1}}}
\tikzset{hatchspread=3pt,
         hatchthickness=0.4pt,
         hatchshift=0pt,
         hatchcolor=black}
\pgfdeclarepatternformonly[\hatchspread,\hatchthickness,\hatchshift,\hatchcolor]
   {custom north west lines}
   {\pgfqpoint{\dimexpr-2\hatchthickness}{\dimexpr-2\hatchthickness}}
   {\pgfqpoint{\dimexpr\hatchspread+2\hatchthickness}{\dimexpr\hatchspread+2\hatchthickness}}
   {\pgfqpoint{\dimexpr\hatchspread}{\dimexpr\hatchspread}}
   {
    \pgfsetlinewidth{\hatchthickness}
    \pgfpathmoveto{\pgfqpoint{0pt}{\dimexpr\hatchspread+\hatchshift}}
    \pgfpathlineto{\pgfqpoint{\dimexpr\hatchspread+0.15pt+\hatchshift}{-0.15pt}}
    \ifdim \hatchshift > 0pt
      \pgfpathmoveto{\pgfqpoint{0pt}{\hatchshift}}
      \pgfpathlineto{\pgfqpoint{\dimexpr0.15pt+\hatchshift}{-0.15pt}}
    \fi
    \pgfsetstrokecolor{\hatchcolor}
    \pgfusepath{stroke}
   }

\usepackage{ifthen}
\usepackage{lscape}

\usepackage{todonotes}
\usepackage{mathrsfs}

\usepackage{framed}

\newcommand{\Transfer}{\textit{Transfer}}
\newcommand{\pathwidth}{\textit{pathwidth}}
\newcommand{\Eval}{\textit{Eval}}
\newcommand{\inp}{\textit{in}}
\newcommand{\out}{\textit{out}}
\newcommand{\mO}[1] {\mathcal{O}\left( #1 \right)}
\newcommand{\lo}[1] {o\left( #1 \right)}
\newcommand{\Om}[1] {\Omega\left( #1 \right)}
\newcommand{\thet}[1] {\Theta\left( #1 \right)}
\newcommand{\ceil}[1] {\left\lceil #1 \right\rceil}
\newcommand{\lr}[1] {\left( #1 \right)}
\newcommand{\lrC}[1] {\left\{ #1 \right\}}
\newcommand{\sH} {\mathscr{S}}
\newcommand{\wL} {\mathbb{W}}
\newcommand{\eL} {\mathbb{E}}

\newcommand{\fg}{[k_1]\times\dots\times[k_n]}
\newcommand{\sStencil}[1]{S_s\left(#1\right)}
\def\ac{\!\!\!\!\!\!}
\newcommand{\size}[1]{\left|#1\right|}
\newcommand{\para}[1]{\paragraph{\textbf{#1}}}
\newcommand{\algo}{\mathcal{A}}
\newcommand{\sweepSeq}{\mathcal{X}}

\newcommand{\argmin}{\operatornamewithlimits{argmin}}
\newcommand{\argmax}{\operatornamewithlimits{argmax}}

\newcommand{\drawEllBall}[5]{
	\foreach \y in {1,...,#3}{
		\foreach \x in {1,...,#3}{
			\draw[#5,fill] (#1+\x-1-\y+1,#2+#3-\x+1-\y+1) circle (#4);
			}
		}

	\foreach \y in {2,...,#3}{
		\foreach \x in {2,...,#3}{
			\draw[#5,fill] (#1+\x-2-\y+2,#2+#3-\x+2-\y+2-1) circle (#4);
			}
		}

	\draw[#5, thick] (#1, #2+#3) -- (#1+#3-1, #2+1) -- (#1, #2-#3+2) -- (#1-#3+1, #2+1) -- cycle;
}

\newcommand{\drawSquare}[5]{
	
	\foreach \y in {0,...,#3}{
		\foreach \x in {0,...,#3}{
			\draw[#5,fill] (#1+\x,#2+\y) circle (#4);
			}
		}

	\draw[#5, very thick] (#1, #2) -- (#1+#3, #2) -- (#1+#3, #2+#3) -- (#1, #2+#3) -- cycle;
}

\newcommand{\drawEllBallCluster}[2]{
			\drawEllBall{0+#1}{0+#2}{3}{0.2}{yellow} 
			\drawEllBall{2+#1}{3+#2}{3}{0.2}{yellow!33!blue}
			\drawEllBall{3+#1}{-2+#2}{3}{0.2}{yellow!66!blue} 
			\drawEllBall{5+#1}{1+#2}{3}{0.2}{blue}
}

\begin{document}

\title{Tight Bounds for Low Dimensional Star Stencils in the Parallel External Memory Model
\thanks{The final publication (focusing on the lower bounds) is available at link.springer.com: Philipp Hupp and Riko Jacob. \textit{Tight bounds for low dimensional star stencils in the external memory model}. Algorithms and Data Structures, volume 8037 of Lecture Notes in Computer Science, pages 415--426. Springer Berlin Heidelberg, 2013. http://dx.doi.org/10.1007/978-3-642-40104-6\_36
 }
}


\titlerunning{Tight Bounds for Low Dimensional Star Stencils in the PEM Model}        

\author{Philipp Hupp         \and
        Riko Jacob 
}


\institute{P. Hupp \at
%
ETH Z\"urich,
Institute of Theoretical Computer Science,
Universit\"atstrasse 6,
8092 Zurich,
Switzerland\\
              Tel.: +41-44-6337022, 
              Fax: +41-44-6321399, 
              \email{philipp.hupp@inf.ethz.ch}           
           \and
           R. Jacob \at
ETH Z\"urich,
Institute of Theoretical Computer Science,
Universit\"atstrasse 6,
8092 Zurich,
Switzerland\\
}

\date{Received: date / Accepted: date}

\maketitle

	\begin{abstract}
		Stencil computations on low dimensional grids are kernels of many scientific applications including finite difference methods used to solve partial differential equations.
          On typical modern computer architectures, such stencil computations are limited by the performance of the memory subsystem, namely by the bandwidth between main memory and the cache.
		This work considers the computation of star stencils, like the 5-point and 7-point stencil, in the external memory model and parallel external memory model and analyses the constant of the leading term of the non-compulsory I/Os.
		While optimizing stencil computations is an active field of research, there has been a significant gap between the lower bounds and the performance of the algorithms so far. 
		In two dimensions, this work provides matching constants for lower and upper bounds closing a multiplicative gap of 4.
		In three dimensions, the bounds match up to a factor of $\sqrt{2}$ improving the known results by a factor of $2 \sqrt{3}\sqrt{B}$, where $B$ is the block (cache line) size of the external memory model. 
		For dimensions $d\geq 4$, the lower bound is improved between a factor of $4$ and $6$.
		For arbitrary dimension~$d$, 	the first analysis of the constant of the leading term of the non-compulsory I/Os is presented. 
		For $d\geq 3$ the lower and upper bound match up to a factor of $\sqrt[d-1]{d!}\approx \frac{d}{e}$. 
%
%
%
%
%
%
%

	\end{abstract}

	\section{Introduction}
		Stencil computations are the most performance critical component for many tasks in scientific computing.
		In particular, they appear when Partial Differential Equations (PDEs) are solved.
		In order to solve a PDE with numerical methods, the space needs to be discretized and a standard discretization method for low dimensional Euclidean spaces are regular grids.
		The differential operator can then be turned into a linear function of a grid point and its neighbors by a finite difference method.
		Such a linear function is also called stencil and results in a very regular sparse system of linear equations.
		To make use of the sparsity, such systems are typically solved with iterative solvers like the Jacobi or Gauss-Seidel method.
		The kernel of these methods is the evaluation of the underlying stencil. 

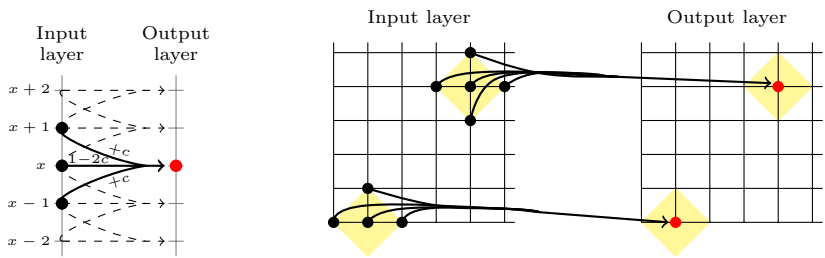
\begin{figure}
	\centering
		\begin{tabular}{ccc}
			\begin{tikzpicture}[rotate = -90,scale=.5,every text node part/.style={align=center}]
				\draw[step=1cm,gray,thin] (-2.4,-.2) grid (2.4,.2);
				\draw[step=1cm,gray,thin] (-2.4,2.8) grid (2.4,3.2);
				\foreach \x in {-2,...,2} {

					\draw[->,dashed] (\x,0) -- (\x,2.7);
				}
				\foreach \x in {-2,...,1} {
						\draw[dashed] plot [smooth,tension=1.5] coordinates{(\x,0) (\x+.5, 0.6) (\x + 1, 2.2)};
					\draw[dashed] plot [smooth,tension=1.5] coordinates{(\x+1,0) (\x+.5, 0.6) (\x , 2.2)};

				}
				\draw[->,thick] (0,0) -- (0,2.7);
				\draw[thick] plot [smooth,tension=1.5] coordinates{(1,0) (.5, 0.6) ( 0, 2.2)};
				\draw[thick] plot [smooth,tension=1.5] coordinates{(-1,0) (-.5, 0.6) (0 , 2.2)};

				\draw[fill, black] (-1,0) circle (.15cm);
				\draw[fill, black] (0,0) circle (.15cm);
				\draw[fill, black] (1,0) circle (.15cm);
				\draw[fill, red] (0,3) circle (.15cm);

				\draw (-.15,.7) node[] {$\scriptscriptstyle 1-2c$};
				\draw (-.4,1.5) node[rotate = -15] {$\scriptscriptstyle+c$};
				\draw (.4,1.5) node[ rotate = 15] {$\scriptscriptstyle+c$};
			\tikzstyle{every node}=[font=\tiny]
				\draw (0,-.2) node[left] {$x$};
				\draw (-1,-.1) node[left] {$x+1$};
				\draw (-2,-.1) node[left] {$x+2$};
				\draw (1,-.1) node[left] {$x-1$};
				\draw (2,-.1) node[left] {$x-2$};
			\tikzstyle{every node}=[font=\scriptsize]
			\draw (-3.2,0) node[] {Input \\layer};
			\draw (-3.2,3) node {Output \\layer};

			\end{tikzpicture}&	\hspace{5ex} &
			\begin{tikzpicture}[scale = 0.45,every text node part/.style={align=center}]
			 	\tikzstyle{every node}=[font=\scriptsize]

				\draw[fill,yellow!50] (0,0) -- (1,1) -- (2,0) -- (1,-1) -- cycle;
				\draw[fill,yellow!50] (3,4) -- (4,5) -- (5,4) -- (4,3) -- cycle;

				\draw (0,0) grid (5.3,5.3);
				\draw (2.5,6) node[] {Input layer};
				
				\draw[fill] (4,4) circle (0.15cm);
				\draw[fill] (5,4) circle (0.15cm);
				\draw[fill] (3,4) circle (0.15cm);
				\draw[fill] (4,5) circle (0.15cm);
				\draw[fill] (4,3) circle (0.15cm);

				\draw[fill] (1,0) circle (0.15cm);
				\draw[fill] (1,1) circle (0.15cm);
				\draw[fill] (2,0) circle (0.15cm);
				\draw[fill] (0,0) circle (0.15cm);

				\begin{scope}[xshift = 9cm]
					\draw[fill,yellow!50] (0,0) -- (1,1) -- (2,0) -- (1,-1) -- cycle;
					\draw[fill,yellow!50] (3,4) -- (4,5) -- (5,4) -- (4,3) -- cycle;
					\draw (0,0) grid (5.3,5.3);
					\draw[fill,red] (4,4) circle (0.15cm);
					\draw (2.5,6) node {Output layer};
					\draw[fill,red] (1,0) circle (0.15cm);
				\end{scope}

				\draw[thick] plot [smooth, tension = 0.8] coordinates{(4,4) (5, 4.4) (8 , 4.3) (8,4.3)};
				\draw[thick, ->] (8, 4.3) -- (12.8,4.1);
				\draw[thick] plot [smooth, tension = 0.8] coordinates{(5,4) (6, 4.4) (8 , 4.3) };
				\draw[thick] plot [smooth, tension = 0.8] coordinates{(4,5) (6, 4.4) (8 , 4.3) };
				\draw[thick] plot [smooth, tension = 0.8] coordinates{(3,4) (4,4.4) (7, 4.4) (8 , 4.3) };
				\draw[thick] plot [smooth, tension = 0.8] coordinates{(4,3) (4.7, 4.2) (6.5, 4.4) };

				\draw[thick] plot [smooth, tension = 0.8] coordinates{(1,0) (2,0.4) (5, 0.4) (6, 0.3) };
				\draw[thick, ->] (6,0.3) -- (9.8,0);
				\draw[thick] plot [smooth, tension = 0.8] coordinates{(2,0) (2.7,0.4) (5, 0.4) (6, 0.3) };
				\draw[thick] plot [smooth, tension = 0.8] coordinates{(1,1) (3,0.5) (5, 0.4) (6, 0.3) };
				\draw[thick] plot [smooth, tension = 0.8] coordinates{(0,0) (1,0.5) (5, 0.4) (6, 0.3) };

			\end{tikzpicture}
		\end{tabular}
		\caption{The computation graph implied by the 1-star stencil in 1D (left) and 2D (right).}
		\label{fig:layeredGraphsAndStencil}
\end{figure}

		To clarify how stencils are used to solve PDEs we give a simple example. 
		Consider the one-dimensional heat equation which describes the variation of temperature on a pole over time. 
		For a function $u(t,x)$ describing the temperature of the pole at time $t$ and position $x$, this problem can formally be written as the PDE $\frac{\partial u}{\partial t} = \frac{\partial^2u}{\partial x^2}$. 
		We approximate the pole by a one-dimensional grid where each grid point stores the temperature of the pole at the respective point. 
		Using an explicit finite difference method, the temperature of the grid points at time $t+1$ can be computed given the temperature at time $t$ . 
		The PDE is approximated by $\frac{u(t+1,\,x)-u(t,x)}{\Delta t} = \frac{u(t,\,x-1)-2u(t,\,x)+u(t,\,x+1)}{\left(\Delta x\right)^2}$. 
		Abbreviating $c := \frac{\Delta t}{\left(\Delta x \right)^2},$ this  solves to $u(t+1, \, x) = c \cdot u (t,x-1) + (1-2c) \cdot u(t, \, x) + c \cdot u(t,\,x+1)$ which in turn gives rise to the one-dimensional 1-star stencil.
		Hereby the stencil states which neighboring vertices 
		of a grid point are necessary to update the grid point.
		The task is to recompute the values at the vertices of the grid according to the stencil.
		Another well known example is the linear approximation of the Laplacian on a regular two dimensional grid as given by
		$
		\Delta u (x,\,y) \; \dot{=} \;\frac{1}{h^2}\big[u((x-h),\,y) +u((x+h),\,y)+ u(x,y-h)+u(x,y+h)-4u(x,\,y)\big]
		$.
		This defines the so called 5-point or 1-star stencil depicted in Fig. \ref{fig:layeredGraphsAndStencil}.
		This stencil can then be used in an Jacobi iteration to compute one time step for the 2-dimensional heat equation as exemplified in Alg.~\ref{alg:2dHeatAlgo}.
		
		\begin{algorithm}[tbp]
			\For {$i\leftarrow 1$ \KwTo $k_1$} {
				\For {$j\leftarrow 1$ \KwTo $k_2$} {
					$B(i,j) =  -4\cdot A(i,j) +\left(A(i-1,j)+A(i+1,j)+A(i,j-1)+A(i,j+1)\right)$
				}
			}
			\caption{One Jacobi iteration for the 2-dimensional 1-star stencil on the input array $A$ and the output array $B$. Truncation of the stencil at the boundary is disregarded.}
			\label{alg:2dHeatAlgo}
		\end{algorithm}

		Stencil computations are typically memory bound as they perform relatively few floating point operations on the data.
		The theoretically available peak floating point performance cannot be achieved because the memory system is the bottleneck limiting the speed. 
		Hence, optimizing the memory access has become the main focus when designing high performance stencil-code.
		We employ the external memory model~\cite{av88} and the parallel external memory model~\cite{pemACM} to count the number of memory accesses.

		For stencil computations, the classical asymptotic analysis is too coarse to give interesting insights  as the majority of the I/O operations is already needed for reading the input and writing the output.
		In fact, many simple algorithms for the 5-point stencil are within a factor of $5$ of this lower bound.
		I/O operations related to the initial read of the input and the final write of the output are called \emph{compulsory} I/Os or \emph{cold} cache misses. 
		All other I/Os are called \emph{non-compulsory} I/Os or \emph{capacity} misses (because they are unnecessary for sufficiently large main memory $M$).


		This work examines the constant of the leading term of the non-compulsory I/Os caused by one update of the grid according to the $s$-star stencil in the external memory model and the parallel external memory model.
		Any na\"ive algorithm evaluating a stencil achieves the correct asymptotics for the total number of I/Os.
		Analyzing the constant of the total number of I/Os allows to distinguish na\"ive algorithms and whether they exploit locality at all.
		But to quantify how well locality is exploited, we need to drill down to the leading term of non-compulsory I/Os.
		The asymptotics of the leading term of the non-compulsory I/Os determines whether an algorithm exploits the data layout it works on (as the presented \emph{memory efficient band algorithms}) or is not able to do so (as standard blocked algorithms working on a row- or column-major data layout).
		Examining the constant of the leading term of the non-compulsory I/Os allows to determine the efficiency of different data layouts and the corresponding algorithms.
			
		In two dimensions, matching lower and upper bounds are given closing a multiplicative gap of 4. 
		In three dimensions the provided bounds match up to a factor of $\sqrt{2}$ improving the known results by a factor of $2 \sqrt{3}\sqrt{B}$.
		For dimensions $d\geq 3$, the lower bound is improved between a factor of $4$ and $6$.
		For arbitrary dimension~$d$, 
		the first analysis of the constant of the leading term of the non-compulsory I/Os is presented. 
		For $d\geq 3$ the lower and upper bound match up to a factor of $\sqrt[d-1]{d!}$.
		For high dimensions~$d$, this can be approximated as $\sqrt[d-1]{d!} \approx \frac{d}{e}$. 


		The lower bound combines a round argument with an isoperimetric inequality to bound the progress that can be achieved in each round.
		The isoperimetric results needs to be deduced and adapted carefully, using a pathwidth argument amongst others, such that no constant factors are lost in the analysis.
		
		To analyze the upper bounds in a uniform manner, the framework of (memory efficient) band algorithms is introduced.
		In summary, a $d-1$ dimensional sweep shape is swept through the grid creating work bands which are evaluated one after another.
		To be able to evaluate the whole grid these work bands need to overlap.
		This overlap gives rise to the data layout which is employed to optimize the constant of the leading term of the non-compulsory I/Os.
		In particular, the data layout is organized by so called $k$-intersections which store vertices that are used in succession in contiguous memory.

		Stencil computations can be thought of as a sparse-matrix vector multiplication where the large, sparse matrix is defined by the stencil.
		The structure of the sparse matrix determined by the stencil can be exploited to find memory efficient algorithms performing this multiplication.
		Hereby, it is crucial to find permutations of the matrix that create dense blocks as dense blocks can be applied efficiently.
		The algorithms and data layouts presented in this paper provide such reorderings.


		The implications of the theoretical results of this paper for actual stencil computations are limited due to three main reasons:
		First, the compulsory misses dominate the non-compulsory misses by a factor of $\thet{\sqrt[d-1]{M}}$, i.e. for every non-compulsory cache miss there are $\thet{\sqrt[d-1]{M}}$ compulsory misses.
		Hence, the presented optimizations may result in only marginal performance improvements 
		unless the cache is very small, e.g. a register.
		
		Second, the implications of the results are limited as we examine one single update of the grid according to the stencil. 
		Optimizing multiple stencil passes at once is common in practice and can be modeled by introducing a temporal dimension.
		When a temporal dimension is included the compulsory I/Os no longer dominate the non-compulsory ones  as the number of computations is the product of the spatial and temporal dimensions whereas the number of compulsory I/Os solely depends on the spatial dimensions. 
		Hence, for multiple stencil passes, increasing temporal (and spatial) locality can speedup the code significantly.
		We do not analyze a time step setup in this paper as it introduces a directed dimension and hence changes the structure of the computation graph, the stencil defining the neighborhood of a set and hence the isoperimetric sets and inequalities. 
		To transfer the lower bounds to a time step setting an isoperimetric inequality for the directed, multi-layer time step computation graph needs to be derived.
		With this new inequality, the rest of the argument can be applied as before.
		The best upper bounds in two (Diagonal Band Algorithm) and three dimensions (Hexagonal Band Algorithm) should easily be transferable to a time step setting as their structure is compatible with the setting of one temporal and one respectively two spatial dimensions because of the alternating sweep sequence.
		Parallelizing the algorithms is, however, more difficult when there is a temporal dimension.
		All in all, the techniques for the lower as well as the upper bounds derived in this paper are also applicable in a time step setup.

		
		Third, the implications of the results are limited as we focus on the theoretical I/O behavior of stencil computations. 
		We limit the analysis to two levels of the memory hierarchy and assume, as usual for the external memory model, a fully associative cache.
		When implementing stencil computations, one may need to be careful about theoretic I/O behavior, which this work studies, and optimizations that improve runtime on current computer architectures.
		Some of the data layouts presented in this work, in particular the 3-dimensional hexagonal band layout, are complex and may not be suitable for implementation as they could require sophisticated padding schemes or might interfere with prefetching, etc.
		In general, they could increase runtime although the number of cache misses is reduced.
		Implementing the presented algorithm is out of the scope of this work and we limit ourselves to improving the theoretic I/O behavior of stencil computations. 
		However, diagonal hyperspace cuts, similar to the ones proven optimal in this work, are often employed in empirical work to select suitable substructures for computation.

		The paper proceeds by defining the theoretical model and problem precisely, presents the results and discusses related work.
		In \S\ref{sec:lowerBounds} the lower bounds are derived.
		The lower bound section first proofs an isoperimetric result and analyzes the isoperimetric sets.
		After mentioning the relevant concepts of pathwidth these findings are assembled to the lower bound.
		The algorithmic framework for the upper bounds as well as the notation required for the upper bounds is given in \S\ref{sec:upperBoundsFramework}.
		Then, the upper bounds which are inspired by the lower bounds are described in \S\ref{sec:upperBoundsAlgo}.
		The paper concludes in \S\ref{sec:discussion} with a discussion of alternative theoretical models and a summary of the results and problems that remain open.


		\subsection{Problem Definition}
			
			\label{sec:problemDef}
			
			\subsubsection{Computational Model}
			The \emph{computational model} we consider is the external memory  (EM) model or I/O model of Aggarwall and Vitter~\cite{av88}.
			This model is the generalization of Hong and Kung's red-blue pebble game~\cite{hk81} to arbitrary block size $B$.
			There are two levels of memory, an \emph{external memory} of infinite size on which all data is stored initially, and an \emph{internal memory} of size $M$ to which the data has to be loaded to perform computations.
			The external memory is organized in \emph{blocks} of size $B$.
			An I/O operation is the transfer of one block of data of size $B$ from external to internal memory (read) or from internal to external memory (write).

			We classify the I/Os into \emph{compulsory I/Os} (cold misses), which account for the first access to a block and writing the final output, and \emph{non-compulsory I/Os} (capacity misses).
			Non-compulsory I/Os are due to the limited size of the internal memory. 
			In the I/O model the cache is always assumed to be fully associative and hence conflict misses do not occur.

			In the I/O model, the internal memory is managed explicitly.
			This has three major implications.
			First, the cache replacement strategy can be specified by the user is hence assumed to be optimal.
			Second, it is up to the user how data is evicted from internal memory.
			The data can either be stored back to external memory or deleted within internal memory without causing an I/O operation.
			Third, data can be written directly to blocks in external memory, without loading these blocks to internal memory first.

			We further assume that all I/Os are simple, i.e. data elements are moved instead of copied between internal and external memory. 
			While this facilitates the derivation of our bounds, this assumption is not crucial and matching bounds assuming simple I/Os translate to matching bounds using non-simple I/Os as we discuss in \S\ref{sec:upperBoundsAndDiscussion}.



			\subsubsection{Task}
			The task we consider is t perform one update of all values of a grid according to the $s$-star stencil.
			For the basic notation let $[k]$ abbreviate $\{0,\dots, k-1\}$, let $[k_1] \times \; \dots \; \times [k_d]$ denote the $d$-dimensional \emph{grid} and $\mathbb{Z}_{k_1} \times \; \dots \; \times \mathbb{Z}_{k_d}$ the $d$-dimensional \emph{torus} of side lengths $k_i$.
			Denote by $||\cdot ||_1$ the $\ell^1$-norm which is defined as usual for the grid.
			For an element $v~\in~\mathbb{Z}_{k_1} \times \; \dots \; \times \mathbb{Z}_{k_d}$
			of the torus it is given by $||v||_1 = \sum_{i=1}^d \min\{$ \mbox{$(-v_i \mod k_i),$} $ (  v_i \mod k_i )\}$ (assuming $(v_i \mod k_i) \in \{0, \, \dots \, , \, k_i-1\})$.
%
%
%
			Denote by $V$ the vertices of the grid.

			We consider out-of-place computations, hence there is an input layer $V_{\inp} := V \times \{\inp\, \}$ and an output layer $V_{\out} := V \times \{\out\,\}$ of the grid. 
			Initially, each vertex of the input layers stores a value while the output layer is empty.
			At the end of the computation, the values updated according to the stencil have to be stored in the output layer $V_{\out}$.
			The function which maps the values of $V_\inp$  to $V_\out$ is described by a stencil.
			The task is to evaluate the stencil for all points of the output layer, i.e. to compute all values of the output grid and to write these results to external memory.
			
			We consider so called \emph{$s$-star stencils}.
			Denote by $v_{\inp} \in V_{\inp}$ and $v_{\out}\in V_{\out}$ corresponding vertices of the input and output layer, i.e. the first $d$ coordinates of these vertices of the $d$-dimensional grid or torus are identical.
			The $s$-star stencil $S_s$ for a vertex $v_\out\in V_\out$ is defined as all vertices within distance $s$ from $v_\inp$, $S_s(v_\out) := \{ w \in V_\inp:\, ||w-v_\inp||_1 \leq s\}$.
			This also implies that the stencils are cut off at the boundary of the grid as shown in Fig.~\ref{fig:layeredGraphsAndStencil}.
			For the asymptotic notation we assume throughout the paper that $s$ is a small constant.
			
			The computation graph  $(V_{\inp}  \mathbin{\dot{\cup}}  V_{\out} ,\, E)$ for the $s$-star stencil is obtained by connecting the input layer to the output layer by adding edges $(w,v_\out)$ for all vertices $w \in S_s(v_\out)$ and repeating this process for all vertices in $v_\out \in V_\out$, $E := \left\{ (w,\, v_{out})\in V_{in} \times V_{out}:\, ||w - v_{in}||_1 \leq s \right\}$.

			The 1-star stencils are the most common stencils.
			Since upper (lower) complexity bounds for the $s$-star stencil induce upper (lower) bounds for all stencils which are subsets (supersets) of the $s$-star stencil meaningful choices also include $s=2$ and $s=3$. 
			
			Working out-of-place on an input and output layer of the grid is not essential for neither the lower nor the upper bounds but simplifies their analysis.
			When we later argue about the stencil computations, the distinction between input and output layer is less strict. 
			We say to evaluate a vertex $v$ of the grid (torus) $V$ when we compute the stencil for $v_\out$ and have the input $S_s(v_\out) \subset V_\inp$ in internal memory.
			\S\ref{sec:upperBoundsAndDiscussion} discusses the implications when we want to work in-place.

			We consider computing the value for one grid point $v_{out}\in V_{out}$ as an \emph{atomic} operation.
			This means that all input required to compute $f(v_\out)$, namely $S_s(v_\out)$, needs to reside in internal memory to do the calculation and partial computations are not allowed. 
			Refer to \S\ref{sec:upperBoundsAndDiscussion} for a discussion of this assumption.

		\subsection{Results}
			\label{sec:results}
			This work examines the leading term of the non-compulsory I/Os of the $s$-star stencil.
			In two dimensions, matching lower and upper bounds are given closing a multiplicative gap of 4.
			In three dimensions the provided bounds match up to a factor of $\sqrt{2}$ improving the known results by a factor of $2 \sqrt{3}\sqrt{B}$.
			For dimensions $d$ bigger than three, the lower bounds are improved between a factor of $4$ and $6$.
			For arbitrary dimension~$d$, 
			the first analysis of the constant of the leading term of the non-compulsory I/Os is presented. 
			For $d\geq 3$ the lower and upper bound match up to a factor of $\sqrt[d-1]{d!}$.
			For high dimensions~$d$, this can be approximated as $\sqrt[d-1]{d!} \approx \frac{d}{e}$.

			We use the following assumptions for the asymptotic analysis.
			The dimension $d$ is assumed to be fixed.
			Given $d$, we assume that there is an abstract parameter $n$ governing the size of our problem.
			In particular, $n$ is the parameter which goes to infinity in the $\mathcal{O}$-notation.
			All other parameters of the problem, the grid sizes $k_i$ ($1 \leq i \leq d$), the size of the internal memory $M$ and the block size $B$ are going to depend on $n$.
			Hence, when we write $k_i$ we actually mean $k_i(n)$.
			The same holds for $M$ and $B$ and we assume that $k_i(n)$, $M(n)$ and $B(n)$ are all positive, non-decreasing functions.
			The grid sizes $k_i(n)$ are assumed to ordered by size, i.e. $k_1(n) \geq k_2(n) \geq \dots \geq k_d(n)$.
			Further, we assume $\frac{k_d(n)}{M(n)} \stackrel{ n\to \infty}{\longrightarrow} \infty$ and a weak tall cache assumption, namely $\frac{M(n)}{B(n)}\stackrel{ n\to \infty}{\longrightarrow} \infty$.
			In other words, $M(n) = \lo{k_d(n)}$ and $B(n)= \lo{M(n)}$.
			We regard everything that grows slower than the leading term of the non-compulsory I/Os as lower order terms.
			Terms that solely depend on $d$ or $s$ are regarded  constant.

%
			
			Denote by $C_s(k_1,\, \dots , k_d)$ the number of simple I/Os to evaluate the $s$-point stencil on $[k_1] \times \, \dots \, \times [k_d]$.
			Then the following holds in the serial case: 
 			\begin{align*}
				C_s(k_1,\,k_2) &=  \left( 2+\frac{4s^2}{ M}
				\cdot\left.
				\begin{cases}
					\;1+\;\mathcal{O}\left( \frac{B}{M} +\frac{M}{k_1}\right)\\
					\vspace{-1em} \\ 
					\;1-\;\mathcal{O}\left( \frac{1}{M} +\frac{M}{k_2}\right) 
				\end{cases}\ac \right\} \right) \cdot \frac{k_1k_2}{B}\\
				%
				C_s(k_1,\,k_2,\, k_3) &=    \left( 2+ \frac{8}{\sqrt{3}}\cdot\frac{s^{3/2}}{ \sqrt{M}}
				\cdot \left.
				\begin{cases}
					\; \sqrt{2} + \mathcal{O}\left( \sqrt{\frac{B}{M}} \right) \\
					\vspace{-1em}\\
					\; \; \; \, 1 \, - \mathcal{O}\left( \frac{1}{\sqrt{M}}+\frac{\sqrt{M}}{k_3}\right) 
				\end{cases}\ac \right\}  \right) \cdot \frac{k_1k_2k_3}{B}  \\
				%
				C_s(k_1, \, \dots \, , \, k_d) &= \\ &\!\!\!\!\ac \ac \ac \ac \ac = \left(2+ \frac{4s \cdot\sqrt[d-1]{2s}\cdot  (d-1)}{\sqrt[d-1]{M}}\cdot 
                                  \left.\begin{cases} \quad 1 \quad \, +\mathcal{O}\left(\sqrt[d-1]{\frac{B}{M}}\right)  \\
							\vspace{-1em} \\
                                                   \frac{1}{\sqrt[d-1]{d!}} -\mathcal{O}\left(\frac{1}{\sqrt[d-1]{M}} \!+\!\frac{\sqrt[d-1]{M}}{k_d}\right) 
											 \end{cases} \right\} \right) \cdot 
							  \frac{\prod_{i=1}^d k_i}{B}  \; .
			\end{align*}
			The bounds consist of three parts. The first part is the constant 2 accounting for the compulsory I/Os. The second part is the leading term of the non-compulsory I/Os on which this work focuses. The third part characterizes lower order terms that we do not explore further.
			The best 3-dimensional upper bound is only proven for $s\in \lrC{1,2,3}$ but should generalize to arbitrary $s \in \mathbb{N}$.

			Parallelization of both, the lower and upper bounds, to the CREW (concurrent read, exclusive write) parallel external memory model (PEM)~\cite{pemACM} is simple when we work not-in-place and the number~$P$ of processors is of order  $\mO{\frac{1}{M}\prod_{i=1}^{d-1} k_i}$. 
			Then, the complexities are reduced by a factor of~$P$.
			The lower bound of this paper is derived for $B=1$ and works in the parallel setting just as well:
			as we assume that stencil evaluations are atomic there are no intermediate results.
			Hence, we can simulate any parallel algorithm using $P$ processors on a single processor increasing the total number of I/Os by at most a factor of $P$.\footnote{
			Unlike with classical computational complexity (i.e. on PRAM), taking advantage of the combined internal memory of the PEM model of size~$P\cdot M$ enables speedups above $P$ for certain tasks.
%
			}
			This simulation implies that the lower bound in the parallel setting is by at most a factor of $P$ weaker then the serial lower bound.
%
%
			For the simulation, simply execute all computations of processor $p_1$ first, followed by all computations of processor $p_2$ and so forth until we finish with processor $p_P$.
			This serialization requires one modification which does not change the total number of I/Os.
			As we assume simple I/Os, a processor may need to store a vertex back to external memory such that a processor which is simulated later can access this vertex.
			Consider any particular vertex $x$ of the grid and say it is read $k$ times in the serialized algorithm.
			The vertex needs to be transfered back to external memory the first $k-1$ times it is evicted from internal memory.
			Hence, this vertex causes 1 compulsory read and $k-1$ non-compulsory reads  and $k-1$ non-compulsory writes.
			The same holds for the number of I/Os this vertex causes in the  parallel version of the algorithm.
			Just the processor that have to perform the non-compulsory writes change.
			As in the serial setting, the parallel lower bound can be generalized to arbitrary $B$  by the simple observation that one I/O operation affects at most $B$ elements.




			Regarding the algorithms: as we are working not-in-place, i.e. on an input and an output copy of the grid, all evaluations of stencils are independent from each other and can hence be done in parallel.
			Therefore, we use the proposed serial algorithms and merely split the computation into~$P$ contiguous parts. 
			For instance, the work band list $\wL$ can be split in parts that contain an equal number of work bands and each processor evaluates the evaluation bands corresponding to its part of the work band list. 
			The only additional non-compulsory I/Os are used to initially fill the local memory.  
			Assuming $P = \mO{\frac{1}{M}\prod_{i=1}^{d-1} k_i}$, this is a lower order term, namely the one that we analyze as the difference between the torus and the grid.


	\subsection{Related Work}

The external memory (EM) or I/O model for $B=1$, focusing on temporal locality,  was introduced by Hong and Kung as ''red-blue pebble game''~\cite{hk81}.
Aggarwal and Vitter generalized Hong and Kungs model to arbitrary $B$ taking spatial locality into account~\cite{av88}.
There are also various approaches to extend the external memory model to several layers of the memory hierarchy~\cite{av87layersBlock,alpern94umh,sen02layers}.
Arge et al. generalized the EM model to the \emph{parallel external memory model} (PEM model)~\cite{pemACM}. 
The prior models have in common that they are cache-aware:
the cache size $M$ as well as the block size $B$ are known to the algorithm and can be exploited to select subproblems that are small enough to be handled efficiently.
In contrast, Frigo et. al. introduced the cache-oblivious model~\cite{prokop99cacheObl,frigo99cacheOblivious} in which the parameters $M$ and $B$ are not known to the algorithm.
The idea is to design algorithms that work efficiently for any $M$ and $B$ and hence efficiently across different machines.

The pathwidth of graphs was studied intensely in the series of papers on graph minors by Robertson and Seymour~\cite{robertson83excludingForest}.
Pathwidth can also be modeled by a robber-cop game~\cite{seymour93}.
For more details about pathwidth and treewidth refer to Bodlaender's survey~\cite{bodlaender98apartial}.
Pathwidth is of interest to us (see~\S~\ref{sec:pathwidth}) as evaluating star-stencils on a graph has to cause non-compulsory I/Os if the pathwidth is at least $M$ (see Lem.~\ref{lem:evalStar}).
This allows to split the algorithm into rounds by the number of non-compulsory I/Os and apply an isoperimetric result to each of these rounds.

%
%
%
%
%
%
%
%
%
%
%
%
%
%

The idea of splitting an algorithm into rounds and applying a sort of isoperimetric results goes back to Hong and Kung~\cite{hk81}.
Hong and Kung use dominator sets to determine how much input has to be loaded to compute a certain round.
With that approach they derive the first I/O bounds for products of graphs. 
In particular, they proof a lower bound of $\Theta \left(\frac{1}{\sqrt[n-1]{M}}\cdot \prod_{i=1}^nk_i \right)$ I/Os for the graph that is the product of paths.
This graph, however, differs significantly from the setup we examine as the product of paths has only one single input and one single output vertex and hence the compulsory I/Os are negligible.
Using similar techniques Hong and Kung also derive bounds for problems like the Fast-Fourier-Transform (FFT) and matrix-matrix multiplication.

Hong and Kungs lower bound for matrix-matrix multiplication is extended to a distributed memory setup by Irony et al. \cite{toledo04commLowBounds}.
With this work as starting point, a series of papers studies the relation between communication costs of algorithms and expansion properties of the underlying computation graphs for various problems from numerical linear algebra.
Problems include Strassen's matrix-matrix multiplication~\cite{ballard14commCosts,ballard12graphExp,ballard12strassen}, sparse random matrix-matrix multiplication~\cite{ballard13sparseRandM}, triangular substitution, Gaussian elimination, Krylov subspace methods, LU factorizaion~\cite{solomonik14tradeoffs}, Cholesky factorization, LDLT-factorization, QR-factorizaion, the Gram-Schmidt algorithm, eigenvalue and singular value algorithms~\cite{ballard11minimizing} and in general programs that reference arrays \cite{christ13arrays}.
For further references, refer to these papers and the references therein.
This research includes lower bounds and also examines trade-offs between local computation and the required communication.
The studies focus on the asymptotic complexity of the problem and are often limited to $B=1$.
The expansion properties of the computation graph are closely related to the dominator set of Hong and Kung and the isoperimetric inequality of Bollob\'as and Leader~\cite{bollobas90torus} we use to derive our bounds for stencil computations.
By carefully adapting the isoperimetric inequality we are able to analyze star-stencil computations past the leading term of the compulsory I/Os and prove a lower bound for the constant of the leading term of the non-compulsory I/Os 


As stencil computations can be regarded as multiplying a sparse-matrix with a vector, the complexity of this problem is of particular interest.
The complexities of sparse-matrix vector multiplication have been derived for sparse random matrices of varying densities in the serial~\cite{jacob07spMv} and also the parallel setting~\cite{greiner12diss}.
Further, the complexity of multiplying the same sparse, random matrix with several vectors at once has been analyzed~\cite{greiner10bilinear}.
As one stencil gives rise to one particular sparse matrix, we can exploit the structure of our problem and hence perform better than these bounds which state the worst case complexities over all sparse random matrices of a certain density.

%


The particular I/O complexity of the $1$-star stencil has already been studied independently by Frumkin and Wijngaart \cite{frumkin02tightBounds} and Leopold \cite{leopold02relaxation,leopold02timeLoop,leopold023D} for arbitrary $B$. 
Both lines of work examine the leading term of the non-compulsory I/Os but do not analyze it with the precision presented in this paper.
The different results for the leading term of the non-compulsory I/Os are given in Table \ref{tab:boundsComparison} and have to be multiplied by the number of vertices $\prod_{i=1}^n k_i$. 
Frumkin and Wijngaart consider arbitrary dimensions but focus on the asymptotic behavior of the non-compulsory I/Os.
The lower bound uses an isoperimetric argument similar to the one presented in this article but does not exploit its full strength.
We improve these results by a factor between 4 and 6.
The upper bound focuses on the asymptotic behavior and is an existence results. 
Leopold focuses on the two and three dimensional cases.
Her lower bounds exploit a weak isoperimetric result \cite{leopold02relaxation,leopold023D} which we improve by a factor of 2 and $\frac{4}{\sqrt{3}}$ for two respective three dimensions.
The upper bounds discuss row and column layouts.
By using a data layout suited for our algorithms we decrease the upper bounds by $\frac{1}{2}$ and $\frac{2}{3\sqrt{B}}$ for two and three dimensions.
Leopold also discusses two spatial and one temporal dimension \cite{leopold02timeLoop}, which is out of the scope of this paper (also see \S\ref{sec:upperBoundsAndDiscussion}).
Further, the conference version of this work establish the lower bounds~\cite{hupp13starStencils}.
This work exceeds the previous versions  by developing the framework of \emph{memory efficient band algorithms} (\S\ref{sec:upperBoundsFramework}) to describe and analyze the upper bounds in a uniform manner.

			\begin{table}[tbp]
				\footnotesize
				\centering
				\caption{
				Comparison of the bounds for the leading term of the non-compulsory I/Os for the $1$-star stencil ($s=1$). All to be multiplied with the number of grid points $\prod_{i=1}^d k_i$.
				The best presented result as well as the previously known best result, upon which we improve, are bold.}	
				\label{tab:boundsComparison}
				\renewcommand{\tabcolsep}{0.5pt}
				 	\begin{tabular}{l|c|c|c|c}
						\textbf{Lower Bounds}& Presented Result & Frumkin and Wijngaart & Leopold &\textbf{Improvement}\\
						\hline
						\hline &&& \\[-1ex]
						Lower Bound 2D &  $\boldsymbol{\frac{4}{BM}}$ & $\frac{8}{9}\frac{1}{BM}$ & $\boldsymbol{\frac{2}{BM}}$ &2\\[2ex]
						Lower Bound 3D & $\boldsymbol{\frac{8}{\sqrt{3}}\frac{1}{B \sqrt{M}}}$ & $\frac{2}{\sqrt{3}}\frac{1}{B\sqrt{M}}$ & $\boldsymbol{\frac{2}{B \sqrt{M}}} $& $\frac{4}{\sqrt{3}}$\\[2ex]
						Low. Bnd. Arb. D &  $ \boldsymbol{\frac{4  \cdot \sqrt[d-1]{2}\cdot(d-1)}  {\sqrt[d-1]{d!}}\frac{1}{B\sqrt[d-1]{M}}} $ & $\boldsymbol{ \frac{2\cdot \sqrt[d-1]{2}\cdot d}{3\cdot \sqrt[d-1]{3\cdot (d-1)!}} \frac{1}{B\sqrt[d-1]{M}}}$& n.a.  & $\frac{6\cdot \sqrt[d-1]{3}\cdot (d-1) }{d\cdot\sqrt[d-1]{d}}$\\
						\\[3ex]
						\textbf{Upper Bounds}& Presented Result & Frumkin and Wijngaart & Leopold &\textbf{Improvement}\\\hline\hline 
						&&&& \\[-1ex]
						Upper Bound 2D & $\boldsymbol{\frac{4}{BM}}$& $ \mathcal{O}\left(\frac{1}{M}\right)$ & $\boldsymbol{\frac{8}{BM}}$&2\\[2ex]
						Upper Bound 3D & $\boldsymbol{\frac{8\sqrt{2}}{\sqrt{3}}\frac{1}{B \sqrt{M}}}$ & $ \mathcal{O}\left(\frac{1}{\sqrt{M}}\right)$ & $\boldsymbol{\frac{4\sqrt{6}}{\sqrt{B}\sqrt{M}}}$&$\frac{3}{2}\sqrt{B}$\\[2ex]
						Upp. Bnd. Arb. D & $\boldsymbol{ \frac{4\cdot \sqrt[d-1]{2}\cdot(d-1)}{B\sqrt[d-1]{M}}}$  & $\boldsymbol{\mathcal{O}\left(\frac{1}{\sqrt[d-1]{M}} \right)}$& n.a.	&$\mO{B}$
					\end{tabular}
			\end{table}

The research on optimizing stencil computations, mostly in two and three dimensions, on modern computer architectures is vast and ongoing.
As stencil computations are not compute intensive, this research focuses on improving the I/O behavior of the algorithms.
All known algorithms work on either the standard row- and column-major layout and do not adapt the data layout to the structure of the stencil as done in this paper. 
Typically, the number of cache misses is reduced by tiling.
Tiling is typically done in either the spatial dimensions alone, increasing spatial locality, or in the spatial and time dimensions, increasing spatial as well as temporal locality.

It was observed that spatial tiling alone has less and less effects due to a refined memory hierarchy and may actually interfere with prefetching techniques~\cite{kamil05memSubsystems,datta09review}.
Further, standard spatial blocking in 3 dimensions is  difficult as the blocks would need to be very small to fit in memory~\cite{rivera00tiling3d}.
Hence, Rivera and Tseng suggest to block only the 2 least significant dimensions reducing the dimensionality of the blocks by 1.
This approach is very similar to the one presented in this paper moving $d-1$ dimensional sweep shapes through the grid.
The presented results also suggest that spatial blocking alone can decrease runtime by just a small fraction as the spatial blocking addresses the leading term of the non-compulsory I/Os but not the dominating term of compulsory I/Os.

Optimizing stencil computations with a time step changes the game.
In a time step setup the compulsory I/Os no longer dominate and hence increasing  temporal (and spatial) locality can speedup the code significantly.
Keep in mind, however, that merging time steps is not applicable in many application domains~\cite{kamil05memSubsystems} and if computations should be performed between time steps~\cite{kamil06implExpl,datta09review}.
Hence, improving spatial locality alone is the more general task and always applicable.







Time skewing reorders the computations to enable parallelism and to reduce synchronization points.
The first time skewing approaches include work from Wolfe~\cite{wolfe89itSpaceTiling}, Song and Li~\cite{song99tilingTempLocality} and Wonnacott~\cite{wonnacott00timeSkewing}.
More recently, wavefront approaches have been implemented for multicore chips with shared cache~\cite{wellein09wavefront,treibig11parallelizationIterative}
In these approaches the different processor of the multicore chip update successive temporal wavefronts of the spatial block improving temporal locality.
Also, special hyperplane cuts to enable concurrent startup of the tiles have been derived~\cite{bandishti12maxParallelism}.

%
%
%

%
%
%

%


Cache oblivious algorithms for stencil computations using trapezoidal space-time cuts have been derived and analyzed for $B=1$~\cite{frigo05UPCacheOblivious} as well as implemented \cite{frigo07cacheOblivious,zeiser08cacheOblLatticeBoltzmann}.
The complexity of the cache oblivious algorithm proposed by Frigo and and Strumpen matches the lower bound of Hong and Kung~\cite{hk81} asymptotically.
Also, cache oblivious algorithms based on space filling curves and stacks have been derived and implemented~\cite{mehl06cacheObl,guenther06cacheAwareSpaceFillingCurves}.
As space filling curves can model hierarchical data structures with no memory overhead, they are also very suitable for multigrid methods which are among the most efficients PDE solvers.
%
%
%
%
%
%
%
%
%
%
%
%
%
%
%
Comparing cache aware and cache oblivious approaches it was observed that cache aware algorithms typically outperform their cache oblivious counterparts for stencil computations~\cite{kamil06implExpl,datta09review}.
Further, it was observed that cache oblivious approaches may increase runtime although they decrease memory traffic.

%
%




The literature also includes work on compiler optimization \cite{tang11pochoir,henretty13compiler} and auto-tuning.
General autotuning of sparse-matrices is provided by the ''optimized sparse kernel interface'' (OSKI)~\cite{vuduc05oski}.
For stencil computations, automatic time skewing schemes~\cite{li04automaticTilingIterative}, tiling strategies for parallel startup and execution~\cite{krishnamoorthy07parallelization}, parallelization strategies~\cite{kamil2010auto} exist.
Stencil autotuners also exist for GPUs~\cite{christen11parallelIterativeStencils}.
Datta's autotuner applies a wide range of optimizations including problem decomposition techniques, data allocation schemes, bandwidth optimizations and in-core optimizations~\cite{datta09thesis,datta08autoTune}.
The autotuners are complemented by predictive models~\cite{rahman11understanding,schafer13predictive} which can guide the optimizations.
%
%
%
%
%
%
The autotuners can build upon the polyhedron model\footnote{The polyhdedron model is also called polyhedral model or polytope model.} which provides an abstract framework to represent loop programs as computation graphs.
Applying techniques from linear programming allows to reorder this computation graph to enhance parallel execution, minimize the number of synchronization points and optimize performance.
Refer to~\cite{feautrier11poly} for a recap of the model which was stimulated from the architecture~\cite{karp67polyArchitecture} as well as the software~\cite{lamport74polySoftware} community.
%



Bisseling's survey~\cite{bisseling04parallelScientificComputing} describes tiling techniques very similar to the ones presented in this paper.
In 2 dimensions, he argues that tiling the space with diamonds ($\ell^1$ balls) has a better surface to volume ratio than tiling the space with squares.
In 3 dimensions, he proposes to tile the space with a truncated octahedron to improve the surface to volume ratio over that of cubes. 
We prove that the diamond tiling is in fact optimal in 2 dimensions and provide a lower bound (and algorithm) for the 3 dimensional case.

%
%




All mentioned stencil algorithms of the literature work on the standard row or column major data layouts.
In this paper we present algorithms that work on non-standard layouts to reduce the memory traffic.
While reordering the data to a specific layout may not be worthwhile for a single stencil sweep, reordering should pay off when the stencil is applied repeatedly.
In particular, we show that non-standard data layouts are crucial to optimize memory traffic.

	\section{The Lower Bounds}

		\label{sec:lowerBounds}
		The lower bound is derived by splitting an arbitrary algorithm into rounds of a certain number of non-compulsory I/Os and applying an isoperimetric result combined with a pathwidth argument to each round.
%
%
The lower bound is first deduced assuming that an I/O operation accesses one element ($B=1$) and is then generalized for arbitrary $B$.
		This section first introduces some notation, mainly from \cite{bollobas90torus}, to then state the required isoperimetric result. Thereafter the isoperimetric sets are examined and results concerning pathwidth summarized. These findings are then merged to derive the lower bound. 

		\subsection{Notation: Fractional Sets, Boundary and Core}

			The notation necessary to prove and apply the isoperimetric result includes fractional systems, the notion of weight, boundary and interior of these systems and fractional balls as special systems.
			A fractional system or simply system $f$ is a function from $\mathbb{Z}^d_k$ or $\mathbb{Z}^d$ to the unit interval $[0,\,1]$. For $f:\mathbb{Z}^d \to [0,\,1]$ the function can take non-zero values only for a finite number of grid points. The weight $w$ of a system $f$ is $w(f) = \sum_{x \in \mathbb{Z}^d_k}f(x)$ or $w(f) = \sum_{x \in \mathbb{Z}^d}f(x)$ according to the domain of $f$. A fractional system $f$ on $\mathbb{Z}^d_k$ or $\mathbb{Z}^d$ is therefore a generalization of a subset $S$ of $\mathbb{Z}^d_k$ or $\mathbb{Z}^d$ respectively. If a fractional systems $f$  takes just the values 0 and 1, then $f$ is naturally identified with the set $S = f^{-1}(1)$  and the weight $w(f)$ is the cardinality of $S$.
			The closure $\partial f$ of a system $f$ is given by
			\begin{equation*}
				\partial f(x) = 
				\begin{cases}
					1, &f(x) > 0 \\
					\max_{||x-y||_1 =1}\{ f(y) \}, & f(x) = 0 
				\end{cases} 
				\enspace . 
			\end{equation*}
			Similar to the closure we define the \emph{inner core} $\Delta f$ of $f$ by
			\begin{equation*}
				\Delta f(x)  = 
				\begin{cases} 0 , &  f(x)<1\\
				\min_{||x-y||_1 =1}\{ f(y)\}, &  f(x) =1
				\end{cases}
			\end{equation*}
			 and the \emph{inner-$s$-core} by applying the operator repeatedly, $\Delta_s f = \underbrace{\Delta \dots\Delta}_{s \mbox{ times }} f $. This is now used to define the \emph{inner-$s$-boundary} by $\Gamma_s f(x)=f(x) -\Delta_s f(x)$\enspace. The fractional $\ell^1$--ball $b^{(r,\,\alpha)}_y$ of radius  $r \in \mathbb{N}_0$, $0\leq r \leq \frac{k}{2}$, surplus $\alpha \in (0,\,1)$ and center $y \in \mathbb{Z}^d_k$ is defined as 
			\begin{equation*}
				b^{(r,\,\alpha)}_y(x):= 
				\begin{cases}
					1, &||x-y||_1 \leq r \\
					\alpha, &  ||x-y||_1 = r+1  \\
					0, & ||x-y||_1 > r+1
				\end{cases}	\enspace \enspace .
			\end{equation*}
			For $0 \leq v \leq k^d$ we also use the notation $b^v_y$ which describes the unique ball of weight $v$ and center $y$. For the isoperimetric inequalities the centers of the balls are irrelevant and hence we omit the subscript $y$ when it is not needed.

		\subsection{The Isoperimetric Result}
			\label{sec:isoResult}
			
			The goal of this section is to prove the isoperimetric result given in Theorem~\ref{thm:isoZn}.
			An isoperimetric inequality states how many vertices can be enclosed by a fixed number of boundary vertices. The optimal sets in this sense are called isoperimetric sets and, as proven by Bollob\'as and Leader \cite{bollobas90torus}, the isoperimetric sets in $\mathbb{Z}_k^d$ are (fractional) $\ell^1$--balls.\footnote{It is known that the isoperimetric sets in the continuous domains $\mathbb{R}^d$ are $\ell^2$ balls.}
			Precisely, Bollob\'as and Leader have proven that $\ell^1$ balls have the smallest closure of all systems of the same weight.
			
			\begin{thm}[An isoperimetric inequality on the discrete torus] {\ \\}
				\label{thm:bollobasTorus}
				Let $k \geq 2$ and even, let $f$ be a fractional system on $\mathbb{Z}^d_k$. Then $w(\partial f) \geq w\left(\partial b^{w(f)}\right)$.
			\end{thm}
			\begin{proof}
				The result has been proven by  Bollob\'as and Leader  \cite{bollobas90torus} as Theorem 4.
			\end{proof}

			 We need a version of this result which allows us to bound the number of interior vertices given the number of inner-boundary vertices. This differs in two aspects from the above theorem: First, we want to look at the boundary as part of the set, the inner-boundary, and not add it in addition like in the closure. Second, we need to have a result for systems of all weights but bounded inner-boundary. This will make it necessary to translate Theorem \ref{thm:bollobasTorus} to the infinite grid $\mathbb{Z}^d$ where the boundary of the balls is growing strictly monotonic. The desired result reads:

			\begin{thm}[The boundary bounds the core on $\mathbb{Z}^d$]{\ \\} \label{thm:isoZn}
				Let $s\in \mathbb{N}$ and $f$ be a fractional system on $\mathbb{Z}^d$. For $v \in \mathbb{R}^+_0$ the following holds:
				\begin{equation}
					\left(\; w(\Gamma_{2s} f ) \leq w(\Gamma_{2s} b^v) \;\right) \Rightarrow \left(\; w(\Delta_s f) \leq w(\Delta_s b^v) \;\right) \; .
				\end{equation}
			\end{thm}
			We first prove two lemmata. 
			\begin{lem} \label{lem:boundCore}
				For a fractional system $f$ on $\mathbb{Z}^d$ the following inequality holds:
				\begin{align*}
					(\partial(\Delta f))(x) &\leq f(x)\; .
				\end{align*}
			\end{lem}
			\begin{proof}
				The claim is proven by examining the three different cases carefully.
				If $f(x) = 0$ it follows that $\partial(\Delta f) (x) = 0$ as well since all neighbors of $x$ are set to $0$ by the $\Delta$-operator.
				When $0< f(x)<1 $, $\Delta f (x) = 0$ and for all $y$ such that $||x-y|| = 1$ we have $\Delta f(y) \leq f(x)$ and hence $\partial(\Delta f)(x) \leq f(x)$. When $f(x) =1 $ the claim holds trivially.
			\end{proof}
			\noindent The second lemma states that balls have the largest inner-core of all systems of the same weight.
			\begin{lem}[A version of the isoperimetric inequality]{\ \\} 	\label{lem:fracBolloOutside}
				For even $k$, $s \in \mathbb{N}$ and all fractional systems $f$ on $\mathbb{Z}^d$ it holds that
				\begin{equation*}
					w(\Gamma_s f) \geq w\left(\Gamma_s b^{w(f)}\right)
				\end{equation*}
				which is by definition equivalent to 
				\begin{equation}
					w(\Delta_s f) \leq w\left(\Delta_s b^{w(f)}\right)\; .
					\label{eq:interiorBound}
				\end{equation}
			\end{lem}

			\begin{proof}
				The claim is proven by induction over $s$. First, consider the case $s=1$. If $w(f) \leq 1$ then $w(\Gamma f)  = w(f)$ and $w(\Delta f)=0$ such that the claim holds. Assume there exists some fractional system $f$ with $w(f) > 1$ such that 
				\begin{equation*}
					w(\Gamma f) < w\left(\Gamma b^{w(f)}\right) \qquad \mbox{and hence} \qquad w(\Delta f) > w\left(\Delta b^{w(f)}\right)\; .
				\end{equation*}
					By the latter and the strict monotonicity of $w\left(\partial b^{(\cdot)}\right)$ we get 
				\begin{equation*}
					w\left(\partial b^{w(\Delta f)}\right) > w\left(\partial b^{w\left(\Delta b^{w(f)}\right)}\right) \; .
				\end{equation*}
				To simplify the right hand side we use that the inner core of a ball is itself a ball and hence we can discard building the ball of it. 
				\begin{equation*}
					w\left(\partial b^{w\left(\Delta b^{w(f)}\right)}\right)= w\left(\partial \Delta b^{w(f)}\right) \; .
				\end{equation*}
				For a ball with $w(f) >1$ the closure of the inner core is pointwise equal to the ball itself. Furthermore we employ Lemma \ref{lem:boundCore}.
				\begin{equation*}
					w\left(\partial\Delta b^{w(f)}\right) = w(b^{w(f)})= w(f) \geq w (\partial \Delta f ) \; .
				\end{equation*}
				Reading this sequence of inequalities altogether yields 
				\begin{equation*}
					w\left(\partial b^{w(\Delta f)}\right) > w(\partial \Delta f )\; .
				\end{equation*}
				Since $f$ takes just a finite number of non-zero values, we can find $k$ such that all non-zero values of $f$ are in the grid $\{-k, \dots , k\}^n$ and we can embed $f$ in the torus $\mathbb{Z}_{2k+3}^n$ such that no points of $f$ touch were the grid is closed to a torus. 
				Therefore we can transfer the counterexample to the torus where it contradicts Theorem \ref{thm:bollobasTorus} for $\Delta(f)$ as fractional system and proves the claim for $s=1$.

				Let us now prove the claim for $s$ assuming it holds for $s-1$. Using the induction assumption for $\Delta f$ we arrive at
				\begin{equation*}
					w(\Delta_s f) = w(\Delta_{s-1} \Delta f ) \leq w\left(\Delta_{s-1}b^{w(\Delta f)}\right) \; .
				\end{equation*}
				Noting that $b^\cdot$, $\Delta b^\cdot$ and $\Delta_{s-1} b^\cdot$ are pointwise monotonically increasing yields that $w(\Delta_{s-1} b^\cdot)$ is monotonically increasing. Hence we can apply the result proven for $s=1$ to yield
				\begin{equation*}
					w\left(\Delta_{s-1}b^{w(\Delta f)}\right) \leq w\left(\Delta_{s-1}b^{w\left(\Delta b^{w(f)}\right)}\right) \;.
				\end{equation*}
				But the inner core of a ball is a ball itself, so we can discard building the ball of it and this simplifies to the required result
				\begin{equation*}
					w\left(\Delta_{s-1}b^{w\left(\Delta b^{w(f)}\right)}\right) =  w\left(\Delta_{s-1} \Delta b^{w(f)}\right) = w\left(\Delta_{s} b^{w(f)}\right) \;.
				\end{equation*}
			\end{proof}

			Since the weight of the inner-$s$-core of a ball is monotonically increasing with the weight of the ball, this result can be used to deduce the implication
			\begin{equation*}
				\left( w(f) \leq v \right) \Rightarrow \left(\; w(\Delta_s f) \leq w(\Delta_s b^v) \;\right)  \; .
			\end{equation*}
			Nevertheless, we run into problems when bounding the weight of a ball given that its inner-boundary is bounded. The inner-boundary of balls is only monotonically increasing until $v \approx \frac{k^n}{2}$ and thereafter monotonically decreasing. To overcome this problem, Theorem \ref{thm:isoZn} transfers the results to the infinite grid, where the inner-$s$-boundary of balls is monotonically increasing with respect to the weight of the ball.

			\begin{proof}[Proof of Theorem \ref{thm:isoZn}]
				The proof is split into two parts. $\left(\; w(\Gamma_{2s} f ) \leq w(\Gamma_{2s} b^v) \;\right) \Rightarrow \left(\; w(f) \leq v \;\right) \;$ is first proven by contraposition. Hence, we first prove 
				\begin{equation*}
					\left(\; w(\Gamma_{2s} f ) > w(\Gamma_{2s} b^v) \;\right) \Leftarrow \left(\; w(f) > v \;\right) \enspace.
				\end{equation*}
				From Lemma \ref{lem:fracBolloOutside}, namely $w(\Gamma_s f ) \geq  w(\Gamma_s b^{w(f)})$, and the observation that the weight $w(\Gamma_s b^v)$ is strictly monotonically increasing with respect to $v$ on $\mathbb{Z}^n$ follows
				\begin{equation*}
					w(\Gamma_s f ) \geq  w(\Gamma_s b^{w(f)}) >  w(\Gamma_s b^v)
				\end{equation*}
				Since $s$ was arbitrary it also follows that $w(\Gamma_{2s} f ) >  w(\Gamma_{2s} b^v)$ which establishes the first part.

				Employing Lemma \ref{lem:fracBolloOutside} again and noting that $w(\Delta_s b^v)$ is monotonically increasing with respect to $v$ yields $\left( w(f) \leq v \right) \Rightarrow \left(\; w(\Delta_s f) \leq w(\Delta_s b^v) \;\right)$ and the proof is complete. 
			\end{proof}

		\subsection{The Size of the $\ell^1$--Ball and its Boundary}

			\label{sec:ballAndBound}
			This section derives the asymptotic expansion for the number of vertices of a ball and its inner-boundary  in $\mathbb{Z}^d$ with respect to the radius $r$,
			\begin{equation}
				w\left(b^{(r,0)}\!\right) =  \frac{2^d}{d!}\cdot r^d + \mO{r^{d-1}}
				\label{eq:balls}
			\end{equation}
			and
			\begin{equation}
				w\left(\Gamma_1 b^{(r,0)} \right) =  \frac{2^d}{(d-1)!}\cdot r^{d-1} + \mO{r^{d-2}} \;.
				\label{eq:ballsBound}
			\end{equation}
			The dimensions $d$ is assumed to be constant.
			As long as the sides of the torus or grid are big enough, $k \geq 2(r+1)$, the formulas apply there also.
			Note that all lower order terms have positive coefficients.

			We derive these formulas by recursing over the dimensions. Hence it is useful to introduce the notation $b_d^{(r,0)}$ for the ball of radius $r$ in $d$ dimensions. The $\ell^1$--ball of dimension $d$ consists of smaller balls of one dimension less, namely the level sets in the new dimension:
			\begin{equation}
				b_d^{(r,0)} = b_{d-1}^{(r,0)} + 2 \cdot \sum_{l = 0}^{r-1} b_{d-1}^{(l,0)} \; .
				\label{eq:recBalls}
			\end{equation}
			Another simple fact is $b^{(r,0)}_n = b_n^{(r-1,0)}+\Gamma b_n^{(r,0)}$ which yields when combined with \eqref{eq:recBalls}
			\begin{equation}
				\Gamma b_{d}^{(r,0)}= b_{d-1}^{(r,0)}+b_{d-1}^{(r-1,0)}\; .
				\label{eq:recBound}
			\end{equation}
			Since $w\left(\Gamma b_d^{(0,0)}\right) =1$ for all $d \in \mathbb{N}$ and  $w\left(\Gamma b_1^{(r,0)}\right) =2 $ for $ r \geq 1$ the weight of the one-dimensional balls is given by
			\begin{equation}
				b_1^{(r,0)} = 2r +1 \; . \label{eq:b^1_k}
			\end{equation}
			Recursion \eqref{eq:recBalls} yields that $w\left(b_d^{(r,0)}\right)$ and $w\left(\Gamma b_d^{(r,0)}\right)$ are polynomials in $r$ of degree $d$ and $d-1$ with non-negative coefficients. So they can be written as 
			\begin{align*}
				w\left(b_d^{(r,0)}\right) = \sum_{i = 0}^d  \alpha_{d,  i} \cdot r^i  \quad \mbox{and} \quad
				w\left(\Gamma b_d^{(r,0)}\right) = \sum_{i = 0}^{d-1}  \beta_{d,  i} \cdot r^i  \; .
			\end{align*}
			Examining the leading term $\alpha_{d , d}$ of $w\left( b_d^{(r,0)}\right)$ yields
			\begin{align*}
				\notag w\left( b_d^{(r,0)}\right) 
				&= w\left( b_{d-1}^{(r,0)}\right)+2 \sum_{l = 0}^{r-1} w\left( b_{d-1}^{(l,0)}\right) 
				= \mathcal{O}(r^{d-1}) + 2 \sum_{l = 0}^{r-1} \sum_{i=0}^{d-1} \alpha_{d-1, \; i} \cdot l^i  = \\
				\notag &=  \mathcal{O}(r^{d-1}) +  2 \sum_{i=0}^{d-1} \left(  \alpha_{d-1, \; i} \sum_{l = 0}^{r-1} l^i \right) \leq 
				 \mathcal{O}(r^{d-1}) +  2 \sum_{i=0}^{d-1} \left(  \alpha_{d-1, \; i} \int_{0}^{r} l^i \; dl \right) =\\
				\notag & =  \mathcal{O}(r^{d-1}) +  2 \sum_{i=0}^{d-1} \left(  \alpha_{d-1, \; i} \frac{r^{i+1}}{i+1} \right) 
				= 2 \alpha_{d-1, \; d-1} \frac{r^d}{d} + \mathcal{O} (r^{d-1}) \; .
			\end{align*}
			Comparing the coefficient of the leading terms yields the recursion
			\begin{equation*}
				\alpha_{d,d} = \frac{2}{d} \alpha_{d-1, \; d-1} \;.
			\end{equation*}
			The recursion stops with  \eqref{eq:b^1_k}, namely $\alpha_{1,\;1} = 2$. Hence we get 
			\begin{equation*}
				\alpha_{d,d} = \frac{2^d}{d!} \qquad \mbox{and} \qquad w\left(b_d^{(r,0)}\right) =  \frac{2^d}{d!}\cdot r^d + \mathcal{O}(r^{d-1})\; .
			\end{equation*}
			Now  \eqref{eq:recBound}  yields
			\begin{equation*}
				\beta_{d,d-1} = \frac{2^d}{(d-1)!} \qquad \mbox{and hence} \qquad w\left(\Gamma b_d^{(r,0)}\right) =  \frac{2^d}{(d-1)!}\cdot r^{d-1} + \mathcal{O}(r^{d-2})\; .
			\end{equation*}

		\subsection{Pathwidth}

			\label{sec:pathwidth}
			We employ pathwidth \cite{robertson83excludingForest} to ensure that we are working on the ``inside'' of the torus and can treat it like the infinite grid which allows to apply Theorem~\ref{thm:isoZn}.
			
			\begin{defin}[Pathwidth \cite{robertson83excludingForest}]
				A path decomposition of a graph $G=(V,E)$ is a sequence of subsets of vertices $(X_1, X_2,\; \dots \; ,X_r)$, called bags, such that
				\begin{enumerate}
					\item $\bigcup_{1\leq i \leq r} X_i = V$.
					\item for all edges $(v,w) \in E$ there exists an $i \in \{ 1, \; \dots\; , r \}$ such that $v\in X_i$ and $w \in X_i$.
					\item for all $i,j,k$ such that $1\leq i\leq j\leq k \leq r$ it holds that $X_i \cap X_k \subseteq X_j$.
				\end{enumerate}
				The width of a path decomposition  $(X_1, X_2,\; \dots \; ,X_r)$ is $\max_{1\leq i \leq r}|X_i|-1$. The width of a graph $G$ is the minimum width over all possible path decompositions of $G$.
			\end{defin}
			\noindent Condition (3) implies that a vertex can only be in a consecutive block of bags and not reappear after it has been removed from a bag once. 

			\begin{lem}\label{lem:evalStar}
				Let $G$ be a graph. Denote by $M$ the size of the internal memory. If $\pathwidth(G) \geq M$, then any algorithm evaluating the $s$-star on $G$ has to cause non-compulsory I/Os.
			\end{lem}
			\begin{proof} 
				We will prove the contraposition of the claim: If there exists an algorithm evaluating the $s$-star stencil on $G$ with only compulsory I/Os then $\pathwidth(G) < M$.

				If we can evaluate the $s$-star stencil on $G$ with internal memory of size $M$ and without loading a vertex twice this immediately induces a path decomposition with bags of size at most $M$. The bags are the different sets of elements the internal memory is containing at different stages of the algorithm and hence $\pathwidth(G) \leq M-1$.
			\end{proof}

			\begin{lem} \label{lem:gridTorusNC}
			  Evaluating the $s$-star stencil on a two dimensional grid or torus with $\min \{k_1,k_2\} \geq M$ has to cause non-compulsory I/Os.
			\end{lem}
			\begin{proof}
				Since the two dimensional grid $[k_1]\times[k_2]$ has pathwidth $\min\{k_1,k_2\}$ (Corollary 89 of \cite{bodlaender98apartial}) the claim follows from Lemma \ref{lem:evalStar}.
			\end{proof}
			
			Pathwidth can also be modeled by a \emph{robber and cop game} \cite{seymour93}.
%
				\noindent The robber and cop game is played on an arbitrary undirected graph like the grid or the torus. 
				Initially $p$ cops are placed on the vertices of the graph and afterwards the robber chooses its initial position. 
				The robber is visible to the cops during the game and the game proceeds in rounds. First the cops announce were they want to be placed in the next round. 
				Then every cop that wants to move boards a helicopter.
				While the cops are moving in the air the robber is allowed to move to an arbitrary vertex of the graph if he can reach it without running into a cop. 
				Thereafter the cops land and the robber escapes in that round if no cop lands on the vertex the robber is standing on.
				The game then continues with the next round. 
				If there is a strategy so that the robber is able to escape the cops for an infinite number of rounds we say that the robber wins.
			
			The following implication holds \cite{seymour93}: When the robber cop game is played with $p$ cops on a graph $G$ and if there is a strategy so that the robber wins, $G$ has to have pathwidth bigger than $p-1$.
			\begin{lem} \label{lem:subgraphRowsCols}
				If the subgraph $H$ of a two dimensional grid or torus consists of $p+1$ complete rows and complete columns, then $\pathwidth(H) \geq p$.
				\label{lem:robCopOnComplRowsAndCols}
			\end{lem}
			\begin{proof}
				To prove the claim we give a strategy in the robber and cop game such that the robber wins against $p$ cops for any strategy the cops have. 
				Since there are $p+1$ complete rows and columns in $H$, the robber is free to start in a row which is empty, after the initial placements of the cops. When the cops announce their move, there will be a free column in the next configuration. Since the robber is in a free row, it can move to this column which is free in the next configuration. The game now proceeds with rows and columns interchanged. The robber always escapes from a free row to a free column and vice versa. 
			\end{proof}

			\begin{lem} \label{lem:subgraphNCs}
				Let $M$ be the size of the internal memory. If the subgraph $H$ of a two dimensional grid or torus consists of $M+1$ complete rows and complete columns any algorithm evaluating the $s$-star stencil on $H$ has to cause non-compulsory I/Os.
			\end{lem}
			\begin{proof}
				The claim follows from the combination of Lemmata \ref{lem:subgraphRowsCols} and \ref{lem:evalStar}.
			\end{proof}

		\subsection{Splitting into Rounds and Deducing the Lower Bound}

			To derive the lower bound it is left to describe how to split an algorithm into rounds. Therefore assume an arbitrary algorithm evaluating the $s$-star stencil on $\mathbb{Z}_{k_1} \times \; \dots \; \times \mathbb{Z}_{k_d}$ is given.
			As $k_i(n)= \Om{M(n)} \; \forall i$ it follows that $\min\{k_1, k_2\}\geq M$ for almost all $n$.
			In these cases, the algorithm causes non-compulsory I/O operations by Lemma~\ref{lem:gridTorusNC}.
			We can count these operation and split the algorithm into rounds of $c$ non-compulsory I/Os. $c$ denotes the round length and hence all rounds except the last one cause $c$ non-compulsory I/Os. 
			This approach is similar to the idea presented by Hong and Kung~\cite{hk81} and therefore we call the rounds Hong-Kung rounds. 
			
			To apply the isoperimetric inequality we need to establish a link between the inner-core, the inner-boundary and the rounds. 
			Choose one of the Hong-Kung rounds and denote with $S$ the set of vertices which are in internal memory at some point of this round.
			Let $\Transfer(S)$ be the \emph{transfer vertices} of $S$, i.e. vertices which are also present in internal memory during other rounds. Precisely, a vertex is a transfer vertex if at least one of four cases applies:
			\begin{itemize}
				\item The vertex is transferred from the previous to the current round by residing in internal memory at the beginning of the current round.
				\item The vertex has been written back to external memory in a preceding round and is read again in the current round.
				\item The vertex is written from internal to external memory in the current round to be read again in a subsequent round.
				\item The vertex is transferred from the current to the proceeding round by residing in internal memory at the end of the current round.
			\end{itemize}
			We denote further $\Eval(S)$ as the \emph{evaluated vertices} which are all vertices of $S$ for which the $s$-point stencil is evaluated in the current round. The following two observations relate these sets to the inner-core and the inner-boundary:
			\begin{equation}	\label{eq:subsetTransfer}
				\Gamma_{2s}(S) \subset \Transfer{}(S)
			\end{equation} 
			\begin{center}and\end{center}
			\begin{equation} \label{eq:subsetEval}
				\Eval(S) \subset \Delta_{s}(S) \; .
			\end{equation}
			A vertex can only be evaluated in a round if all its neighbors within distance $s$ are in $S$ as well. $\Delta_s(S)$ consists of exactly these vertices. Equivalently $\Gamma_s(S)$ are the vertices which cannot be evaluated in round $S$. Take any $x\in \Gamma_s(S)$. All vertices which are within distance $s$ from $x$ need to be in the round in which $x$ is evaluated. Hence they need to be transferred. The set of all vertices of $S$ within distance $s$ from any of the vertices of $\Gamma_s(S)$ is $\Gamma_{2s}(S)$. Therefore these vertices are a subset of the transfer vertices.

			Furthermore, we can give an upper bound for the number of transfer vertices of a round. 
			At the beginning and at the end of a round there are at most $M$ vertices in internal memory. Together these account for at most $2M$ transfer vertices.
			The only other way a vertex can be a transfer vertex is that it has been rewritten to external memory in a previous round and is reloaded in the current round or rewritten to external memory in the current round to be reloaded in a subsequent round. So either the reload or write of the vertex causes a non-compulsory I/O. Since there are at most $c$ non-compulsory I/Os per round, the total number of transfer vertices is at most $2M+c$,
			\begin{equation}
				w(\Transfer{}(S)) \leq 2M +c \; .
				\label{eq:transferLimited}
			\end{equation}
			

%
%
%
			We can embed  $S$ in the infinite grid since the torus is assumed to be large.
			Denote by $e_i$ the vector of the $i$.th unit direction.
			From $k_i(n) = \Om{M(n)} \; \forall i $ it follows that $k_1,\,  k_2 \geq 2M+c+(M+1)$ and $k_i \geq 2M+c+1$ for $i \in \{3, \dots , d\}$ for almost all $n$.
			In these cases we know by \eqref{eq:transferLimited} that the vertices of (at least) $M+1$ hyperplanes of normal $e_1$, $M+1$ hyperplanes of normal $e_2$ and one hyperplane of normal $e_i$ ($3\leq i \leq d$) do not belong to $\Transfer(S)$.
			The union $U$ of these hyperplanes forms a connected component in \mbox{$\mathbb{Z}_{k_1} \times \dots \times \mathbb{Z}_{k_d}$}. 
			As a connected component $U$ could either be a subset of $S\setminus \Transfer(S)$ or disjoint from $S$. 
			Assume that $U\subset \left(S\setminus\Transfer(S)\right)$.
			Taking the union of all hyperplanes of normal $e_1$ and normal $e_2$ and intersecting them with all other hyperplanes results in a subset $H \subset U$ of a two dimensional torus of at least $M+1$ complete rows and columns. 
			By Lemma \ref{lem:subgraphNCs} evaluating the $s$-star stencil on $H$ has to cause non-compulsory I/Os. 
			But evaluating the $s$-star stencil for vertices of $S\setminus \Transfer(S)$ does not cause non-compulsory I/Os by definition.
			Hence the case  $U\subset \left(S\setminus\Transfer(S)\right)$ is not possible and it follows that $U$ is disjoint from $S$.
			Therefore, at least one hyperplane of each normal direction $e_i$ ($1\leq i \leq d$) is disjoint from $S$.
			Deleting these hyperplanes allows to embed $S$ in the infinite grid $\mathbb{Z}^d$.

			Treating $S$ as a subset of the infinite grid allows to apply Theorem \ref{thm:isoZn} and yields the lower bound.
			Denote with $v_0$ the weight such that 	
			\begin{equation} \label{eq:boundEq}
				w(\Gamma_{2s}b^{v_0}) = 2M +c \; . 
			\end{equation}
			Combining \eqref{eq:transferLimited} and \eqref{eq:subsetTransfer} reads
			\begin{equation*}
				w\left(\Gamma_{2s}(S) \right)  \leq w\left(\Transfer(S) \right) \leq 2M +c = w(\Gamma_{2s}b^{v_0})\; .
			\end{equation*}
			By Theorem \ref{thm:isoZn} and \eqref{eq:subsetEval} it follows that $w(\Eval(S)) \leq w(\Delta_s S) \leq w(\Delta_s b^{v_0})$.
			Therefore, a lower bound for the evaluation of the $s$-point stencil on $\mathbb{Z}_{k_1} \times \dots \times \mathbb{Z}_{k_d}$ is given by
			\begin{equation}
				\frac{c}{w(\Delta_s b^{v_0})} \cdot \prod_{i=1}^d k_i \; .
				\label{eq:genLowBound}
			\end{equation}

			It is left to determine the round length $c$ that gives the best lower bound. Using the assumption that $s$ is small and constant we simplify \eqref{eq:boundEq} before solving.
			Denote $(r_0, \, \alpha_0)$ the radius and surplus such that $b^{v_0} = b^{(r_0, \, \alpha_0)}$. Using \eqref{eq:ballsBound}, the asymptotic expansion of $w(\Gamma_{2s}b^{v_0})$ is given by
			\begin{align}
				w(\Gamma_{2s}b^{v_0})
				=\sum_{i=0}^{2s-1} \frac{2^d\cdot (r_0-i)^{d-1}}{(d-1)!} + \mathcal{O}\left(r_0^{d-2}\right)
				& = \frac{ 2s \cdot2^d}{(d-1)!}(r_0-2s)^{d-1} +\mathcal{O}\left(r_0^{d-2}\right)\;.
				\label{eq:boundDetRoundLengthAsym}  
			\end{align}
			Since all coefficients in the lower order terms are non-negative, dropping the lower order terms before solving  \eqref{eq:boundDetRoundLengthAsym} increases $r_0$ and $v_0$, increases $w(\Delta_s b^{v_0})$ and hence weakens the lower bound~\eqref{eq:genLowBound}. Solving  \eqref{eq:boundEq} without the lower order terms yields 
			\begin{equation}
				r_0 = \sqrt[d-1]{(d-1)!\frac{2M+c}{2s\cdot 2^d}}+2s \; .
				\label{eq:ballSizeDepOnRoundLength}
			\end{equation} 
			The round length $c$ giving the strongest lower bound is chosen by plugging \eqref{eq:ballSizeDepOnRoundLength} into  \eqref{eq:genLowBound}
			and maximizing over~$c$ by setting the derivative to 0 and checking that the solution is a maximum. As the round length can be chosen arbitrarily we disregard lower order terms whens solving and choose
			\begin{equation*}
				c= 2 (d-1)\cdot M \; .
			\end{equation*}
			Using this round length in \eqref{eq:ballSizeDepOnRoundLength}, we determine an upper bound for the radius of a ball to be handled in one round as
			$
				r_0 = \sqrt[d-1]{\frac{d!}{2^d}\frac{M}{s}} +2s
			$.
			Finally, by plugging this radius into  \eqref{eq:genLowBound} and using \eqref{eq:balls} to simplify,  the lower bound reads
			\begin{align*}
				&\frac{2(d-1)M}{w\left(\Delta_s b^{(r_0, \, 0)}\right)} \cdot \prod_{i=1}^d k_i \geq 
					 \frac{2(d-1)M}{w\left( b^{(r_0, \, 0)}\right)} \cdot \prod_{i=1}^d k_i =\\
				&= \frac{2(d-1)M}{\frac{2^d}{d!}\cdot \left[ \left( \sqrt[d-1]{\frac{d!}{2^d} \cdot \frac{M}{s}}+2s \right)^d  \right]+\mO{\left(\sqrt[d-1]{\frac{d!}{2^d} \cdot \frac{M}{s}}+2s \right)^{d-1} }} \cdot \prod_{i=1}^d k_i = \\
				&= \frac{2(d-1)M}{\frac{2^d}{d!}\cdot \left[ \left( \sqrt[d-1]{\frac{d!}{2^d} \cdot \frac{M}{s}} \right)^d+\mO{M}\right] +\mO{M} } \cdot \prod_{i=1}^d k_i = \\
				&= \frac{2(d-1)M}{\frac{2^d}{d!}\cdot  \left( \sqrt[d-1]{\frac{d!}{2^d} \cdot \frac{M}{s}} \right)^d +\mO{M} } \cdot \prod_{i=1}^d k_i = \\
				&=  \left( 4s\cdot \sqrt[d-1]{2s}\cdot (d-1) \cdot \sqrt[d-1]{\frac{1}{  d!}} \cdot \frac{1}{\sqrt[d-1]{M}+\mO{1}} \right) \cdot  \prod_{i=1}^d k_i =\\
				& =  \left( 4s\cdot \sqrt[d-1]{2s}\cdot (d-1)\cdot \sqrt[d-1]{\frac{1}{  d!}} \cdot \frac{1}{\sqrt[d-1]{M}}-\mathcal{O} \left(\frac{1}{\sqrt[d-1]{M^2}}  \right) \right) \cdot  \prod_{i=1}^d k_i  \; .
			\end{align*}

			This bound was derived on the torus $\mathbb{Z}_{k_1} \times \dots \times \mathbb{Z}_{k_d}$ and we can apply it to the grid $[k_1] \times \dots \times [k_d]$ using a reduction. 
			\begin{lem}
				Any algorithm using internal memory of size $M$ and evaluating the $s$-point stencil on the grid $[k_1] \times \dots \times [k_d]$ induces an algorithm, using internal memory $M$ and evaluating the $s$-point stencil, on the torus $\mathbb{Z}_{k_1} \times \dots \times \mathbb{Z}_{k_d}$ causing at most $\mathcal{O}\left(\prod_{i=1}^{d-1}k_i\right)$ additional I/Os.
			\end{lem}
			\begin{proof}
				When the algorithm for the grid is evaluated on the torus, only the vertices close the boundary of the grid have to be treated differently. If a vertex is within $\ell^1$ distance $s-1$ in a unit direction from a bounding hyperplane of the grid, at most half of the points of the $s$-point stencil, corresponding to that unit direction, have to be read and written additionally for this vertex on the torus. Altogether the number of I/Os is at most 
				\begin{equation*}
					\frac{ b^{(s,0)}}{2}\cdot s \cdot 2 \cdot \sum_{j=1}^d\prod_{\substack{i=1\\i\neq j}}^{d}k_i = \mathcal{O}\left(\prod_{i=1}^{d-1}k_i\right) \;.
				\end{equation*}
			\end{proof}

			Furthermore, the lower bound can be generalized to arbitrary $B$  by the simple observation that one I/O operation affects at most $B$ elements.
			Hence, for the grid the total number of I/Os, including the compulsory ones, is
			\begin{equation*}
				\left( 2+ \frac{4s\cdot \sqrt[d-1]{2s}\cdot(d-1) }{\sqrt[d-1]{  d!}}\cdot \frac{1}{ \sqrt[d-1]{M}}-\mathcal{O} \left(\frac{1}{\sqrt[d-1]{M^2}}+\frac{1}{k_d}  \right) \right)   \frac{\prod_{i=1}^d k_i}{B}
			\;.
			\end{equation*}

	\section{Notation and Algorithmic Framework for the Upper Bounds}

	\label{sec:upperBoundsFramework}
		Algorithms evaluating the $s$-star stencil on the $d$~dimensional grid $[k_1] \times \dots \times [k_d]$ are presented in this section. 
		A single time step is considered, i.e. the whole grid is updated once according to the stencil.
		Hence we limit ourselves to spatial tiling.
		It has been observed that tiling is limited to spatial tiling in many application domains~\cite{kamil05memSubsystems} and if computations should be performed between time steps~\cite{kamil06implExpl,datta09review}.
		As we are working not-in-place, we always keep two copies of the grid, one with the \emph{input} values and one with the \emph{output} values  according to the stencil.
		Also, we consider simple I/Os in the sense that data items are moved between external and internal memory and hence the original value of a vertex has to be stored back if it is accessed at a later point in time.
		

		All the upper bounds have in common that a \emph{sweep shape} is moved through the grid in unit shifts in a \emph{simple sweep sequence} resulting in \emph{work bands}. 
		This approach is similar to the one employed by Leopold~\cite{leopold02relaxation}.
		The data layout is crucial to the design of a memory efficient algorithm: 
		First, the data layout and the sweep shape have to match. 
		Second, vertices that belong to different sets of work bands, these sets are going to be defined as $k$-intersections, have to be stored together.
		This section first introduces the necessary definitions and ideas for the \emph{Band Algorithms} given in \S\ref{sec:upperBoundsAlgo}  and then states and proves the algorithmic framework to analyze their complexity. 

		Many of the involved definitions and constructions are necessary as we want to analyze the constant of the leading term of the non-compulsory I/Os and derive the asymptotic behavior of the lower terms.
		Limiting the analysis to the asymptotic complexity of leading term of the non-compulsory I/Os or disregarding lower order terms would greatly simplify the analysis.
		Further, the \emph{Diagonal Band Algorithm} and the \emph{Hexagonal Band Algorithm}, the best algorithms presented in 2 and 3 dimensions respectively, employ several different unit shifts which further complicates the analysis.
		We think, however, that employing different unit shifts is crucial when matching upper and lower bounds should be designed.
		Hence, we present and analyze these algorithm to disclose design issues  which are essential for matching bounds.


	\subsection{Notation and Setup for the Upper Bounds}

		This section first introduces the necessary definitions for sweep sequences, sweep shapes, work bands, evaluation bands and $k$-intersections.
		Then, we examine the sweep shapes in more detail defining the size parameter $m$ of a sweep shape and the data layout within a sweep shape.
		Finally, we reduce the dimensionality of the problem to $d-1$ by cutting it with hyperplanes.
		Let us start with the definitions.


		The vector of the $i$.th unit direction is denoted by $e_i$.
		For a vertex $w \in \fg$ or $w \in\mathbb{Z}^d$ the $i$.th component is denoted by $w_i$.
		Recall that the definition of the $s$-star stencil of a vertex $w \in \fg$ is given by
		\begin{equation}
			\sStencil{w} = \{v \in \fg: ||v -w ||_1 \leq s \} \;.
		\end{equation}

		A  \emph{simple sweep sequence} or just \emph{sweep sequence} $\sweepSeq$ of length $k$ is the sequence of the first $k$ unit directions $e_i$ ordered by increasing $i$.
		We denote by $\delta_i \in \sweepSeq$ the $i$.th element of the simple sweep sequence. 
		
		A sweep shape $\sH$ is a subset of vertices of the infinite grid $\mathbb{Z}^d$.
		All considered sweep shapes are the integral points of $d-1$ dimensional polygons lying in a hyperplane of normal $\sum_{x\in \sweepSeq}x$.
		In two dimensions the sweep shapes are therefore simply line segments.
		In three dimensions, the employed sweep shapes can be described by squares, diamonds and hexagons.
		Therefore, you can think of the sweep shapes as convex and $d-1$ dimensional objects.
		However, we do not define the notions of convexity and the dimension of a set of points for the discrete setting. 
		The distance of a sweep shape $\sH$ from the origin is defined as the $\ell^2$ distance of the hyperplane containing $\sH$ from the origin.

		For the following discussion and definitions, it is assumed that a simple sweep sequence and a sweep shape have been chosen.

		An infinite work band $W^\infty$ is a subset of the infinite grid $\mathbb{Z}^d$.
		$W^\infty$ results from shifting  a  sweep shape $\sH$ according to the sweep sequence $\sweepSeq$ over and over in positive (or negative) direction.
		If the end of the sweep sequence is reached, we start over with the first element of the sweep sequence.
		We say that a sweep shape $\sH'$ proceeds (precedes) the sweep shape $\sH$ if it results from $\sH$ be applying the next (previous) unit shift of the sweep sequence.
		The infinite work band resulting from $\sH$ and $\sweepSeq$ is given by (assuming that the elements $\delta_i$ of the sweep sequence are indexed from 0 to $\size{\sweepSeq}-1$)
		\begin{equation*}
			W^\infty = \left\{\!\!\! \begin{array}{cl}
							y \in \mathbb{Z}^d: \exists  z \in \sH, \; \exists r \in \mathbb{Z}, \text{ s.t.:}\!\!\!\!
							& \text{if } r \geq 0: y = z + \sum_{i=1}^r \delta_{\lr{(i-1)\!\!\!\! \mod \size{\sweepSeq}}}\\[.5ex]
							& \text{if } r < 0: y = z +\sum_{i=-r}^{-1} \delta_{\lr{i \!\!\!\! \mod \size{\sweepSeq}}}
			                   \end{array}\!\!\!\right\}.
		\end{equation*}
		Each infinite work band $W^\infty$ corresponds to a (finite) work band $W$ containing the vertices of $W^\infty$ that are part of the grid $[k_1]\times\dots\times[k_d]$, $W = W^\infty \cap \lr{ [k_1]\times\dots\times[k_d]}$.

		In addition to the sweep sequence and sweep shape, an algorithm is going to be defined by a list of work bands $\wL$ which are evaluated one by one.
		All work bands of one particular algorithm result from the same, possible shifted, sweep sequence and shape.
		Different work bands are obtained by shifting the sweep shape to a new start position before applying the sweep sequence.

		We associate an \emph{evaluation band} $E_W$ or simply $E$ to each work band $W\in \wL$.
		Fix one particular work band $W$.
		The evaluation band $E_W$ is the set of vertices $w \in W$ for which the $s$-star stencil $S_s(w)$ can be evaluated if all vertices of $W$ would fit into internal memory, 
		\begin{equation}
			E_W = \{w \in W: S_s(w) \subset W \} \; .
		\end{equation}
		For an infinite work band the infinite evaluation $E^\infty$ band is defined in the analogous way.
		Similarly to associating an evaluation band $E_W$ with a work band $W$, we associate a work band $W_E$ with an evaluation band $E$.
		If an algorithm evaluates $s$-star stencil for all grid points, the evaluation bands have to cover the grid.

		As an evaluation band is a true subset of its corresponding work band, the work bands have to overlap if the evaluation bands shall cover the grid.
		Because of this overlap there are going to be vertices that belong to several work bands. 
		These vertices that are part of several different work bands cause the non-compulsory I/Os.

		We introduce the notion of \emph{$k$-intersections} to partition the vertices of the grid according to the work bands and evaluation bands they belong to.
		The $k$-intersections are fundamental for the data layout and allow a simple counting of the non-compulsory I/Os.
		To define the $k$-intersections, let $\wL$ be the set of all work bands that an algorithm works on.
		For $k \in \mathbb{N}$ and two subset $\eL' \subset \wL$ and $\wL'\subset \wL $ such that $\eL' \subset \wL'$ and $|\wL'| = k$, the $k$-intersection $\Phi\lr{\wL', \eL'}$ is the set of all vertices which belong to all $W'$ for $ W' \in \wL'$ and all $E_W$ for $W \in \eL'$, but not to any other work or evaluation bands,
		\begin{align*}
			&\Phi\lr{\wL', \eL'} := \Phi\lr{\wL'} \cap \Phi^E\lr{\eL'} \quad \text{for }\\
			&\Phi\lr{\wL'} :=\lr{\bigcap_{W \in \wL'} W} \setminus \lr{\bigcup_{W\notin \wL'} W} \quad \text{and}\\
			&\Phi^E\lr{\eL'} :=\lr{\bigcap_{W \in \eL'} E_W} \setminus \lr{\bigcup_{W\notin \eL'} E_W}\; .
		\end{align*}
		We call $\Phi\lr{\wL'}$ \emph{work band intersection} and $\Phi^E\lr{\eL'}$ \emph{evaluation band intersection}.
		For a fixed $k \in \mathbb{N}$ and a particular work band $W \in \wL$ the family of all $k$-intersections which contain vertices of $W$ is given by 
		\begin{equation}
			\Phi\lr{W,k} :=\{\Phi\lr{\wL', \eL'}: \wL',\eL' \subset \wL,\; \eL' \subset \wL',\; \size{\wL'}=k \text{ and } W \in \wL' \}
		\end{equation}
		The $k$-intersections describe sets of vertices that are going to be either read or written in sequence to achieve good performance.
		To use the compulsory reads effectively, the data is split according to the $\Phi\lr{\wL'}$.
		To make sure that the compulsory write can store a whole block of vertices to external memory, the data is organized by the $\Phi^E\lr{\eL'}$.
		To avoid overhead with respect to both, compulsory reads and writes, the data is divided according to the $k$-intersections $\Phi\lr{\wL', \eL'}$.
		In the presented algorithms the vertices of the 1-intersections are only going to contribute to the compulsory I/Os and the vertices of the 2-intersections are going to determine the leading term of the non-compulsory I/Os.
		The I/Os caused by vertices in the $k$-intersections for $k\geq 3$ will only amount to lower order terms.
		

			To describe the size of the sweep shape and the resulting work band a parameter $m$ sufficing 
			\begin{equation}\label{eq:sweepSizeDef}
				\size{\sH} = c\cdot m^{d-1}+\mO{m^{d-2}}
			\end{equation}
			for a constant $c\in \mathbb{R}$ is employed.
			For our polygonal sweep shapes a natural choice for $m$ is the width 
			of $\sH$ in one unit direction, e.g.  $m = \max \{|u_1-v_1| : u,\,v \in \sH\}$.
			The size parameter $m$ is going to be specified for each individual sweep shape.
			The size of a work band is going to be chosen so that we can evaluate the corresponding evaluation band by doing one sweep of the work band, i.e. by loading each vertex of the work band exactly once.

			For a  sweep shape $\sH$ and the resulting work band $W$ and evaluation band $E_W$, all vertices of $\sH \cap E_W$ can be evaluated if the $s$ preceding  and $s$ proceeding sweep shapes of $\sH$ are in internal memory.
			Hence an evaluation band can be evaluated by one sweep of the work band when the internal memory can hold $2s+1$ sweep shapes.
			When the vertices within a sweep shape are not evaluated randomly but in lexicographic order, only vertices equivalent to $2s$ full sweep shapes ($\mO{m^{d-1}}$ vertices each) and an overhead of $\mO{m^{d-2}}$ vertices is needed in internal memory instead of $2s+1$ full sweep shapes.
			The vertices of the $s$-th preceding sweep shape can be deleted or written to the external memory as vertices of the $s$-th proceeding sweep shape are loaded. 
			For the leading term of the non-compulsory I/Os only the number sweep shapes is relevant. 
			The overhead of $\mO{m^{d-2}}$ vertices is only going to contribute to lower order terms.

			To prove that the sweep shape and the resulting work band are small enough to evaluate the evaluation band with a single sweep of the work band, the notion of a \emph{work band order} is introduced.
			Fix one work band $W$.
			The vertices $w\in W$ are sorted in two stages:
			\begin{enumerate}
				\item Sweep shape by sweep shape in increasing distance of the sweep shape to the origin.
				\item Within sweep shapes in lexicographic order.
			\end{enumerate}
			The lexicographic order is the lexicographic order with respect to the coordinates of the vertices, $x_d$ being the index changing fastest and $x_1$ the slowest index.
			Formally, 
			\begin{align*}
				w &\leq w' 
					\Leftrightarrow  \\ &\Leftrightarrow \Big(\!\!\lr{\exists j \in \{1,\dots,d\}\!\!: \forall i < j\!: w_i = w_i' \wedge w_j < w_j'} \vee \lr{w_i = w_i' \; \forall i \in \{1,\dots,d\}} \!\!\Big).
			\end{align*}

			For a vertex $w \in W$ its \emph{work band position} or \emph{work band order} $o_W(w)$ is its position in the work band according to this order. 
			For two vertices $w, w' \in W$ their distance in the work band order is $||w-w'||_{W}= |o(w) - o(w')|$, i.e. the difference of their respective positions within this work band $W$.
			Note that vertices which are in $k$-intersections for $k \geq 2$ belong to more than one work band and hence are assigned a work band order for each of the work bands they belong to.
			Also, vertices which do not belong to the same work band are not comparable.
			
			In the very same way as the work band order $o_W(\cdot)$, the \emph{evaluation band order} $o_E(\cdot)$ and the $k$-intersection order $o_{\Phi \lr{\wL', \eL}}(\cdot)$ 
			are defined for all vertices $w \in E$ or $w\in \Phi\lr{\wL',\eL}$ respectively.
			As a consequence, the orders of different work bands, evaluation bands and $k$-intersections are consistent with each other.
			Formally, let $A$ and $B$ be either be work bands, evaluation bands or $k$-intersections.
			The orders of $A$ and $B$ are called consistent if and only if:
			\begin{align*}
				\forall w,w' \in A\cap B:   \ o_A(w) \leq o_A(w')	\Rightarrow o_B(w) \leq o_B(w') \; .
			\end{align*}

			The linear work band order gives rise to the definition of an \emph{interval} of vertices of the work band $W$.
			For $\delta \in \mathbb{N}$,
			\begin{equation*}	
				[ -\delta +w , \; w +\delta]_W := \{w' \in W: ||w'-w||_W \leq \delta \}
			\end{equation*}
			is the interval of midpoint $w$ and width $2\delta$.
			The vertices needed to evaluate a vertex $w\in E_W$ of the evaluation band $E_W$ are contained in the interval
			\begin{equation} \label{eq:Iw}
				I(w) := \left\{ 
								w' \in W: \exists v \in S_s(w), \; \exists v' \in S_s(w) \text{ s.t. } o_W(v) \leq o_W(w') \leq o_W(v') \;.
						\right\}
			\end{equation}
			By definition, $S_s(w) \subset I(w)$.
			Note that if $w \in E_W$ then it also holds that $w\in W$.
			Further, it follows from $w \in E_W$ that the whole $s$-star stencil of $w$ is a subset of $W$, i.e. $S_s(w) \subset W$, and the definition above is well defined.
			Finally, observe that there is a simple characterization of $I(w)$ following directly from the definition.
			Using $w_{\min} = \argmin\{o_W(w): w \in S_s(w) \subset W\} $ (the vertex in $S_s(w)$ of the smallest work band order in $W$) and $w_{\max}\argmax\{o_W(w): w \in S_s(w) \subset W\}$ (the vertex in $S_s(w)$ of the largest  work band order in $W$), $I(w)$ is given by
			\begin{equation}\label{eq:IwCharacterization}
				I(w) = [ o_W(w_{\min}), \; o_W(w_{\max} )]
			\end{equation}
			
	

			To facilitate the analysis, we reduce the dimensionality of the problem to $d-1$ by cutting it with hyperplanes of normal $e_1$.
			In particular, the vertices of the $k$-intersections causing the non-compulsory I/Os are going to be counted layer by layer, hyperplane of normal $e_1$ by hyperplane of normal $e_1$.
			The counting is done in a two-step process.
			First the number of work bands that are in one hyperplane is bounded.
			Then, for each hyperplane and each work band the number of vertices in any $k$-intersection is bounded.
			For both tasks, we need more notation.

			First, let us introduce notation to estimate the number of work bands in a particular hyperplane of normal $e_1$. 
			Denote by $H_h$ the hyperplane of normal $e_1$ at distance $h$ from the origin and let $\wL$ be a list of work bands.
			The work bands that have a vertex in $H_h$ are denoted by $\wL_h$,
			\begin{equation*}
				\wL_h= \{ W \in \wL: W \cap H_h \neq \emptyset \}
			\end{equation*}
			Estimating the number of work bands in $\wL_h$ is the first part of the complexity analysis.
			To carry out this analysis later, we need more notation.
			Let $E^\infty$ be an infinite evaluation band.
			The intersection of the evaluation band at level $h$ is then defined as
			\begin{equation}\label{eq:defEInftyH}
				E^\infty_h = E^\infty \cap H_h \; .
			\end{equation} 
			(In general, for sets $A$ the subscript $h$ is used as a shortcut for $A_h = A \cap H_h$, i.e. for all vertices of $A$ for which  $x_1 = h$ holds.)
			For simple sweep sequences it holds that $E^\infty_{h_1} = E^\infty_{h_2}$ up to translations for all $h_1, h_2 \in \mathbb{Z}$.
			We say that a set of work bands is created by the same sweep shape (and sweep sequence) when only translations of one particular sweep shape are used to create the different work bands. 
			Given that the work bands are created by the same sweep shape and sweep sequence, all corresponding infinite evaluation bands have the same cross-sections up to translations.
			In particular, for two infinite evaluation bands $E^\infty$ and $\lr{E'}^\infty$ whose corresponding work bands are given by the same sweep shape and sweep sequence, the sizes of their level sets at height $h$ and $h'$ are equal,
			\begin{equation*}
				\size{\lr{E^\infty}_h} = \size{\lr{\lr{E'}^\infty}_{h'}} \quad \forall E, E'\quad \forall h,h'\in\mathbb{Z}\;.
			\end{equation*}
			For a fixed sweep shape determining $m$, a fixed sweep sequence and a list of work bands $\wL$ that is created by this sweep shape and sweep sequence, there is a constant $e \in \mathbb{R}, \; e\geq 0$ such that
			\begin{equation}\label{eq:defEhInfty}
				\forall h \in [k_1], \; \forall\, W \in \wL: \size{\lr{E_W}^\infty_h}\geq e \cdot m^{d-1}-\mO{m^{d-2}} \;.
			\end{equation}
			Such an $e$ always exists as $e=0$ may be chosen.
			We are going to derive the value of $e$ in the sections of the different algorithms for the respective sweep shapes and sweep sequences.

			Finally, denote by $l_i$ ($2\leq i \leq d$) the width of $E^\infty_h$ in direction $i$, i.e. 
				\begin{equation}\label{eq:defLi}
					l_i = \max_{ x \in E_h^\infty }\left\{x_i \right\} - \min_{x \in E_h^\infty} \left\{x_i \right\} \; .
				\end{equation}
			Again, given that the same sweep shape and sequence was used to create different work bands, $l_i$ is independent of the actual choice of $E$ and the level $h$.

			It is left to count the vertices in the $k$-intersections of a work band per hyperplane of normal $e_1$. 
			Let $\wL$ be a list of work bands and choose one $ W \in \wL$.
			For $k \in \mathbb{N}$ and $h \in [k_1]$ the vertices of the $k$-intersections of $W$ at height $h$ are given by
			\begin{equation*}
				\Phi_{(W,k,h)}= \Phi_{(W,k)} \cap H_h \; .
			\end{equation*}

		\subsection{The Algorithmic Framework for the Band Algorithms}

		With this notation in hand, we are ready to define the algorithmic framework of the \emph{band algorithms} presented in this paper.
		Recall the definitions and assumptions we need for the asymptotic analysis.
		The dimension $d$ and the stencil size $s$ are assumed to be fixed and constant.
		The grid sizes $k_i(n)$ are ordered by size, i.e. $k_1(n) \geq k_2(n) \geq \dots \geq k_d(n)$, and we assume $M(n) = \lo{k_d(n)}$ and $B(n)= \lo{M(n)}$ for $n \to \infty$.
		
		All algorithms work on two copies of the data, an input and an output grid.
		The input grid stores the initial values of the vertices and these values are never altered.
		The values updated according to the $s$-star stencil are stored in the output grid.
		When we have to evict blocks of the input data from internal memory due to capacity reasons, we store them back to external memory if that block is accessed again by the algorithm.
		If the block is not accessed again by the algorithm, it is discarded and not written back to external memory.
		Output blocks are always stored in external memory.
		The first time we access an output block it does not need to be read, though, as it does not contain any data.

		\begin{defin}[Band Algorithm]\label{def:bandAlgo}
			A \emph{Band Algorithm} $\algo$ evaluating the $s$-star stencil on the $d$-dimensional grid $[k_1]\times\dots\times [k_n]$ is defined by
			\begin{enumerate}
				\item a simple sweep sequence $\sweepSeq$,
				\item a sweep shape $\sH$ and
				\item a list $\wL$ work bands. 
			\end{enumerate}
		All work bands $W \in \wL$ need to be generated by the sweep shape $\sH$ and sweep sequence $\sweepSeq$.
 
		The algorithm works on a data layout organized
			\begin{enumerate}
				\item by $k$-intersections, 
				\item within $k$-intersection in $k$-intersection order (i.e. sweep shape by sweep shape in increasing distance to the origin and within sweep shapes by the lexicographic order of the coordinates).
			\end{enumerate}
			\noindent The algorithm evaluates the vertices in the following order:
			\begin{enumerate}
				\item Evaluation band by evaluation band in the order of the corresponding work bands in the list $\wL$.
				\item Within an evaluation band in the evaluation band order (i.e. sweep shape by sweep shape in increasing distance to the origin and within sweep shapes in lexicographic order).
			\end{enumerate}
			If the $k$-intersection of an evaluation band has already been evaluated, it is not evaluated again. 
	
			If a vertex $w\in E$ of a particular evaluation band $E$ is evaluated, all blocks that store the input values of the vertices of the interval $I(w)$ are loaded to internal memory. 
			The \emph{least recently used} (LRU) cache replacement strategy is used when input blocks have to be evicted from internal memory.
			Output blocks containing the updated values of a vertex are only evicted from internal memory if the updated values of all vertices of the block have been calculated or the end of the work band has been reached.
		\end{defin}

		\begin{defin}[Correct and Memory Efficient Band Algorithm]\label{def:memEffBandAlgo} 
			Let $\algo$  be a Band Algorithm as given in Def.~\ref{def:bandAlgo}.
			Let $m$ be a parameter for the size of the sweep shape that suffices~\eqref{eq:sweepSizeDef}, i.e. 
			\begin{equation*}
			 \size{\sH} = c\cdot m^{d-1}+\mO{m^{d-2}} 
			\end{equation*}
			for a constant $c \in \mathbb{R}_{>0}$.
			We then choose the size $m$ of the sweep shape as 
			\begin{equation} \label{eq:sweepSize}
				m = \sqrt[d-1]{\frac{M}{2s \cdot c}}-\thet{\sqrt[d-1]{B} } \;.
			\end{equation}
			$\algo$ is called \emph{memory efficient} if the following assumptions hold: 
			\begin{enumerate}
				\item \label{assum:evalPoints} \textbf{The interval of a vertex $w\in W$ is small:} $\forall \,W \in \wL$ and all $w \in E_W$ it holds that $I(w) \subset \left[x- \delta ,\; x+\delta \right]_W$ for $\delta =s \cdot \size{\sH}+\mO{M^{\frac{d-2}{d-1}}}$. 
				\item\label{assum:evalBandsCover} \textbf{The evaluation bands} $\{ E_W: W \in \wL \}$ \textbf{cover the grid}:
						$\forall w \in [k_1]\times\dots\times[k_n]: \exists W \in \wL$ such that $w \in E_W$.
				\item\label{assum:evalWidth} \textbf{The width of an evaluation band is small:} 
				$\quad l_i = \mO{m} \qquad \forall i \in \lrC{2,\dots,d}$.
				\item\label{assum:evalSize} \textbf{Size of the evaluation bands:} $\exists e\in\mathbb{R}_{>0}$ such that \eqref{eq:defEhInfty} holds, i.e.
					\begin{equation*}
						\forall h \in [k_1], \; \forall \,W \,\in \wL: \size{\lr{E_W}^\infty_h}\geq e \cdot m^{d-1}-\mO{m^{d-2}} \;.
					\end{equation*} 
				\item\label{assum:workEvalDistance} \textbf{Work band vertices are not separated from the evaluation band:}  $\forall \,W \in \wL$ and $\forall w \in W: \exists v\in E_W$ such that $w_1 = v_1$ and  $||w-v||_1\leq 2s$.
				\item\label{assum:nbrOfWorkBands}  \textbf{The total number of work bands is small:} $\size{\wL} = \mO{\frac{1}{M}\cdot \prod_{i=1}^{d-1}k_i}$ .
				\item\label{assum:workbandConst} \textbf{Any work band overlaps only with a constant number of other work bands.} 
						$\forall\, W \in \wL: \Big|\big\{V \in \wL:(V  \neq W) \wedge (V \cap W \neq \emptyset) \big\}\Big|= \mO{1}$. 
				\item\label{assum:2intersects} \textbf{The 2-intersections determine the leading term of the non-com\-pul\-sory I/Os:}\\ $\exists b \in \mathbb{R}_{>0}$ such that $\forall h \in [k_1], \;\forall W \in \wL:$ $\size{\Phi_{(W,2,h)}}\leq  b \cdot m^{d-2}+ \mO{ m^{d-3}}$. (For $d=2$: $\size{\Phi_{(W,2,h)}}\leq  b $.)
				\item\label{assum:3intersects} \textbf{The $k$-intersections for $k\geq3$ only contribute to lower order terms of the non-compulsory I/Os:} $\forall h \in [k_1], \;\forall\, W \in \wL, \; \text{ for } k \geq 3:$ $\size{\Phi_{(W,k,h)}}= \mO{m^{d-3}}$. (For $d=2$: $\size{\Phi_{(W,k,h)}}=0$.)
			\end{enumerate}			
		\end{defin}

		\begin{thm}[The I/O Complexity of a Memory Efficient Band Algorithm.]\label{thm:algoFrameworkUpperBounds} {\ \\}
			Let $\algo$ be a memory efficient band algorithm as defined by Defs.~\ref{def:bandAlgo} and~\ref{def:memEffBandAlgo}.
			Then, an upper bound for the non-compulsory I/Os performed by $\algo$ is given by
			\begin{align}
				&d = 2: \quad  \frac{b\cdot c}{e} \cdot 2s  \cdot \frac{k_1k_2}{B \cdot M} +\mO{\frac{k_1k_2}{ M^2} }+\mO{\frac{k_1}{B}}\;, 
					\\[2ex]
				&d \geq 3: \quad  \frac{b\cdot \sqrt[d-1]{c}}{e} \cdot \sqrt[d-1]{2s } \cdot \frac{\prod_{i=1}^d k_i}{B \cdot \sqrt[d-1]{M}} +\mO{ \frac{\prod_{i=1}^d k_i }{\sqrt[d-1]{B^{d-2} \cdot M^2}}} \; .
			\end{align}
		\end{thm}
			For any dimension $d$, the number of compulsory I/Os is, by definition, $2\cdot \frac{1}{B}\cdot \prod_{i=1}^d k_i$.

		For $d\geq 3$, the leading error term $\mO{ \prod_{i=1}^d k_i \big/\sqrt[d-1]{B^{d-2} \cdot M^2}}$ is due to reserving separate blocks for each $k$-intersection.
		For $d=2$, the additional error term $\mO{\frac{k_1}{B}}$ error is due to the estimation of the number of work bands and the last, probably incomplete, work band that the algorithm works on.

		We claim that this analysis is tight and that Thm.~\ref{thm:algoFrameworkUpperBounds} hence gives the complexity of a memory efficient band algorithm.
%
		To prove the theorem, we first prove several Lemmas.
		Once the theorem is proven, we present different band algorithms in~\S\ref{sec:upperBoundsAlgo} and show that these are memory efficient.	
		Hence, their complexity is given by Thm.~\ref{thm:algoFrameworkUpperBounds}.

		\begin{lem}[The number of non-empty $k$-intersections of each work band is bounded] \label{lem:kIntersectsBounded}{\ \\}
			Given the setup of Def.~\ref{def:bandAlgo}, i.e. a sweep shape $\sH$, a sweep sequence and a list of work bands $\wL$, assume that Assumption~\ref{assum:workbandConst} of Def.~\ref{def:memEffBandAlgo} holds. 
			Then, for any $W\in \wL$ the number of $k$-intersections containing vertices of $W$ is bounded by a constant which only depends on the dimension~$d$ and the size of the stencil~$s$.
		\end{lem}

		\begin{proof}
		 	We need to prove that the family 
				$\{ A: A \in \Phi\lr{\wL',\eL'} \text{ for } \wL' \in \wL,\; \eL' \in \wL \text{ and } A \cap W \neq \emptyset \}$
			contains only a constant number of sets. 
			Denote by $O_W$  the set of work bands which overlap with $W$ non-trivially, including $W$ itself.
			Denote by $\mathcal{P(O_W)}$ the power set of $O_W$.
			By Assumption~\ref{assum:workbandConst} of Def.~\ref{def:memEffBandAlgo} one work band overlaps non-trivially with only a constant number of other work bands.
			This constant does only depend on the dimension $d$ and the size of the stencil $s$.
			Hence, $\size{O_W} = \theta(1)$ and therefore it also holds that $\size{\mathcal{P(O_W)}}=\theta(1)$.
			For any non-empty $k$-intersection $\Phi\lr{\wL',\eL'}$ containing at least one vertex of $W$ it follows that $\wL' \in \mathcal{P(O_W)}$.
			As an evaluation band $E$ is always a subset of its work band, $E \subset W_{E}$, it also follows that $\eL' \in \mathcal{P(O_W)}$.
			Hence, their is just a constant number of non-empty $k$-intersections $\Phi\lr{\wL',\eL'}$ that contain vertices of $W$.
%
%
%
%
%
		\end{proof}

		\begin{lem}[The interval $I(w)$ of a vertex  $w$ and a constant number of output blocks fit into internal memory] \label{lem:sizeOfSweepShape}{\ \\}
			Given the setup of Def.~\ref{def:bandAlgo}, i.e. a sweep shape $\sH$, a sweep sequence and a list of work bands $\wL$.
			Assume that \ref{assum:evalPoints} and~\ref{assum:workbandConst} of Def.~\ref{def:memEffBandAlgo} hold.
			Then, for any work band $W \in \wL$ and any vertex $w \in W$, 
			the vertices of the interval of $w$ fit into internal memory together with a constant number of output blocks of size $B$, 
			\begin{equation*}
				I(w) = M -   \Om{1}\cdot B \;.
			\end{equation*}
		\end{lem}

		\begin{proof}
		By Assumption~\ref{assum:evalPoints} of Theorem~\ref{thm:algoFrameworkUpperBounds} we got the inclusion 
				$I(w) \subset \left[x- \delta ,\; x+\delta \right]_W$
		for $\delta =s \cdot \size{\sH}+\mO{M^{\frac{d-2}{d-1}}}$.
		The interval $\left[x- \delta ,\; x+\delta \right]_W$ itself contains $2\delta+1$ vertices.
		However, the interval is split into several chunks  of contiguous memory blocks by the $k$-intersection.
		Denote by $\Phi\lr{\wL',\eL'}$ a $k$-intersection that contains a vertex of $I(w)$.
		Although each $k$-intersection is contiguous in memory, all $k$-intersections of $W$ are most likely not contiguous in memory together.
		Hence, we do not only need to account for the vertices in  $\left[x- \delta ,\; x+\delta \right]_W$ but also for all vertices which are in the same blocks 
		as any of these vertices.
		As the work band and the $k$-intersection orders are consistent, the vertices of $I(w)\cap \Phi\lr{\wL',\eL'}$ are contiguous in memory. 
		Therefore, in at most 2 blocks of $I(w)\cap \Phi\lr{\wL',\eL'}$, the first and the last block, are also vertices not from $I(w)$.
%
%
		Hence, per $k$-intersection there are less than $2$ blocks in internal memory which contain vertices not in $I(w)$.
		By Lemma~\ref{lem:kIntersectsBounded}, there the number of $k$-intersections per work band is constant.	
		Hence, the union of all blocks containing vertices of $I(w)$ consist of at most $(2\cdot \delta+1)+\mO{B}$ vertices altogether. 
		Hence proving
			$2\cdot \delta +1 + \mO{B}  =M-   \Om{B}$
		establishes the lemma.
		\begin{align*}
				&2\delta +1 +\mO{B} = 
			 2\cdot \lr{s \cdot \size{\sH}+\mO{M^{\frac{d-2}{d-1}}}}+1+\mO{B} 
			= \\
			&\stackrel{\eqref{eq:sweepSizeDef}}{=} 2 \cdot s \cdot\lr{ c\cdot m^{d-1}+\mO{m^{d-2}}} + \mO{M^{\frac{d-2}{d-1}}} +\mO{B}=\\
			&\stackrel{\eqref{eq:sweepSize}}{=} 2 \cdot s \cdot c\cdot  \lr{ \sqrt[d-1]{\frac{M}{2s \cdot c}}-\thet{\sqrt[d-1]{B} }}^{d-1}+ \\&\qquad \qquad +\mO{M^{\frac{d-2}{d-1}}} + \mO{M^{\frac{d-2}{d-1}}} +\mO{B}= \\
			&= M - \thet{M^{\frac{d-2}{d-1}}\cdot B^{\frac{1}{d-1}}} + \mO{M^{\frac{d-2}{d-1}}}+\mO{B}  = \\ 
			&=M - \thet{\!\sqrt[d-1]{M^{d-2}\cdot B}} = M-   \Om{B}\; .
		\end{align*}
		\end{proof}

		\begin{lem}[Number of work bands per hyperplane of normal $e_1$]\label{lem:nbrOfWorkBandsArb}{\ \\}
			Given the setup of Def.~\ref{def:bandAlgo}, i.e. a sweep shape $\sH$, a sweep sequence and a list of work bands $\wL$.
			Assume that Assumptions~\ref{assum:evalWidth}, \ref{assum:workEvalDistance}, \ref{assum:workbandConst}, \ref{assum:2intersects} and~\ref{assum:3intersects} 
			of Def.~\ref{def:memEffBandAlgo} hold.
			Then, for all $h \in [k_1]$, the hyperplane of distance~$h$ from the origin and normal~$e_1$ contains vertices of at most 
			\begin{align*}
				\frac{ \prod_{i=2}^d\lr{k_i+\mO{m}}}{\size{E^\infty_h}-\mO{m^{d-2}} }
			\end{align*}
			different work bands.	
		\end{lem}

	\begin{proof}
			Recall the definition of $E^\infty_h$ and $l_i$, namely \eqref{eq:defEInftyH} and \eqref{eq:defLi}.
			In particular note that the values of neither of them depends on the choice of the work band $W_E$ or the level $h$. 
			Let $K$ be a level set of the grid $[k_1]\times\dots\times[k_n]$ in $e_1$ direction, $K = \lrC{h} \times [k_2]\times \dots \times [k_d]$.
			Hence, $\size{K} = \prod_{i=2}^d k_i$.

			To estimate the number of work bands in a hyperplane of normal $e_1$ we make a detour to the evaluation bands.
			We first want try to estimate the number of evaluation bands needed to cover $K$ by $\size{K} \big/ \size{E^\infty_h}$: 
			dividing $\size{K}$ by the number of vertices that can be evaluated per work band and hyperplane of normal $e_1$, namely $\size{E^\infty_h}$.
			This fraction would underestimate the number of work bands that have a vertex in $K$ for three reasons.
			First, if the simple sweep sequence also contains a shift different from $e_1$, it is possible that a hyperplane of normal $e_1$ contains a vertex of the work band but none of the corresponding evaluation band.
			Second, the straight boundary of the grid cannot be reproduced if $E^\infty$ does not have the same straight boundary, i.e. whenever the sweep shape is more complicated.
			Third, the irregular structure of $E^\infty$ could make it impossible to align different evaluation bands without overlap.
			$\size{K} \big/ \size{E^\infty_h}$ would assume that perfect partitioning is possible and would allow fractions of evaluation bands to cover $K$.
			However, we are interested in the number of evaluation needed and are not allowed to add fractional evaluation bands together to form a single one.
			
			We therefore enlarge the grid in two steps to get an upper bound for the number of work bands involved in each hyperplane of normal $e_1$.
			To address the first issue, pad $2s$ grid points at the beginning and end of each coordinate direction $x_i$ for $2 \leq i \leq d$ of the grid.
			By Assumption~\ref{assum:workEvalDistance}, this ensures that for every vertex of the work band in the original grid there is also a vertex of the evaluation band  in the same hyperplane of normal $e_1$ in the extended grid.
			Hence, by counting the evaluation bands in the extended grid for a hyperplane of normal $e_1$, we get a lower bound for the work bands in the original grid in this hyperplane.

			To address the second issue, pad another $l_i$ grid points at the beginning and end of unit direction $x_i$ for  all $2\leq i \leq d$. 
			Hence, if there is a vertex of an evaluation  band $E$ in the hyperplane of normal $e_1$ before this second enlargement, then all of $E^\infty_h$ is in the twice enlarged grid.
			Therefore, each fractional evaluation band from before is now completely in the grid.

			Lastly, to address the third issue, we do not divide $\size{K}$ by $\size{E^\infty_h}$ but by the number of vertices in $\size{E^\infty_h}$ that belong to the 1-intersection and hence are not part of any other evaluation band.
			By Assumption~\ref{assum:2intersects} at most $b\cdot m^{d-2}+\mO{m^{d-3}}$ vertices of $\size{E^\infty_h}$  belong to 2-intersections (at most $b$ for $d=2$
			).
			By Assumption~\ref{assum:workbandConst} at most a constant number of work bands, say $f$ work bands, overlap and hence only $k$-intersections for $k$ up to $f$ can be non-empty.
			Using Assumption~\ref{assum:3intersects}, the number of vertices of $\size{E^\infty_h}$ which are in $k$-intersections for $k\geq3$ is hence bounded by $ f\cdot \mO{m^{d-3}} = \mO{m^{d-3}}$ (and is 0 for $d=2$
			).
			Altogether, subtracting all $k$-intersections for $k\geq 2$, the 1-intersection of $\size{E^\infty_h}$ still contains at least 
			\begin{align*}
				d =2: \quad &  \size{E^\infty_h}- b = \size{E^\infty_h}-\mO{m^{d-2}}\\
				d \geq 3: \quad & 	 \size{E^\infty_h}- \lr{\lr{b\cdot m^{d-2}+\mO{m^{d-3}}} + \mO{m^{d-3}}} = \size{E^\infty_h}-\mO{m^{d-2}}
			\end{align*}
			vertices.

			With these three modifications we ensure that the number of work bands containing a vertex of $K$ is overestimated by 
			\begin{equation*}
				\lr{\prod_{i=2}^d\lr{k_i+2\cdot l_i+4s}}\cdot\frac{1}{\size{E^\infty_h}-\mO{m^{d-2}} } \stackrel{(\text{Assum. \ref{assum:evalWidth}})%
					}{=}\frac{ \prod_{i=2}^d\lr{k_i+\mO{m}}}{\size{E^\infty_h}-\mO{m^{d-2}} }  \;.
			\end{equation*}
		\end{proof}

		We are now ready to proof Thm.~\ref{thm:algoFrameworkUpperBounds}.

		\begin{proof}[Proof of Theorem~\ref{thm:algoFrameworkUpperBounds}]
			The proof proceeds in the following steps:
			first, the correctness of a memory efficient band algorithm is proven, making sure all vertices are evaluated and that the internal memory stores the relevant information to evaluate the current vertex.
			Then, we analyze the I/O complexity of a memory efficient band algorithm.
			We first determine that a memory efficient band algorithm evaluates an evaluation band by sweeping through the $k$-intersections of the corresponding work band simultaneously, loading each vertex of the work band exactly once.
			From that it follows that the $k$-intersections determine the number of non-compulsory I/Os that a vertex causes.
			Hence, we estimate the total number of vertices in the $k$-intersections for different $k$.
			After accounting for incomplete blocks at the beginning and end of each $k$-intersection, we use these results to establish the upper bound.


			The correctness of algorithm $\algo$ is verified easily. 
			For any $W \in \wL$ choose any $w\in E_W$.
			First, the interval $I(w)$ contains all vertices necessary to compute the $s$-star stencil of $w$ as $S_s(w) \subset I(w)$  by the definition of $I(w)$.
			Second, Lemma~\ref{lem:sizeOfSweepShape} shows that the internal memory is large enough to hold all blocks of $I(w)$ and a constant number of output blocks.
			By Lemma~\ref{lem:kIntersectsBounded}, the number of $k$-intersections of $W$ is constant and hence, as we need at most one output block per $k$-intersection, also the number of output blocks.
			Hence, all blocks containing vertices of $I(w)$ and the output blocks fit in internal memory together.
			Finally, by Assumption \ref{assum:evalBandsCover} the evaluation bands cover the grid.
			As $\algo$ works through all evaluation bands it also evaluates all vertices of the grid and hence performs one update of the grid according to the $s$-star stencil.

			Let us now analyze the number of I/Os the algorithm performs.

			First, let us establish that $\algo$ evaluates an evaluation band $E$ loading each vertex of $W_E$ exactly once.
			As whole blocks of size $B$ are always loaded, the first and last block of each $k$-intersection of $W_E$ can also contain vertices not in $W$.
			Hence, evaluating $E$ can also load vertices $w\notin W$. 
			We disregard these vertices $w\notin W$ for the moment and account for them separately in \eqref{eq:IOsStartEndKintersect}.
			The algorithm $\algo$ evaluates the vertices of $E$ in the evaluation band order.
			To evaluate a vertex $w\in E$ the algorithm $\algo$ loads the interval $I(w)\subset W$ into internal memory.
			We have already seen in the correctness proof of $\algo$ that the internal memory is large enough to hold $I(w)$ and the constant number of output blocks needed.
			As the orders of the work band, the evaluation band and the $k$-intersections are consistent with each other, this means that $\algo$ sweeps over the $k$-intersections of $W$ simultaneously.
			Therefore evaluating the next vertex in $E$ results in loading subsequent vertices of the work band order and evicting prior vertices.
			Formally, the following holds.
			\begin{align*}
				\text{For } w, \; w' \in E \text{ with } I(w) = [a,b] \text{ and } I(w') = [a',b'] \\
				\text{it holds that: } o_E(w)\leq o_E(w') \Rightarrow \lr{a\leq a' \wedge b \leq b'}
			\end{align*}
			Using the cache replacement strategy described in Def.~\ref{def:bandAlgo} ensures that neither input nor output blocks are evicted from memory when they are still needed for the evaluation of $E$.
			Altogether, although these sweeps happen simultaneously, each vertex of $W$ needs to be read exactly once to evaluate all vertices of $E$.
			We say, that one one sweep of the work band $W$ suffices to evaluate all vertices of $E$.

			The $k$-intersections determine how often a vertex needs to be read. 
			Recall that a vertex of a $k$-intersection is part of exactly $k$ work bands.
			As the vertex needs to be accessed by these $k$ work bands, it causes the first, compulsory read operation followed by a sequence of $k-1$ non compulsory writes and reads.
			As well, the updated value of the vertex is stored by the compulsory write.
			Altogether, a vertex in a $k$-intersection takes part in 2 compulsory and $2(k-1)$ non-compulsory I/Os. 
			This are less than $2$ non-compulsory I/Os for each work band the vertex is part of.
			It is also possible that a vertex is still in internal memory from the previous work band when it is accessed by the next work band.
			This, however, would only decrease the number of I/Os performed and is hence disregarded in the analysis.

			By now we know that vertices in the 1-intersections cause 0 non-compulsory I/Os, vertices in the 2-intersections cause 1 non-compulsory I/O per work band they belong to and vertices in $k$-intersections for $k\geq 3$ cause less than 2 non-compulsory I/O per work band they belong to.
			Hence, it is left to estimate the number of vertices in the $k$-intersections for  $k\geq 2$.
			We count a vertex that belongs to $k$ different work bands $k$ times as the vertex causes 1 non-compulsory I/O per work band if $k=2$ and  less than 2 non-compulsory I/O per work band if $k\geq3$.
%
%
%
			Recall that the number of work bands in $H_h$ is bounded by Lem.~\ref{lem:nbrOfWorkBandsArb} as 
			\begin{equation*}\label{eq:recallWorkBands}
				\frac{\prod_{i=2}^d\lr{k_i+\mO{m}}}{\size{E^\infty_h}-\mO{m^{d-2}} } \;.
			\end{equation*}

			First, let us bound the number of vertices in all $2$-intersections.
			For $d\geq 3$ Assumption~\ref{assum:2intersects} of Def.~\ref{def:memEffBandAlgo} yields that a hyperplane $H_h$ contains at most  
%
			\begin{equation*}
			\lr{b \cdot m^{d-2}+\mO{m^{d-3}} }\cdot  \frac{\prod_{i=2}^d\lr{k_i+\mO{m}}}{\size{E^\infty_h}-\mO{m^{d-2}} } 
			\end{equation*}
			vertices in 2-intersections, counting them once for each work band they belong to.
			As this holds for all $h \in [k_1]$,  the grid $[k_1]\times\dots\times[k_d]$ contains at most
			\begin{align}\label{eq:2intersectsGrid}
					k_1\cdot \lr{\lr{b \cdot m^{d-2}+\mO{m^{d-3}} }\cdot \frac{\prod_{i=2}^d\lr{k_i+\mO{m}}}{\size{E^\infty_h}-\mO{m^{d-2}} } }
			\end{align}
			vertices in 2-intersections for $ d \geq 3$, counting them once for each work band they belong to.
			For $d=2$ the same argument yields that there are at most
			\begin{equation}\label{eq:2intersectsGrid2D}
				k_1\cdot \lr{\lr{b \cdot m^{d-2} }\cdot \frac{\prod_{i=2}^d\lr{k_i+\mO{m}}}{\size{E^\infty_h}-\mO{m^{d-2}} } }
			\end{equation}
			vertices in 2-intersections counting them once for each work band they belong to.

			Similarly, the vertices in the $k$-intersections for $k\geq3$ can be bounded.
			Fix a $k\geq 3$.
			For $d=2$ these $k$-intersections are empty by Assumption~\ref{assum:3intersects} of Def.~\ref{def:memEffBandAlgo}.
			So consider $d\geq 3$.
			Similarly to the 2-intersections, Assumption~\ref{assum:3intersects} of Def.~\ref{def:memEffBandAlgo} yields that the grid $[k_1]\times\dots\times[k_d]$ contains at most
			\begin{equation}\label{eq:3intersectsGrid}
				k_1\cdot  \mO{m^{d-3}}\cdot \frac{\prod_{i=2}^d\lr{k_i+\mO{m}}}{\size{E^\infty_h}-\mO{m^{d-2}} } 
			\end{equation}
			vertices in the $k$-intersection for a fixed $k \geq 3$, counting each vertex once for each work band it belongs to.

			Now, consider the I/Os necessary to load the first and last block of each $k$-intersection of $W_E$ in general containing vertices $w\notin W$.
			We disregarded loading these $w\notin W$ vertices up to now when we said that $E$ can be evaluated by one sweep of the work band $W_E$. 
			All in all, these are are less than 2 blocks or $2\cdot B$ vertices for each $k$-intersection of a work band.
			For each block we account for 2 non-compulsory I/Os, one read and one write operation.
			The number of work bands is given by Assumption~\ref{assum:nbrOfWorkBands} as $\size{\wL} = \mO{\frac{1}{M}\cdot \prod_{i=1}^{d-1}k_i}$.
			By Lemma~\ref{lem:kIntersectsBounded} the number of $k$-intersections per work band is constant. 
			Hence, the total number of non-compulsory I/Os caused by incomplete blocks at the beginning and end of the $k$-intersections is bounded by
			\begin{equation}\label{eq:IOsStartEndKintersect}
				\size{\wL} \cdot 2 \cdot 2 \cdot \mO{1} = \mO{\frac{1}{M}\cdot \prod_{i=1}^{d-1}k_i} \;.
			\end{equation}
			
			We are now ready to establish the upper bound for the non-compulsory I/Os of the memory efficient band algorithm~$\algo$.
			Accounting for the extra non-compulsory I/Os for incomplete blocks at the beginning and end of each $k$-intersection with \eqref{eq:IOsStartEndKintersect}, we can assume that we evaluate $E$ by sweeping through all $k$-intersections of  $W_E$ simultaneously.
			As we access the vertices of the $k$-intersections in their $k$-intersection order, we can always make use of the full block of data loaded.
%
			Recall that vertices of $2$-intersections take part in 1 non-compulsory I/O for their work band and the vertices of $k$-intersections for $k\geq3$ take part in less than 2 non-compulsory I/Os for each of their work bands.
			Also, by Assumption~\ref{assum:workbandConst} only $k$-intersections for $k$ up to some constant $f$ can be non-empty.

			If a vertex belongs to several work bands, it is already counted multiple times in \eqref{eq:2intersectsGrid}, \eqref{eq:2intersectsGrid2D} and \eqref{eq:3intersectsGrid}. 
			Hence, an upper bound for the non-compulsory I/Os of algorithm $\algo$ for $d \geq 3$ is given by 
			\begin{flalign*}
				&1\cdot \frac{1}{B}\cdot k_1\cdot \lr{b \cdot m^{d-2}+\mO{m^{d-3}} }\cdot\frac{\prod_{i=2}^d\lr{k_i+\mO{m}}}{\size{E^\infty_h}-\mO{m^{d-2}} }  +\\
				&\qquad +\sum_{j=3}^f \lr{2\cdot \frac{1}{B}\cdot k_1\cdot  \mO{m^{d-3}}\cdot \frac{\prod_{i=2}^d\lr{k_i+\mO{m}}}{\size{E^\infty_h}-\mO{m^{d-2}} }} +	\mO{\frac{1}{M}\cdot \prod_{i=1}^{d-1}k_i}= 
			\end{flalign*}
			\begin{flalign*}
				&\stackrel{\lr{\text{Assum.~\ref{assum:evalSize}}}}{=} \frac{k_1}{B}\cdot \frac{\prod_{i=2}^d\lr{k_i+\mO{m}}}{e\cdot m^{d-1}-\mO{m^{d-2}} } \cdot \lr{b\cdot m^{d-2}+\mO{m^{d-3}}}+ \mO{\frac{1}{M}\cdot \prod_{i=1}^{d-1}k_i}= \\
				&\stackrel{(\text{Assum. \ref{assum:evalWidth})}
					}{=} \frac{k_1}{B}\cdot \frac{\prod_{i=2}^d k_i}{e\cdot m^{d-1}-\mO{m^{d-2}} } \cdot b\cdot m^{d-2}+ \mO{\frac{k_1}{B}\cdot\frac{m \cdot \prod_{i=2}^{d-1} k_i}{e\cdot m^{d-1} } \cdot  b\cdot m^{d-2}}+\\
				&\qquad\qquad\qquad+ 
				\mO{\frac{k_1}{B}\cdot\frac{\prod_{i=2}^d\lr{k_i+\mO{m}}}{e\cdot m^{d-1}-\mO{m^{d-2}} } \cdot  m^{d-3}}
				+ \mO{\frac{1}{M}\cdot \prod_{i=1}^{d-1}k_i}=\\
				& =\frac{b\cdot \prod_{i=1}^d k_i}{B } \cdot \frac{1}{e\cdot m-\mO{1}} +\mO{\frac{ \prod_{i=1}^{d-1} k_i}{B}}+ \mO{\frac{\prod_{i=1}^d k_i}{B\cdot m^2} } + \mO{\frac{1}{M}\cdot \prod_{i=1}^{d-1}k_i}=\\
				&\stackrel{
					\eqref{eq:sweepSize}}{=} \frac{b\cdot \prod_{i=1}^d k_i}{B } \cdot \frac{1}{e\cdot \lr{\sqrt[d-1]{\frac{M}{2s \cdot c}}-\thet{\sqrt[d-1]{B} }}}+\mO{\frac{ \prod_{i=1}^{d-1} k_i}{B}}+ \mO{\frac{\prod_{i=1}^d k_i}{B\cdot \sqrt[d-1]{ M^2}} }=\\
				&= \frac{b\cdot \prod_{i=1}^d k_i}{B } + \lr{\frac{1}{ e\cdot \sqrt[d-1]{\frac{M}{2s \cdot c}} }+\thet{\frac{\sqrt[d-1]{B}}{\sqrt[d-1]{M^2}}}} +\\&\qquad\qquad\qquad+ \mO{\frac{ \prod_{i=1}^{d-1} k_i}{B}}+ \mO{\frac{\prod_{i=1}^d k_i}{B\cdot \sqrt[d-1]{ M^2}} }=\\
				&
				= \frac{b\cdot \sqrt[d-1]{c}}{e} \cdot \sqrt[d-1]{2s } \cdot \frac{\prod_{i=1}^d k_i}{B \cdot \sqrt[d-1]{M}} +\mO{ \frac{\prod_{i=1}^d k_i }{\sqrt[d-1]{B^{d-2} \cdot M^2}}}	\;.
			\end{flalign*}
			For $d=2$, the bound can be deduced by very similar calculations as (just the $\mO{m^{d-3}}$ terms are missing and the last step combining lower order terms is not possible)
			\begin{flalign*}
				&1\cdot \frac{1}{B}\cdot k_1\cdot b\cdot\frac{k_2+\mO{m}}{\size{E^\infty_h}-\mO{1} }  + \mO{\frac{ k_1}{M}}= \frac{b\cdot c}{e} \cdot 2s  \cdot \frac{k_1k_2}{B \cdot M} +\mO{\frac{k_1k_2}{ M^2} }+\mO{\frac{k_1}{B}}\; .
			\end{flalign*}
		\end{proof}

\section{The Upper Bounds}
	\label{sec:upperBoundsAlgo}
	This section gives several memory efficient band algorithms for 2 dimensions, 3 dimensions and arbitrary dimensions.
	The complexity of these algorithms is given by Thm.~\ref{thm:algoFrameworkUpperBounds}.
	The leading term of the non-compulsory I/Os of the presented memory efficient band algorithms is asymptotically optimal.
	The constant of the leading term of the non-compulsory I/Os depends on the choice of the sweep shape and sweep sequence.
	For 2 dimensions the constant of the leading term of the non-compulsory I/Os of the diagonal band algorithm matches the lower bound.
	In 3 dimensions, the best constant is by a factor of $\sqrt{2}$ worse than the lower bound and in arbitrary dimensions $d$ the best constant differs by a factor of $\sqrt[d-1]{d!} \stackrel{d \text{ large} }{\approx} \frac{d}{e}$.
	In addition, we present two standard algorithms, that work on the standard row- and column-major data layout, which are also blocked and show that their performance is asymptotically worse.

	\subsection{General Upper Bounds for Arbitrary Dimensions}
	\label{sec:algosArbD}

	This section discusses algorithms that work in arbitrary dimensions.
	First, we discuss the hypercube band algorithm which takes advantage of the non-standard data layout specified in Thm.~\ref{thm:algoFrameworkUpperBounds}.
	Then, we analyze two standard algorithms, the row and the column algorithm.
	These algorithms are standard in the sense that they work on the standard row- or column-major layout of the data, respectively.
	They both work with basic blocks given by the work bands just as the band algorithms do.
	Due to the standard data layout, however, the Row and the Column Algorithm cannot take advantage of the blocks with respect fastest changing dimension.
	As the Row and the Column Algorithms do not work on the band layout, we cannot take advantage of Thm.~\ref{thm:algoFrameworkUpperBounds} to analyze their complexity.
	The leading terms of the non-compulsory I/Os for the Row and Column algorithms as well as for the Hypercube Band Algorithm are summarized in Tab. \ref{tab:dDUpperBounds}.
	The Row Algorithm is factor of $\thet{\sqrt[d-1]{B}}$ worse then the lower bound and the Column Algorithm by a factor of $\thet{B}$.
	The Hypercube Band Algorithm achieves the correct asymptotic of the leading term of the non-compulsory I/Os but is by a constant factor of $\sqrt[d-1]{d!}$ worse than the lower bound.
	As all three algorithms work in particular in the 2- and 3-dimensional setting, they are also fundamental algorithms for the low-dimensional cases.


\begin{table}[htbp]
		\footnotesize
		\centering
		\renewcommand{\tabcolsep}{2pt}
		\caption{
		Leading term of the non-compulsory I/Os for different algorithms in arbitrary dimension $d$.
		} 
		\label{tab:dDUpperBounds}
			\begin{tabular}{c|c|c|c}
				\textbf{$d$D Algorithms} & Sweep Shape  & Sweep Seq. & Non-Compulsory I/Os \\
				\hline & & & \\[-1ex]
				Row Algorithm & 	Hypercube   & $e_1$ &	$\Om{\frac{1}{\sqrt[d-1]{B^{d-2}\cdot M}}\cdot \prod_{i=1}^d k_i}$ \\[1.5ex]
				Column Algorithm & 	Hypercube   & $e_1$ &	$\Om{\frac{1}{\sqrt[d-1]{ M}}\cdot \prod_{i=1}^d k_i}$ \\[1.5ex]
				Hypercube Band Algorithm & Hypercube& $e_1$ & $   \frac{4s \cdot \sqrt[d-1]{2s }\cdot (d-1) }{B \cdot \sqrt[d-1]{M}} \cdot \prod_{i=1}^d k_i $  \\[1.5ex]
				\hline
				\hline&&&\\[-1.5ex]
				\textbf{Lower Bound} & n.a. & n.a. &$	\frac{1}{\sqrt[d-1]{d!} }\cdot\frac{4s\cdot \sqrt[d-1]{2s} \cdot(d-1)   }{B \cdot \sqrt[d-1]{M}} \cdot   \prod_{i=1}^d k_i$
			\end{tabular}
\end{table}

	\subsubsection{The Hypercube Band Algorithm: Asymptotically Optimal for Arbitrary Dimensions}
\begin{figure}[tb]
			\centering
				\includegraphics[width=.45\columnwidth, bb=5cm 18.5cm 12.3cm 23.8cm]{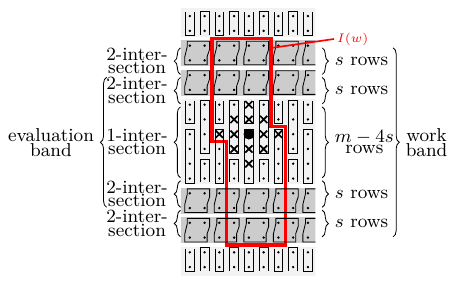} \hspace{.5cm}
				\includegraphics[width=.45\columnwidth, bb=5cm 17.8cm 12.3cm 23.8cm]{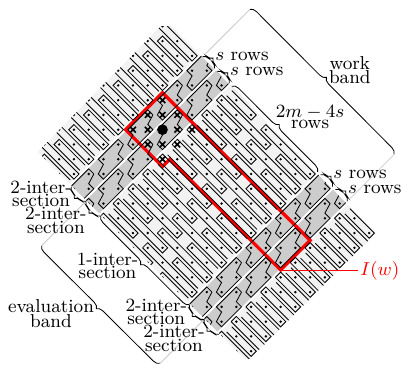}
			\caption{In two dimensions: the Hypercube Band Algorithm (lower left) and the Diagonal Band Algorithm (lower right) for $s=2$. Vertices currently in internal memory in red.}
			\label{fig:2DUpperBoundsHyperCubeDiagonal}
		\end{figure}

		This section discusses the \emph{Hypercube Band Algorithm} which is a simple, asymptotically matching upper bound for arbitrary dimension $d$. 
		As the Hypercube Band Algorithm works on a data layout supporting its particular access, its complexity is matching the lower bound asymptotically.
		The Hypercube Band Algorithm is within the framework of memory efficient band algorithms and hence we can apply Theorem~\ref{thm:algoFrameworkUpperBounds}.
		For two dimensions, this algorithm is depicted in Fig.~\ref{fig:2DUpperBoundsHyperCubeDiagonal}.

		\begin{figure}[tb]
			\centering
				\includegraphics[width=.45\columnwidth, bb=5cm 19.5cm 12.3cm 23.7cm]{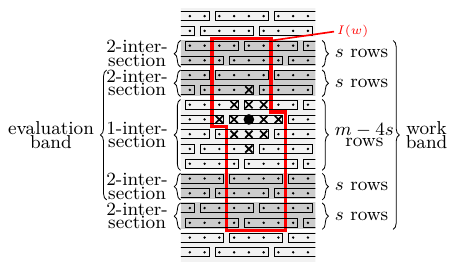} \hspace{.5cm}
				\includegraphics[width=.45\columnwidth, bb=5cm 19.5cm 12.3cm 23.7cm]{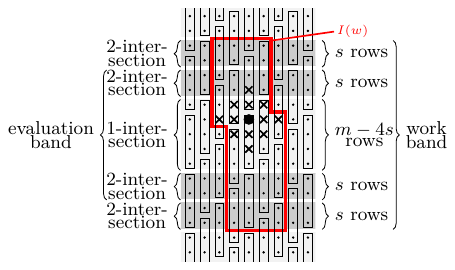}
			\caption{In two dimensions: the Row Algorithm (left) and the Column Algorihtm (right) for $s=2$. Vertices currently in internal memory in red.}
			\label{fig:2DUpperBoundsRowCol}
		\end{figure}

		As sweep sequence $\sweepSeq$ only the single unit shift $e_1$ is employed.
		The sweep shape $\sH$ is a  $d-1$ dimensional hypercube lying in a hyperplane of normal $e_1$ with $m$ vertices in each of the remaining $d-1$ directions,
		\begin{equation*}
			\sH = \left\{x  = 
				\lr{x_1,\dots,x_d}\in [k_1]\times \dots \times [k_d]: x_1 =1 \wedge\lr{ 1\leq x_i\leq m \text{ for } 2 \leq i \leq d }\right\} 
		\end{equation*}
		the sweep shape consists of 
			 $\size{\sH} =m^{d-1} $
		vertices in total.
		The constant $c$ describing the relation between the size parameter $m$ and the number of vertices in $\sH$ in \eqref{eq:sweepSizeDef} is therefore $c=1$.
		The intersection of a resulting infinite evaluation band with a hyperplane of normal $e_1$ is a hypercube of side length $m-2s$.
				Therefore 
		\begin{equation*}
			\size {E^\infty_h} = (m-2s)^{d-1} = m^{d-1}+\mO{m^{d-2}}
		\end{equation*}
		and hence $e=1$ (Assumption~\ref{assum:evalSize}).
		The evaluation bands are characterized by $E^\infty_h$ being a $d-1$ dimensional hypercube of side length $m-2s$ lying in the center $W^\infty_h$, i.e. $s$ units away from every side of the work band.
		Hence Assumptions \ref{assum:evalWidth} and~\ref{assum:workEvalDistance} of Def.~\ref{def:memEffBandAlgo} is fulfilled.
		
\begin{figure}
			\centering
			\begin{tabular}{cp{1em}c}
				\begin{tikzpicture}[scale=.3]
					\draw[gray,thick,densely dotted] (0,0) grid (11.5,7.5);

					\drawSquare{0}{0}{3}{0.2}{yellow}
					\drawSquare{4}{0}{3}{0.2}{yellow!33!blue}
					\drawSquare{8}{0}{3}{0.2}{yellow}

					\drawSquare{0}{4}{3}{0.2}{yellow!33!blue}
					\drawSquare{4}{4}{3}{0.2}{yellow}
					\drawSquare{8}{4}{3}{0.2}{yellow!33!blue}

					\draw [decorate,decoration={mirror, brace},thick] (0,-.4) -- (3,-.4) node [below,midway] {$m-2s$};
					\draw [decorate,decoration={ brace},thick] (-.4,0) -- (-.4,3) node [left,midway] {$m-2s$};
				\end{tikzpicture}&&

				\begin{tikzpicture}[scale=.3]
					\draw[fill, green!30!white] (3,8) rectangle (8,3);

					\draw[fill, yellow!30!white] (-1,-1) rectangle (4,4);

					\draw[fill, yellow!50!green] (3,3) rectangle (4,4);
					\draw[fill, yellow!50!green] (-1,3) rectangle (0,4);

					\draw[fill, yellow!50!brown] (0,3) rectangle (3,4);

					\fill[pattern=custom north west lines, hatchspread=10pt, hatchcolor=yellow!66!blue,hatchthickness=2pt] (-1,3) rectangle (4,8);
					\fill[pattern=north east lines, hatchcolor=yellow!66!blue] (3,-1) rectangle (8,4);

					\draw[gray,thick,densely dotted] (-1.5,-1.5) grid (8.5,8.5);
	
				 	\drawSquare{0}{0}{3}{0.2}{yellow}
					\drawSquare{4}{0}{3}{0.2}{yellow!33!blue}
					\drawSquare{0}{4}{3}{0.2}{yellow!66!blue}
					\drawSquare{4}{4}{3}{0.2}{yellow!33!green}

					\draw [decorate,decoration={mirror, brace},thick] (-1.4,-1.4) -- (4.4,-1.4) node [below,midway] {$m$};
					\draw [decorate,decoration={ brace},thick] (-1.6,-.4) -- (-1.6,3.5) node [left,midway] {$m-2s$};
					\draw [decorate,decoration={ brace},thick] (-1.6,-1.4) -- (-1.6,-.6) node [left,midway] {$s$};
					\draw [decorate,decoration={ brace},thick] (-1.6,3.6) -- (-1.6,4.3) node [left,midway] {$s$};

					\draw [decorate,decoration={brace},thick] (0.6,8.6) -- (2.4,8.6) node [above,midway] {$m-4s$};

					\draw [decorate,decoration={ mirror, brace},thick] (8.7,2.6) -- (8.7,4.4) node [right,midway,distance=10cm] {$2s$};

				\end{tikzpicture}
			\end{tabular}
			\caption{For $d=3$, $m=4$ and $s=1$: \emph{Left:} Covering a hyperplane of $H_h$ (2-dimensional) with the evaluation bands resulting from sweeping a $d-1$-dimensional hypercube along $e_1$.
					\emph{Right:} the size of the $k$-intersections determined by the work bands surrounding the evaluation bands per $H_h$.
					}
			\label{fig:rowAlgoTilingEvalBands}
		\end{figure}
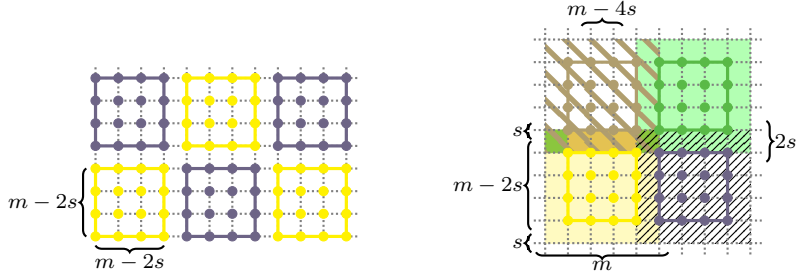

		As the sweep sequence consists only of the unit shift $e_1$, covering the grid $[k_2]\times\dots\times[k_n]$ with sets  $E^\infty_h$ yields a cover for the $d$ dimensional grid once the sweep sequence is applied.
		As $E^\infty_h$ is also a hypercube, it is easy to cover the $d-1$ dimensional grid.
		The grid $[k_2]\times\dots\times[k_n]$ is covered by the evaluation bands if an evaluation band is placed every $m-2s$ vertices for the last $d-1$ unit directions in a lattice-like structure as shown in Fig.~\ref{fig:rowAlgoTilingEvalBands}. 
		This specifies the list of work bands $\wL$.
		The number of work bands per $H_h$ equals the total number of work bands and is 
		\begin{equation*}
			\size{\wL} = \prod_{i=2}^d{\left\lceil \frac{k_i}{m-2s}\right\rceil} = \thet{\prod_{i=2}^d\frac{k_i}{m}}\stackrel{\eqref{eq:sweepSize}}{=} 
			\mO{\frac{1}{M}\cdot\prod_{i=2}^d k_i} \;. 
		\end{equation*}
		Hence, Assumption~\ref{assum:nbrOfWorkBands} is satisfied.
%
		By construction, the evaluation bands cover the grid (Assumption~\ref{assum:evalBandsCover}). 
%
		One work band $W\in \wL$ overlaps  with at most $3^{d-1}-1= \thet{1}$ other work bands and hence Assumption~\ref{assum:workbandConst} is met. 

		Lemma~\ref{lem:distSweepX1} ensures that Assumption~\ref{assum:evalPoints} is met.
		
		\begin{lem}\label{lem:distSweepX1} 
			Given the $d$ dimensional grid $[k_1]\times\dots\times[k_d]$ and the setup of Def.~\ref{def:memEffBandAlgo} assuming that the sweep sequence is $\sweepSeq = \{e_1\}$.
			It then holds for any $W \in \wL$ and any $w \in E_W$ that
			\begin{equation*}
				I(w) = [x-\delta, x+\delta] \quad \text{ with } \quad \delta = s\cdot \size{\sH} \;.
			\end{equation*}
		\end{lem}

		\begin{proof}
			By \eqref{eq:IwCharacterization} we know that $I(w) = [o_W(w_{\min}), o(w_{\max})]$ with $w_{\min} = \argmin\{o_W(w)\!: w \in S_s(w)\} $ and $w_{\max}\argmax\{o_W(w): w \in S_s(w) \}$.
			Denote by $\sH$ the sweep shape of~$W$ that contains $w$ and by~$\sH_k$ the sweep shape that is preceding ($k <0$) or proceeding ($k>0$) $\sH$ in the work band by $k$ shifts.
			As we are only sweeping in $x_1$-direction, the only vertex of $S_s(w)$ which is in $\sH_{-s}$ is $w - s\cdot e_1$.
			Also, $S_s(w) \cap \sH_{k}= \emptyset$ for $k < -s$.
			Hence $w_{\min} = w - s\cdot e_1$.
			As $w_{\min}$ and $w$ have the same lexicographic position within their respective sweep shape we get $||w_{\min} - w||_W = s \cdot \sH$.
			Similarly $w_{\max} = w + s\cdot e_1$ and $||w_{\min} + w||_W = s \cdot \sH$ and hence the claim follows.
		\end{proof}				

		It is left to estimate the size of the $k$-intersections per hyperplane of normal $e_1$, $\size{\Phi_{(W,k,h)}}$ (see Fig.~\ref{fig:rowAlgoTilingEvalBands}).
		For one work band, there are $2(d-1)$ different 2-intersections, one for each of the faces of the $d-1$ dimensional hypercube sweep shape.
		Per hyperplane of normal $e_1$, any 2-intersection contains $2s\cdot (m-4s)^{d-2}$ vertices. 
		In total these are
		\begin{equation*}
			2(d-1) \cdot 2s\cdot (m-4s)^{d-2}  = 4 s (d-1) m^{d-2}+\mO{m^{d-3}}
		\end{equation*}
		vertices if $ d\geq 3$. For $d=2$ these are
			$2\cdot (2-1) \cdot 2s\cdot (m-4s)^{2-2}  = 4 s$ 
		vertices.
		Hence, $b =  4 s\cdot (d-1)$ in both cases and Assumption~\ref{assum:2intersects} holds.

		If three or more work bands intersect, at least one of them has to be offset from $W$ in two different unit directions. 
		The intersection of two work bands contains $2s$ vertices in direction $i$ if the work bands are at a offset in this direction and $m-4s$ vertices if they are not at offset in this direction.
		Hence, the number of vertices in any $k$-intersection for $k\geq 3$ is $\mO{k_1 \cdot m^{d-3}}$ or $\mO{m^{d-3}}$ vertices per hyperplane of normal $e_1$.
		As there are only a constant number of non-empty $k$-intersections per work band by Lemma~\ref{lem:kIntersectsBounded}, also $\size{\Phi_{(W,k,h)}} = \mO{m^{d-3}}$ for any $k \geq 3$.
		If $d=2$, only up to two work bands overlap and the $k$-intersections for $k\geq 3$ are empty.
		This yields that Assumption~\ref{assum:3intersects} is also satisfied.

		Hence, Theorem~\ref{thm:algoFrameworkUpperBounds} can be applied to the Hypercube Band Algorithm with $ c = e =1$ and $ b =  4 s\cdot (d-1)$.
		Therefore, the number of non-compulsory I/Os of the Hypercube Band Algorithm is upper bounded by 
		\begin{align*}
			 4s  \cdot \sqrt[d-1]{2s }\cdot (d-1) \cdot \frac{\prod_{i=1}^d k_i}{B \cdot \sqrt[d-1]{M}} +\mO{ \frac{\prod_{i=1}^d k_i }{\sqrt[d-1]{B^{d-2} \cdot M^2}}} \; .
		\end{align*}
		For $d=2$ and $d=3$ this bound reads
		\begin{align*}
			&d = 2: \quad    8s^2  \cdot \frac{k_1k_2}{B \cdot M} +\mO{\frac{k_1k_2}{ M^2} }+\mO{\frac{k_1}{B}}\; \text{ and}\\[2ex]
			&d \geq 3: \quad  8\cdot\sqrt{2}\cdot s^{3/2} \cdot \frac{k_1k_2k_3}{B \cdot \sqrt{M}} +\mO{ \frac{k_1k_2k_3}{\sqrt{B \cdot M^2}}} \; .
		\end{align*}

	\subsubsection{Standard Row and Column Algorithms: Baselines With Non-Optimal Asymptotic Behavior}

		This section analyses two standard approaches to evaluate stencils in arbitrary dimension called Row Algorithm and Column Algorithm.
		Both algorithms are depicted in Fig.~\ref{fig:2DUpperBoundsRowCol} for two dimensions.
		Both algorithms cause an asymptotically non-optimal number of non-compulsory I/Os.
		Hence, the analysis focuses on the asymptotic number of non-compulsory I/Os only and does neither derive the constant of the leading term nor lower order terms.

		The algorithms are standard in the sense that they work on the common data layouts used to stored multidimensional arrays, i.e. grids, and not the layout specified in Theorem~\ref{thm:algoFrameworkUpperBounds}.

		These standard layouts are the row- and the column-major layout.
		To be specific, the first coordinate $x_1$ is changing fastest  in the row-major layout and the second coordinate $x_2$ is changing fastest  in the column-major layout.

		The Row Algorithm and the Column Algorithm are going to sweep a $d-1$ dimensional hypercube through the grid in $x_1$-direction.
		Sweeping the hypercube in $x_1$-direction ($x_2$-direction) in a row layout is the same as sweeping the same hypercube in $x_2$-direction ($x_1$-direction) in a column layout. 
		Hence only sweeps in $x_1$-direction will be discussed but both, row and column, layouts. 
		Also, layouts in which the $i$.th coordinate is changing fastest could be discussed.
		It only matters, however, whether we sweep along the fastest changing index or not.
		Hence discussing row and column layouts covers all cases.

		The Row Algorithm and the Column Algorithm are specified by their sweep sequence, sweep shape, list of work bands and the data layout they work on.
		Apart the data layout, the algorithms work precisely as described in Thm.~\ref{thm:algoFrameworkUpperBounds}.
		However, as the Row and the Column Algorithm use a different data layout than specified in Thm.~\ref{thm:algoFrameworkUpperBounds} we cannot use this theorem to derive the number of non-compulsory I/Os for these algorithms.
		The structure and analysis of the algorithms is, however, very similar to the one carried out in Thm.~\ref{thm:algoFrameworkUpperBounds}.

		\para{Row Algorithm}
		
			The Row Algorithm is based on a standard row-major layout of the vertices of the grid.
			For two dimensions, the Row Algorithm is depicted in Fig.~\ref{fig:2DUpperBoundsRowCol}.
			The sweep sequence is $\sweepSeq = \{ e_1 \}$ and the sweep shape a $d-1$ dimensional hypercube of side length $m$ lying in a hyperplane of normal $e_1$.
			As the sweep shape and sweep sequence are the same as for the Hypercube Band algorithm, the work bands, evaluation bands  and list of work bands $\wL$ are identical to those of the Hypercube Band Algorithm.
			The sweep shape consists of $m^{d-1}$ vertices.
			As we sweep solely in $x_1$-direction, the work and evaluation bands extend in this direction.
			The evaluation bands are characterized by $E^\infty_h$ being a $d-1$ dimensional hypercube of side length $m-2s$ lying in the center $W^\infty_h$.
			Hence the grid is covered by the evaluation bands if an evaluation band is placed every $m-2s$ vertices for the last $d-1$ unit directions as shown in Fig.~\ref{fig:rowAlgoTilingEvalBands}. 
			This specifies the list of work bands $\wL$.
			The number of work bands per $H_h$ equals the total number of work bands and is 
			\begin{equation*}
				\size{\wL} = \prod_{i=2}^d{\left\lceil \frac{k_i}{m-2s}\right\rceil} = \thet{\prod_{i=2}^d\frac{k_i}{m}} \;. 
			\end{equation*}
			The evaluation order of the vertices as well as the data that is kept in internal memory is as specified in Thm.~\ref{thm:algoFrameworkUpperBounds}. 

	
			Let us determine the maximum size $m$ of the sweep shape.
			Consider the input first.
			As the blocks of the data layout extend in $x_1$-direction we need to keep at least one block of data in internal memory for each of the $m^{d-1}$ rows of the sweep shape.
			We assume $B \geq 2s+1$ and hence one block of data covers $2s+1$ sweep shapes allowing the middle sweep shape to be evaluated.\footnote{
			The worst case would be a block aligned data layout, i.e. all blocks start at the same indices of $x_1$ for all rows.
			In a block aligned data layout it would be necessary to keep the $2s$ previous sweep shapes in internal memory in addition to the $m^{d-1}$ new blocks that are loaded when the end of a set of blocks is reached.
			}
			For the output we need one block of data for each row of the evaluation band, i.e. $(m-2s)^{d-1}$ blocks.
			Hence, at least $B \cdot \lr{m^{d-1} + (m-2s)^{d-1}}= \thet{B \cdot m^{d-1}}$ vertices have to be in internal memory at once.
			This means that the sweep shape size has to be chosen in the order of 
			\begin{equation*}
				m = \mO{\sqrt[d-1]{\frac{M}{B}}}\;. 
			\end{equation*}

			To determine the number of non-compulsory I/Os consider the $k$-in\-ter\-sec\-tions.
			The data layout is not organized by $k$-intersections but the $k$-intersections still describe the number of non-compulsory I/Os a vertex (or block of vertices) has to cause.
			A work band exceeds its evaluation band by $2s$ vertices ($s$ vertices before and $s$ vertices after the evaluation band) in each of the $d-1$ coordinates $x_i$ for $2 \leq i \leq d$.
			As the evaluation bands cover the grid without overlap, this means that the 1-intersection of each work band (besides the first and last work band in each direction) is characterized by $\Phi_{(W,1,h)}$ being a $d-1$ dimensional hypercube of side length $(m-4s)$ lying in the middle of $W$.
			Hence per $H_h$ the vertices of a work band shared with other work bands is given by
			\begin{align*}
				\bigcup_{k=2}^\infty \Phi_{(W,k,h)} = m^{d-1}- \lr{m-4s}^{d-1} =
				\thet{m^{d-2}} \;.
			\end{align*}
			With the conservative assumption that all these vertices are only shared by two work bands, each of them causes one non-compulsory I/O for each of the work bands (see also the similar discussion in the proof of Thm.~\ref{thm:algoFrameworkUpperBounds}).
			Each row of the grid is contiguous in memory and so we have one I/O every $B$ vertices of each row, or $k_1 \big/ B$ I/Os per row.
			For the asymptotic analysis we ignore blocks that contain two different rows as this effects 
			only lower order terms.

			Hence, the Row Algorithm causes at least
			\begin{align*}
				\thet{\prod_{i=2}^d\frac{k_i}{m}}\cdot \frac{k_1}{B}\cdot \thet{m^{d-2}} = \thet{\frac{\prod_{i=1}^d k_i}{B} \frac{1}{m}} = 
			\Om{ \frac{\prod_{i=1}^d k_i}{\sqrt[d-1]{B^{d-2}\cdot M}} }
			\end{align*}
			non-compulsory I/Os. Although we do not provide a proof, we claim that this analysis is tight, i.e. the Row Algorithm causes that many non-compulsory I/Os and the effects disregard only concern lower order terms.
			For $d=2$ and $d=3$ this bound reads
			\begin{align*}
				d= 2 : \quad \Om{\frac{k_1k_2}{M}} \qquad \text{and} \qquad d=3 : \quad \Om{\frac{k_1k_2k_3}{\sqrt{B\cdot M}}}\; .
			\end{align*}

			The Row Algorithm causes an asymptotically non-optimal number of non-com\-pul\-so\-ry I/Os as the row-major data layout forces the algorithm  to keep a whole block of $B$ vertices in internal memory for each row of the sweep shape.
			To evaluate the current sweep shape, however, at most the $s$ pre- and $s$ proceeding sweep shapes are required in memory and hence only a constant number of vertices (at most $2s+1$) per row.
			This forces the Row Algorithm to choose a relatively small sweep shape size $m$ and hence decreases the interior-to-boundary ratio of the work bands.

	\para{Column Algorithm}

			The Column Algorithms is based on a standard column-major layout of the vertices of the grid.
			For two dimensions, the Column Algorithm is depicted in Fig.~\ref{fig:2DUpperBoundsRowCol}.
			This means that blocks extend in $x_2$ direction.
			The sweep sequence is $\sweepSeq = \{ e_1 \}$ and the sweep shape a $d-1$ dimensional hypercube of side length $m$ consisting of $m^{d-1}$~vertices.
			As the sweep sequence and sweep shape are identical to the Row Algorithm and the Hypercube Band Algorithm, the work bands, evaluation bands and list of work bands $\wL$ are also identical to those algorithms.
			Hence, the number of work bands is $\size{\wL} =\thet{\prod_{i=2}^d\frac{k_i}{m}}$ as for the Row Algorithm.
			The evaluation order of the vertices as well as the data that is kept in internal memory is as specified in Thm.~\ref{thm:algoFrameworkUpperBounds}. 
	
			Let us determine the maximum size $m$ of the sweep shape.
			For each of its $m^{d-2}$ columns, a sweep shape consists of $\ceil{\frac{m}{B}}$ blocks of data.
			Hence, the number of vertices contained in the blocks of a sweep shape is
			\begin{equation*}
				B \cdot \ceil{\frac{m}{B}} \cdot m^{d-2} = \thet{m^{d-1}} \;. 
			\end{equation*}
			The algorithm needs to keep at least $2s-1 = \thet{1}$ complete sweep shapes in internal memory for the input, i.e. the sweep shape currently being evaluated as well as the $s-1$ preceding and $s-1$ proceeding sweep shapes.
			For the asymptotic analysis we disregard blocks of the output that have to reside in internal memory.\footnote{	
			If the vertices are evaluated in lexicographic order one output block per column of the sweep shape, hence $m$ blocks in total, should be kept in memory.
			This can be reduced to one output block if the vertices are evaluated according to columns, i.e. with $x_2$ and not $x_d$ being the fastest changing index for the evaluation order
			}
			Hence, the number of vertices that have to be in internal memory at once is lower bounded by $(2s-1)\cdot \thet{m^{d-1}}= \thet{m^{d-1}}$.
			We therefore know that the size of the sweep shape is
			\begin{equation*}
				m = \mO{\sqrt[d-1]{M}} \;.
			\end{equation*}

			To determine the number of non-compulsory I/Os consider the $k$-intersections.
			As for the Row Algorithm, the 1-intersection of each work band (besides the first and the last work band in each direction) is characterized by $\Phi_{(W,1,h)}$ being a $d-1$ dimensional hypercube of side length $(m-4s)$ lying in the middle of $W$.
			Hence per $H_h$ and work band $W$ the section above and below the evaluation band in $x_2$ direction is a $d-1$ dimensional hypercube of $2s$ vertices in $x_2$ direction and $(m-2s)$ for all directions $x_i$ for $i \in \lrC{3,\dots,d}$. 
			Hence, there are $(m-2s)^{d-2}$ columns per work band and $H_h$ that contain vertices that are also shared with other work bands.
			This means that there are at least $ 2\cdot (m-2s)^{d-2} = \thet{m^{d-2}}$ blocks per work band and $H_h$ causing non-compulsory I/Os.
			With the conservative assumption that all these blocks are only shared by two work bands, each of them causes one non-compulsory I/O for each work band (see also the similar discussion in the proof of Thm.~\ref{thm:algoFrameworkUpperBounds}).
			We disregard all other $2 \cdot (d-2)$ faces of the evaluation band that also contain vertices of the $k$-intersections for $k \geq 2$.
			These vertices only increase the number of non-compulsory I/Os but would not change the asymptotic behavior of the leading term of the non-compulsory I/Os.

			Hence, the Column Algorithm causes at least
			\begin{align*}
				\thet{\prod_{i=2}^d\frac{k_i}{m}}\cdot k_1 \cdot \thet{m^{d-2}} =  \thet{ \frac{\prod_{i=1}^d k_i}{m}} = 
					\Om{ \frac{\prod_{i=1}^d k_i}{\sqrt[d-1]{M}} }
			\end{align*}
			non-compulsory I/Os. Although we do not provide a proof, we claim that this analysis is tight, i.e. the Column Algorithm causes that many non-compulsory I/Os and the effects disregarded only concern lower order terms.
			For $d=2$ and $d=3$ this bound reads
			\begin{align}
				d= 2 : \quad \Om{\frac{k_1k_2}{M}} \qquad \text{and} \qquad d=3 : \quad \Om{\frac{k_1k_2k_3}{\sqrt{M}}}\; .
			\end{align}

			The Column Algorithm causes an asymptotically non-optimal number of non-compulsory I/Os as the column-major data layout forces the algorithm to perform one non-compulsory I/O per column of the sweep shape to load and store  the $k$-intersections for $k\geq 2$.
			Per column, however, there are only a constant number of $2s$ vertices in the $k$-intersection above and below the evaluation band (besides for the outermost columns).
			This forces the Column algorithm to perform a non-compulsory I/O for each column and each step in $x_1$-direction instead of performing an non-compulsory every $2s \big/ B$ steps in $x_1$-direction and each column.


%
%


\subsection{Upper Bounds in 2 Dimensions}
	\label{sec:2DLayouts}
	
	The upper bounds in two dimensions are summarized in Tab. \ref{tab:2DUpperBounds} and the layouts depicted in Fig.~\ref{fig:2DUpperBoundsRowCol} and~\ref{fig:2DUpperBoundsHyperCubeDiagonal}.
	The Row and the Column Algorithm are both by a factor of $\Om{B}$ worse than the optimum.
	The two dimensional Hypercube Band Algorithm achieves the correct asymptotics but the leading term of the non-compulsory I/Os is by a factor $2$ worse than the lower bound.
	In \cite{leopold02relaxation} Leopold uses a mixed row/column layout which achieves the same constant at the leading term of the non-compulsory I/Os as the Hypercube Band Algorithm.
	In contrast, the constant of the lower bound is matched by the Diagonal Band Algorithm which is depicted in the right part of Fig.~\ref{fig:2DUpperBoundsHyperCubeDiagonal}.
	The key observation is that
	shifting a sweep shape  which lies in a hyperplane of normal $(1,\,-1)$ in the two unit directions $(1,\,0)$ and $(0,\,1)$ alternately doubles the vertices of an evaluation band while the number of vertices of the $2$-intersections stays constant per evaluation band .

\begin{table}[htbp]
		\footnotesize
		\centering
				\caption{	Leading term of the non-compulsory I/Os for different  algorithms in two dimensions.
		} 
		\label{tab:2DUpperBounds}
			\begin{tabular}{c|c|c|c}
				\textbf{2D Algorithms} & Sweep Shape  & Sweep Seq. & Non-Compulsory I/Os \\
				\hline & & & \\[-1ex]
				Row Algorithm & 	Vertical   & $e_1$ &	$\Om{\frac{1}{M}\cdot k_1k_2}$ \\[1.5ex]
				Column Algorithm & 	Vertical   & $e_1$ &	$\Om{\frac{1}{M}\cdot k_1k_2}$ \\[1.5ex]
				Hypercube Band Algorithm & Vertical  & $e_1$ & $ \frac{8s^2}{BM}\cdot k_1k_2$  \\[1.5ex]
				Diagonal Band Algorithm & 	Diagonal & $e_1$, $e_2$&	$\frac{4s^2}{BM} \cdot k_1k_2$  \\[1.5ex]
				\hline
				\hline&&&\\[-1.5ex]
				\textbf{Lower Bound} & n.a. & n.a. &$	\frac{4s^2}{BM}\cdot k_1k_2$
			\end{tabular}
\end{table}

	\subsubsection{Diagonal Band Algorithm (see Fig. \ref{fig:2DUpperBoundsHyperCubeDiagonal} - right): Optimal in 2~Dimensions}

\begin{figure}
	\centering
		\begin{tabular}{cc}
		\footnotesize
		\centering
			\begin{tikzpicture}[scale =.125]

				\draw[blue!20, fill] (0,12) -- (3.3,12) -- (0,8.7);

				\draw[red!20, fill] (0,8.3) -- (3.7,12) -- (7.3,12) -- (0,4.7) -- cycle;
				\draw[blue!20, fill] (0,4.3) -- (7.7,12) -- (11.3,12) -- (0,.7) -- cycle;
				\draw[red!20, fill] (0,.3) -- (11.7,12) -- (15.3,12) -- (3.3,0) -- (0,0) -- cycle;
				\draw[blue!20, fill,xshift=-1cm] (4.7,0) -- (16.7,12) -- (20.3,12) -- (8.3,0) -- cycle;
				\draw[red!20, fill,xshift =7 cm] (.7,0) -- (12.7,12) -- (16.3,12) -- (4.3,0) -- cycle;
				\draw[blue!20, fill,xshift = 7 cm] (4.7,0) -- (16.7,12) -- (20.3,12) -- (8.3,0) -- cycle;
				\draw[red!20, fill,] (15.7,0) -- (27.7,12) -- (30,12) -- (30,10.7) -- (19.3,0) --  cycle;
				\draw[blue!20, fill] (19.7,0) -- (30,10.3) -- (30,6.7) -- (23.3,0) -- cycle;
				\draw[red!20, fill,] (23.7,0) -- (30,6.3) -- (30,2.7)  -- (27.3,0) --  cycle;
				\draw[blue!20, fill] (27.7,0) -- (30,2.3) -- (30,0) -- cycle;

				\draw[gray] (0,0) grid (30,12);

				\draw[very thick, blue,dashed, ->] (.2,10.5) -- (1.5,11.8);
				\draw[very thick, red,dashed, ->] (.2,6.5) -- (5.5,11.8);
				\draw[very thick, blue,dashed, ->] (.2,2.5) -- (9.5,11.8);
				\draw[very thick, red,dashed, ->] (1.5,.2) -- (13.5,11.8);
				\draw[very thick, blue,dashed, ->] (5.5,.2) -- (17.5,11.8);
				\draw[very thick, red,dashed, ->] (9.5,.2) -- (21.5,11.8);
				\draw[very thick, blue,dashed, ->] (13.5,.2) -- (25.5,11.8);
				\draw[very thick, red,dashed, ->] (17.5,.2) -- (29.5,11.8);
				\draw[very thick, blue,dashed, ->] (21.5,.2) -- (29.8,8.5);
				\draw[very thick, red,dashed, ->] (25.5,.2) -- (29.8,4.5);
				\draw[very thick, blue,dashed, ->] (29.5,.2) -- (29.8,0.5);

				\draw[thick, densely dotted] (7.7,0) -- (11.3,0) -- (9.5, 1.8) -- cycle;
				\draw[thick, densely dotted] (11.3,0) -- (9.5, 1.8) -- (11.5, 3.8) -- (13.3,2) --  cycle;
				\draw[thick, densely dotted,xshift=2cm,yshift=2cm] (11.3,0) -- (9.5, 1.8) -- (11.5, 3.8) -- (13.3,2) --  cycle;
				\draw[thick, densely dotted,xshift=4cm,yshift=4cm] (11.3,0) -- (9.5, 1.8) -- (11.5, 3.8) -- (13.3,2) --  cycle;
				\draw[thick, densely dotted,xshift=6cm,yshift=6cm] (11.3,0) -- (9.5, 1.8) -- (11.5, 3.8) -- (13.3,2) --  cycle;
				\draw[thick, densely dotted,xshift=8cm,yshift=8cm] (11.3,0) -- (9.5, 1.8) -- (11.5, 3.8) -- (13.3,2) --  cycle;
				\draw[thick, densely dotted] (19.5,11.8) -- (21.3,10) -- (23.3,12)  -- (19.7, 12) -- cycle;

				\draw (16,12) node[above] {\scriptsize adjacent $\ell^1$ balls};
				\draw[->] (15.5,12.8) -- (15.5,6);
				\draw[->] (15.5,12.8) -- (17.5,8);
				\draw[->] (15.5,12.8) -- (19.5,10);
				\draw[->] (15.5,12.8) -- (21,11.5);
				\draw[->] (15.5,12.8) -- (13.5,4);
				\draw[->] (15.5,12.8) -- (11.5,2);
				\draw[->] (15.5,12.8) -- (10,.5);

			\end{tikzpicture}&

			\scalebox{0.79}{\begin{tikzpicture}[scale=.3,every text node part/.style={align=center}]
				\node at (10.15,1.3) {
					\begin{tikzpicture}[scale=0.3]
					\clip (-5.2,-0.2) rectangle (25.2,2.7);
					\draw[gray,densely dotted] (-5, 0) grid (25,4);
					\draw[black, very thick] (0,0) grid (20,10);

					\foreach \width in {0,...,4}{
						\foreach \y in {0,...,10}{
						\draw[fill,red] (-12+\width+\y,\y) circle (0.2);
						}
					}

					\foreach \width in {0,...,4}{
						\foreach \y in {0,...,10}{
						\draw[fill,blue!40!white] (-7+\width+\y,\y) circle (0.2);
						}
					}

					\foreach \width in {0,...,4}{
						\foreach \y in {0,...,10}{
						\draw[fill,red] (-2+\width+\y,\y) circle (0.2);
						}
					}

					\foreach \width in {0,...,4}{
						\foreach \y in {0,...,10}{
						\draw[fill,blue!40!white] (3+\width+\y,\y) circle (0.2);
						}
					}

					\foreach \width in {0,...,4}{
						\foreach \y in {0,...,10}{
						\draw[fill,red] (8+\width+\y,\y) circle (0.2);
						}
					}

					\foreach \width in {0,...,4}{
						\foreach \y in {0,...,10}{
						\draw[fill,blue!40!white] (13+\width+\y,\y) circle (0.2);
						}
					}

					\foreach \width in {0,...,4}{
						\foreach \y in {0,...,10}{
						\draw[fill,red] (18+\width+\y,\y) circle (0.2);
						}
					}

					\foreach \width in {0,...,4}{
						\foreach \y in {0,...,10}{
						\draw[fill,blue!40!white] (23+\width+\y,\y) circle (0.2);
						}
					}

				\end{tikzpicture}
				} ;

				\draw[decorate,decoration={brace,amplitude=3pt}] (0,2.9) -- (20,2.9) node[midway, above,yshift =2 ] {$k_1 $};
				\draw[decorate,decoration={brace,amplitude=3pt}] (-5,2.9) -- (-1,2.9) node[midway, above,yshift =2 ] {$l_1 $};
				\draw[decorate,decoration={brace,amplitude=3pt}] (21,2.9) -- (25,2.9) node[midway, above,yshift =2 ] {$ l_1 $};

				\draw[decorate,decoration={brace,amplitude=3pt,mirror}] (0,-0.2) -- (2,-0.2) node[midway, below, yshift = -2] {$\scriptscriptstyle \leq \size{E^{x_1}_0} $};
				\draw[decorate,decoration={brace,amplitude=3pt,mirror}] (3,-.2) -- (7,-.2) node[midway, below, yshift = -2] {$\scriptscriptstyle = \size{E^{x_1}_0} $};
				\draw[decorate,decoration={brace,amplitude=3pt,mirror}] (8,-.2) -- (12,-.2) node[midway, below, yshift = -2] {$\scriptscriptstyle = \size{E^{x_1}_0} $};
				\draw[decorate,decoration={brace,amplitude=3pt,mirror}] (13,-.2) -- (17,-.2) node[midway, below, yshift = -2] {$\scriptscriptstyle = \size{E^{x_1}_0} $};
				\draw[decorate,decoration={brace,amplitude=3pt,mirror}] (18,-.2) -- (20,-.2) node[midway, below, yshift = -2] {$\scriptscriptstyle \leq \size{E^{x_1}_0} $};

				\draw[decorate,decoration={brace,amplitude=3pt,mirror}] (-2,-2.2) -- (2,-2.2) node[midway, below, yshift = -2] {$\scriptscriptstyle = \size{E^{x_1}_0} $};
				\draw (-2,-.3) -- (-2, -2.2);
				\draw[decorate,decoration={brace,amplitude=3pt,mirror}] (18,-2.2) -- (22,-2.2) node[midway, below, yshift = -2] {$\scriptscriptstyle = \size{E^{x_1}_0} $};
				\draw (22,-.3) -- (22, -2.2);

				\draw[decorate,decoration={brace,amplitude=3pt,mirror}] (23,-.2) -- (25,-.2) node[midway, below, yshift = -2] {$\scriptscriptstyle \leq \size{E^{x_1}_0} $};
				\draw[decorate,decoration={brace,amplitude=3pt,mirror}] (-5,-.2) -- (-3,-.2) node[midway, below, yshift = -2] {$\scriptscriptstyle \leq \size{E^{x_1}_0} $};

			\end{tikzpicture}
			}
		\end{tabular}
		\caption{In 2 dimensions. Depicted are the evaluation bands. \emph{Left:} covering the grid with adjacent $\ell^1$ balls results in diagonal work and evaluation bands. \emph{Right:} At most  $\lr{k_1+2\cdot l_1}/\size{E^{x_1}_0}$ work bands are needed to cover the side $[k_1]$ of the grid $[k_1]\times[k_2]$ with  evaluation bands. }
		\label{fig:2DNBrOfWorkBands}
\end{figure}
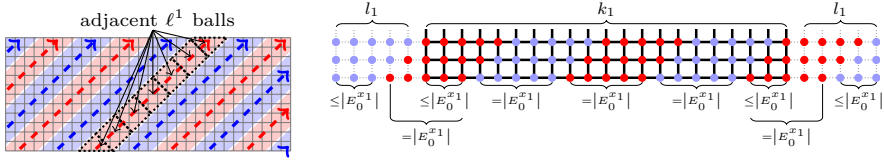

		The Diagonal Band Algorithm is optimal with respect to the leading term of the non-compulsory I/Os.
		As suggested by the lower bound, the algorithm evaluates the grid $\ell^1$-ball by $\ell^1$-ball, by sweeping through adjacent $\ell^1$-balls (see Fig.~\ref{fig:2DNBrOfWorkBands}).
		The algorithm fits the framework of Theorem~\ref{thm:algoFrameworkUpperBounds} and is depicted in detail in Fig.~\ref{fig:2DUpperBoundsHyperCubeDiagonal}.

%

		The sweep sequence $\sweepSeq$ consists of both unit directions, $\sweepSeq = \{ e_1, e_2\}$.
		The sweep shape $\sH$ is a diagonal line segment of $m$ points, 
		\begin{equation}
			\sH = \{x \in \mathbb{Z}^2: \lr{x_1+x_2 = 0} \wedge \lr{x_2 \geq 0} \wedge \lr{x_2 < m} \} \;.
		\end{equation}
		By definition, $b = 1$ holds.
		The sweep shape $\sH$ can be regarded as the intersection of an $\ell^1$-ball with a hyperplane, i.e. a line, of normal $(1,1)$ through the center of that ball.
		This means that we are sweeping an $\ell^1$-ball diagonal per diagonal. 

		The work bands that result from this construction are also diagonal and extend along the direction $(1,1)$ of the sweep.
		One work band consists of $m$ vertices per diagonal hyperplane of normal $(1,1)$.
		Using two different shifts for the sweep sequence together with a diagonal sweep shape, however, doubles, in comparison to the hypercube band algorithm, the vertices of a work band per hyperplane $H_h$ to $2m$.
	
		The evaluation bands are also diagonal bands.
		They consists of $m-s$ vertices per hyperplane of normal $(1,1)$ and $2m-2s$ vertices per $H_h$.
		Hence, $\size{E^\infty_h}= 2m-2s$ and $e=2$ (Assumption~\ref{assum:evalSize}).
		Further, an evaluation band also consists of $2m-2s$ vertices per hyperplane of normal $e_2$ and hence Assumption~\ref{assum:evalWidth} holds.
		The evaluation bands lie in the middle of the work bands.
		In particular, consider the intersection of a work band $W$ with a $H_h$.
		First, there are  $s$ vertices which belong to $W\setminus E_W$, followed by $2m-2s$ vertices of the evaluation band $E_W$ and another $s$ vertices of $W\setminus E_W$.
		Hence, Assumption~\ref{assum:workEvalDistance} is satisfied.

		It is easy to cover the grid with non-overlapping evaluation bands as their structure is very simple.
		Shift the infinite evaluation bands by $2m-2s$ vertices in direction $e_1$ until the whole grid is covered (Assumption~\ref{assum:evalBandsCover}).
		By this approach a work band overlaps with at most two other work bands, the work bands above and below it, and hence Assumption~\ref{assum:workbandConst} is satisfied.
		This covering also immediately yields that at most 
		\begin{equation*}
			\frac{k_1+k_2}{2m-2s}\stackrel{\eqref{eq:sweepSize}}{=} \frac{k_1+k_2}{\frac{M}{2s\cdot c }- \mO{B}}= \mO{\frac{k_1}{M}}
		\end{equation*}
		evaluation bands and hence also work bands are needed to cover the grid.
		Therefore Assumption~\ref{assum:nbrOfWorkBands} is satisfied.

		Let us now analyze the $k$-intersections.
		By the arrangement of the work bands, all $k$-intersections for $k\geq 3$ are empty and Assumption~\ref{assum:3intersects} satisfied.
		For the $2$-intersections consider the intersection of a work band $W$ with a hyperplane $H_h$ of normal $e_1$.
		We have seen that, per $H_h$, the vertices of the evaluation band are adjacent to $s$ vertices of $W\setminus E_W$ on each side of the evaluation band.
		Also, the adjacent work bands, intrude by exactly $s$ vertices into the current work band.
		Hence, $4s$ vertices of the evaluation band, $2s$ vertices on each side, are part of the $2$-intersections for each $H_h$.
		Hence,
			$\Phi_{(W,2,h)} = 4s $
		and Assumption~\ref{assum:2intersects} holds with $b= 4s$.

		We verify the remaining Assumption~\ref{assum:evalPoints} with Lem.~\ref{lem:interval2DDiag}.
		\begin{lem}\label{lem:interval2DDiag}
			Given $d=2$ and the setup of Def.~\ref{def:memEffBandAlgo}.
			Given the Diagonal Band Algorithm specified in this section:
			the sweep sequence is $\sweepSeq = \{ e_1, e_2\}$, the sweep shape $\sH$ is a diagonal line segment of $m$ points and the list of work bands $\wL$ is as specified in this section.
			Then, for any $W \in \wL$ and any $w \in E_W$ the following holds:
			\begin{equation*}
				I(w) = [o_W(w)-\delta, o_W(w)+\delta] \quad \text{with} \quad \delta = s \cdot \size{\sH} + \mO{1}\;.
			\end{equation*}
		\end{lem}
		\begin{proof}
			As \eqref{eq:IwCharacterization} characterizes $I(w)$ as $I(w) = [o_W(w_{\min}), o_W(w_{\max})]$ it is left to determine $w_{\min}$ and $w_{\max}$ as well as their distance to $w$ in the work band order.
			Let $\sH$ 
			be the sweep shape containing $w\in E_W$. 
			First consider $w_{\max}$.
			This vertex has to be in the sweep shape $\sH_s$ proceeding $\sH$ by $s$ shifts.
			Of all vertices in $\sH_s\cap S_s(w)$ the vertex $w+(s,0)$ is the one of maximum lexicographic order, hence $w_{\max}= w+(s,0)$.
			Denote by $v$ the vertex $v$ to which $w$ is shifted in the next $s$.
			It holds that $||v-w||_W = s \cdot \size{\sH}$ and $v \in \sH_s\cap S_s(w)$.
			The vertices in $\sH_s\cap S_s(w)$ form a contiguous, diagonal line segment of $s+1$ vertices.
			Hence, the work band distance between $v$ and $w_{\max}$ is bounded by $s$.
			Hence
			\begin{align*}
				||w - w_{\max}||_W \leq ||w - v||_W +||v - w_{\max}||_W \leq s\cdot \size{\sH} +\mO{1}\;.
			\end{align*}
			Similarly, it can be shown that $w_{\min}= w+(-s,0)$ and that $||w - w_{\min}||_W  \leq s\cdot \size{\sH} +\mO{1}$.
			Hence, the claim follows.
		\end{proof}

		As all assumptions are satisfied, we can apply Theorem~\ref{thm:algoFrameworkUpperBounds} to yield an upper bound for the non-compulsory I/Os of the Diagonal Band Algorithm ($c=1$, $e=2$ and $b=4s$ ).
		The upper bound is
		\begin{equation*}
			4s^2\cdot \frac{k_1k_2}{BM}+\mO{\frac{k_1k_2}{M^2}}+\mO{\frac{k_1}{B}} \; .
		\end{equation*}

\subsection{Upper Bounds in 3 Dimensions}
	\label{sec:3DLayouts}
	The upper bounds in three dimensions are summarized in Tab. \ref{tab:3DUpperBounds}. 
	In three dimensions, the Hypercube Band Algorithm achieves optimal asymptotic behavior and outperforms the Row Algorithm by $\Om{\frac{1}{\sqrt{B}}}$ and the Column Algorithm by $\Om{\frac{1}{B}}$.
	Using a two dimensional $\ell^1$-ball  (Diamond Band Algorithm) instead of a square as sweep shape  improves the leading term of the non-compulsory I/Os by a factor of $\frac{1}{\sqrt{2}}$ compared to the Hypercube Band Algorithm. 
	The best new upper bound improves this by another factor of $\frac{1}{\sqrt{3}}$, leaving a gap of $\sqrt{2}$ to the lower bound (Hexagonal Band Algorithm -- $s \in \lrC{1,2,3}$ only).
	The Hexagonal Band Algorithm shifts a hexagonal sweep shape, resulting from the intersection of a three dimensional $\ell^1$-ball with a plane of normal $(1,\,1,\,1)$, and  alternates the three unit shifts.
	
	One reason why the constants of the lower and upper bound do not yet match is the fact that the grid cannot be tiled with $\ell^1$-balls in three dimensions.
	Hence, imitating the lower bound by tiling the grid with $\ell^1$-balls fails in three dimensions while it was possible in two dimensions.

\begin{table}[htbp]
		\footnotesize
		\centering
				\caption{
		Leading term of the non-compulsory I/Os for different algorithms in three dimensions. The Hexagonal Band Algorithm is analyzed only for $s \in \lrC{1,2,3}$.
				} 
		\label{tab:3DUpperBounds}	
			\begin{tabular}{c|c|c|c}
			\textbf{3D Algorithms} & Sweep Shape  & Sweep Seq. & Non-Compulsory I/Os \\
				\hline & & & \\[-1ex]
				Row Algorithm & Square & $e_1$ & $\Om{\frac{1}{\sqrt{B}\cdot\sqrt{ M}}\cdot k_1k_2k_3}$\\[1.5ex]
				Column Algorithm & Square & $e_1$ &  $\Om{\frac{1}{\sqrt{ M}}\cdot k_1k_2k_3}$\\[1.5ex]
				Hypercube Band Algorithm & Square & $e_1$& $\frac{8\sqrt{2}s^{3/2}}{B \sqrt{M}} \cdot k_1k_2k_3$ \\[1.5ex]
				Diamond Band Algorithm & $\ell^1$-ball & $e_1$& $ \frac{8 s^{3/2}}{B \sqrt{M}} \cdot k_1k_2k_3$\\[1.5ex]
				Hexagonal Band Algorithm& Hexagonal & $e_1$, $e_2$, $e_3$ &$ \frac{8 \sqrt{2}s^{3/2}}{\sqrt{3}B \sqrt{M}}\cdot k_1k_2k_3$\\[1.5ex]
				\hline
				\hline&&&\\[-1.5ex]
				\textbf{Lower Bound} & n.a. & n.a. &$ \frac{8s^{3/2}}{\sqrt{3}B \sqrt{M}}\cdot k_1k_2k_3$
			\end{tabular}
	\end{table}

\subsubsection{Diamond Band Algorithm}
		The three dimensional Diamond Band Algorithm improves upon the three dimensional (Hyper-)Cube Band Algorithm by a factor of $\frac{1}{\sqrt{2}}$.
		Instead of a two dimensional hypercube (square) it uses a diamond, i.e. $\ell^1$-ball, as sweep shape.
		It has been proven in~\S\ref{sec:isoResult} that the $\ell^1$-ball has a better interior to boundary ratio than the cube.
		Hence, it is also advantageous as sweep shape.
		Although the Hexagonal Band Algorithm presented in~\S\ref{sec:hexaAlgo} further reduces the number of non-compulsory I/Os, the Diamond Band Algorithm maybe better for implementation as the complexity of its data layout is still manageable.


		Theorem~\ref{thm:algoFrameworkUpperBounds} can be used to analyze the Diamond Band Algorithm.
		The sweep sequence consists of the first unit shift, $\sweepSeq = \{ e_1 \}$.
		The sweep shape is a two dimensional $\ell^1$-ball of radius $m-1$, i.e. side length $m$, lying in a hyperplane of normal $e_1$,
		\begin{equation*}
			\sH= b^{(m-1,0)}_2=\{ (x_1,x_2,x_3) \in\mathbb{Z}^3: (x_1 = 0) \wedge \lr{|x_2|+|x_3| \leq m-1 }\} \; .
		\end{equation*}
		A sweep shape consists of $\size{\sH} \stackrel{\text{\S\ref{sec:ballAndBound}}}{=} m^2+(m-1)^2= 2m^2-2m+1$ vertices and hence $c =2$.
		
		The work and evaluation bands are simple to describe as we are only sweeping in $x_1$-direction,
		\begin{equation*}
			W^\infty_h= \sH= b^{(m-1,0)}_2 \qquad \text{and} \qquad E^\infty_h = b^{(m-1-s,0)}_2 \;.				 
		\end{equation*}
		An evaluation band lies in the center of its corresponding work band.
		In particular, Assumptions~\ref{assum:evalWidth} and~\ref{assum:workEvalDistance} hold. 

\begin{figure}[tb]
			\centering
			\setlength{\tabcolsep}{0pt}
			\begin{tabular}{cccc}

				\begin{tikzpicture}[scale = 0.13]
					\draw[gray] (0,-0) grid (20,13);

					\drawEllBallCluster{0}{0}
					\drawEllBallCluster{10}{2}
					\drawEllBallCluster{4}{6}

					\drawEllBall{8}{-1}{3}{0.2}{yellow!33!blue} 
					\drawEllBall{11}{-3}{3}{0.2}{blue}

					\drawEllBall{16}{-2}{3}{0.2}{yellow}
					\drawEllBall{21}{-1}{3}{0.2}{blue}

					\drawEllBall{18}{1}{3}{0.2}{yellow!33!blue}

					\drawEllBall{-1}{5}{3}{0.2}{blue}

					\drawEllBall{1}{8}{3}{0.2}{yellow!66!blue}
					\drawEllBall{3}{11}{3}{0.2}{blue}

					\drawEllBall{-2}{10}{3}{0.2}{yellow}
					\drawEllBall{0}{13}{3}{0.2}{yellow!33!blue}

					\drawEllBall{20}{4}{3}{0.2}{yellow!66!blue}
					\drawEllBall{20-3}{4+2}{3}{0.2}{yellow}
					\drawEllBall{20-2*3}{4+2*2}{3}{0.2}{yellow!66!blue}
					\drawEllBall{20-3*3}{4+3*2}{3}{0.2}{yellow}
					\drawEllBall{20-4*3}{4+4*2}{3}{0.2}{yellow!66!blue}
					\drawEllBall{20-5*3}{4+5*2}{3}{0.2}{yellow}

					\drawEllBall{19+3}{9-2}{3}{0.2}{yellow!33!blue}
					\drawEllBall{19}{9}{3}{0.2}{blue}
					\drawEllBall{19-3}{9+2}{3}{0.2}{yellow!33!blue}
					\drawEllBall{19-2*3}{9+2*2}{3}{0.2}{blue}

					\drawEllBall{18+3}{14-2}{3}{0.2}{yellow!66!blue}
					\drawEllBall{18}{14}{3}{0.2}{yellow}
				\end{tikzpicture}
				&
				\includegraphics[width =.25\columnwidth]{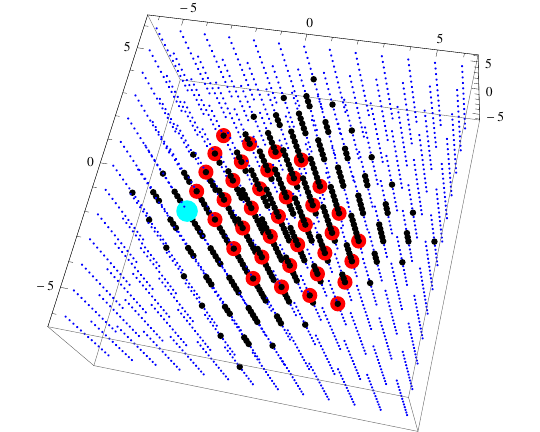}
				&
				\begin{tikzpicture}[scale=0.3]
    \tikzstyle{every node}=[font=\scriptsize]
				\draw[gray] (-3.2,-3.2) grid (3.2,3.2);
				\foreach \x in {0,...,2} {
					\foreach \y in {0,...,2} {
					\draw[fill] (\x, \y) circle (.2);
					\draw[fill] (-\x, -\y) circle (.2);
					}
				}
				\draw[fill] (-1, 1) circle (.2);
				\draw[fill] (1, -1) circle (.2);

				\draw[red,thick] (2.5,0) -- (0,2.5) -- (-2.5,0) -- (0,-2.5) -- cycle ;

				\draw[green,very thick, xshift = 1cm, yshift =1cm ] (1.5,0) -- (0,1.5) -- (-1.5,0) -- (0,-1.5) -- cycle ;
				\draw[green,very thick, xshift = -1cm, yshift =-1cm ] (1.5,0) -- (0,1.5) -- (-1.5,0) -- (0,-1.5) -- cycle ;

				\draw[purple,thick, xshift = 2cm, yshift =2cm ] (0.5,0) -- (0,0.5) -- (-0.5,0) -- (0,-0.5) -- cycle ;
				\draw[purple,thick, xshift = -2cm, yshift =-2cm ] (0.5,0) -- (0,0.5) -- (-0.5,0) -- (0,-0.5) -- cycle ;

				\draw[purple] (3,2) node[right] {$k=2$};
				\draw[green] (3,1) node[right] {$k=1$};
				\draw[red] (3,0) node[right] {$k=0$};
				\draw[green] (3,-1) node[right] {$k=-1$};
				\draw[purple] (3,-2) node[right] {$k=-2$};

				\draw[gray!50!black, densely dotted, thick] (-2.5,-2.5) -- (0,-2.5) -- (2.5,0) -- (2.5,2.5) -- (0,2.5) -- (-2.5,0) -- cycle;
				\end{tikzpicture}
				&
				\includegraphics[width = 0.2\columnwidth]{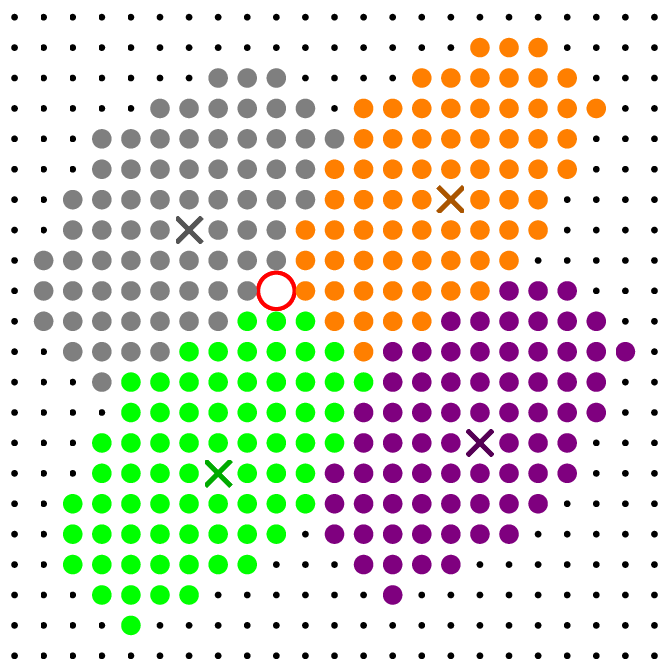}
			\end{tabular}
			\caption{\textbf{Diagonal Band Algorithm:} \emph{Left:} covering the two dimensional grid $[k_2]\times[k_3]$ with $E^\infty_h$, i.e. two dimensional $\ell^1$-balls. 
					\textbf{Hexagonal Band Algorithm:}
					\emph{Middle left:} for m=3, intersecting a 3-dimensional $\ell^1$-ball (black) with plane of normal $(1,\,1,\,1)$ results in  $\sH'$ (red). Additional vertex of $\sH$ for $s=1$ in cyan.
					\emph{Middle right:} projection $P_2$ of 2-star stencil $S_2$. The projection $P_s$ of the stencil consist of the different level sets of the original stencil $S_s$.
					\emph{Right:} trying to cover the two dimensional grid $[k_2]\times[k_3]$  with the $\{w\in W: \;  P_s(w) \subseteq W \}_h \subset E^\infty_h$ resulting from the original hexagonal sweep shapes $\sH'$ for $m=3$ and $s =1$. One vertex (circled) cannot be covered when the  $E^\infty_h$ do not overlap.
					}
			\label{fig:3DDiagonalBand}
			\label{fig:gapForEvalBands}
			\label{fig:3DUpperBounds}
			\label{fig:stencilProjection}
		\end{figure}
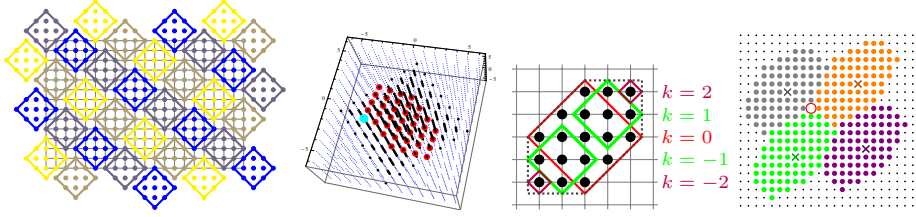

		It follows that $\size{E^\infty_h} = (m-1-s)^2+(m-1-s-1)^2 = 2 m^2+\mO{m}$ and $ e =2 $ (Assumption~\ref{assum:evalSize}).
		Sweeping only in $x_1$-direction means that if we cover $[k_2]\times[k_3]$ with the $ E^\infty_h$, then the three dimensional grid is covered by the resulting evaluation bands.
		In fact, it is easy to cover the two dimensional grid with two dimensional $\ell^1$-balls without overlap (Assumption~\ref{assum:evalBandsCover}).
		See Fig~\ref{fig:3DDiagonalBand} and also \cite{bisseling04parallelScientificComputing} for an example covering.
		This covering gives rise to the list of work bands $\wL$.
		Any work band overlaps only with up to the $8=3^d-1$  work bands adjacent to it and hence Assumption~\ref{assum:workbandConst} holds.

%
%
%
	


		Let us now consider the $k$-intersections.
		Pick one work band $W$.
		The $(E^\infty_W)_h$ is an $\ell^1$-ball of radius $m-1-s$ at the same center as $W^\infty_h = \sH = b^{(m-1,0)}_2$.
		Therefore, a work band $W$ is larger than $E_W$ by $s$ layers of vertices.
		As the evaluation bands cover the grid without overlap, a work band $W' \neq W$ can only reach the $s$ outermost layers of $E_W$. 
		Hence only the outermost $2s$ layers of $W$ may be shared with other work bands.
		Therefore,
		\begin{align}
				\size{\Phi_{W,2,h}} \leq \size{\Gamma_{2s} (b^{(m-1,0)}_2)} \leq \sum_{r=1}^{2s} 4(m-r) = 8s\cdot m -\mO{1} \;.
		\end{align}
		This means that Assumption~\ref{assum:2intersects} holds with $b= 8s$.
		It also follows from the shape and position of the evaluation and work bands that the $k$-intersections are limited to constant size per $H_h$ whenever $k\geq 3$.
		By Lem.~\ref{lem:kIntersectsBounded} it follows that
			$\size{\Phi_{W,k,h}} = \mO{1} $ for 
		$k \geq 3$.
		Hence, also Assumption~\ref{assum:3intersects} is satisfied.

		The number of work bands used in total is the same as the number of work bands per $H_h$ as we are only sweeping in $x_1$-direction.
		Hence, as the necessary assumptions are satisfied, we can apply Lemma~\ref{lem:nbrOfWorkBandsArb} to yield  Assumption~\ref{assum:nbrOfWorkBands}:
		\begin{align*}
			\size{\wL}  \leq \frac{(k_2+\mO{m})(k_3+\mO{m})}{\size{E^\infty_h}-\mO{m}} =  \frac{(k_2+\mO{m})(k_3+\mO{m})}{(2 m^2+\mO{m})-\mO{m}}  
			\stackrel{\eqref{eq:sweepSize}}{=}
			\mO{\frac{k_2k_3}{M}} \;.
		\end{align*}

		Finally, we can apply Lemma~\ref{lem:distSweepX1} as we are only sweeping in $x_1$-direction to show that Assumption~\ref{assum:evalPoints} holds.
		Therefore all assumptions of Def.~\ref{def:memEffBandAlgo} are met and Thm.~\ref{thm:algoFrameworkUpperBounds} can be applied.
		Hence, the number of non-compulsory I/Os of the 3-dimensional Diamond Band Algorithm is upper bounded by ($c =e =2$, $b = 8s$)
		\begin{equation*}
			8 s^{3/2}\cdot \frac{k_1k_2k_3}{B\cdot \sqrt{M}}+\mO{\frac{k_1k_2k_3}{\sqrt{B}\cdot M}} \;.
		\end{equation*}

	\subsubsection{Hexagonal Band Algorithm for 3 Dimensions: Alternate Sweeps Improve the Constant}
		\label{sec:hexaAlgo}

		The best algorithm presented in three dimensions is the \emph{Hexagonal Band Algorithm}.
		Essential to the Hexagonal Band Algorithm is the concept of alternating shifts which has already proven useful in two dimensions for the Diagonal Band Algorithm.
		The Hexagonal Band Algorithm is asymptotically optimal and improves the constant of the leading term of the non-compulsory I/Os by a factor of $\sqrt{3}$ over the three dimensional Hypercube Band Algorithm and a factor of $\frac{\sqrt{3}}{\sqrt{2}}$ over the three dimensional Diamond Band Algorithm.


		The Hexagonal Band Algorithm is presented to show that alternating shifts are very likely necessary if the lower and upper bounds should match. 
		Using alternating shifts, however, makes the analysis and implementation of the algorithm more difficult.
		Due to the sloped traversal of the grid and the irregularity of the $k$-intersections implementing this data layout requires sophisticated logic and index computations to determine the $k$-intersections.
		The overhead created by the irregular $k$-intersections may cancel the performance gained by reducing the non-compulsory I/Os.
		Hence, it is unsure whether implementing this algorithm is worthwhile.
		Regarding the analysis, some of the concepts introduced in~\S\ref{sec:upperBoundsFramework} are solely necessary to cope with a sweep sequence containing different shifts.
		As the Hexagonal Band Algorithm is therefore more of theoretical interest, we present its main ideas and concepts but do not proof all details in a rigorous manner.
		In particular we only analyze this algorithm for $ s \in \{ 1,2,3\}$ and claim that the modifications necessary to the sweep shape are identical for all $s' = s +3\cdot k$ for $k \in \mathbb{N}_0$.
		The case distinction between different $s$ is necessary to show that the evaluation bands cover the grid in the desired manner.
		Also, we only show that the evaluation bands can cover the grid without overlap by presenting a picture and do not explicitly construct the work band list $\wL$  
		by giving the offsets between the work bands. 
		In addition, counting the vertices of a work band that cannot be evaluated (see \eqref{eq:countNotEval}) and those that are part of $k$-intersections for $k\geq2$ (see \eqref{eq:countKIntersects}) is done by proof by picture.
		We leave it to the reader to verify these details in a rigorous manner for all $m$.
		Instead, we describe the essential results and build upon the intuition gained from the other algorithms.


		The Hexagonal Band Algorithm is a memory efficient band Algorithm.
		To describe it, fix the sweep sequence $\sweepSeq$ to consist of the three unit vectors in their natural order,
			$\sweepSeq = \{ e_1, e_2, e_3 \}$.
		The sweep shape $\sH'$ of size $m$ is the intersection of the 3-dimensional $\ell^1$-ball of radius $2m$ centered at the origin, $b^{(2m,0)}_3(0)\subset \mathbb{Z}^d$, with the plane $H$ through the origin and normal $(1,\,1,\,1)$,
		\begin{equation*}
			\sH' = b^{(2m,0)}_3(0) \cap H^{(1,1,1)}_0 \qquad \text{(see Fig. \ref{fig:3DUpperBounds})}\;.
		\end{equation*}
		The vertices in $\sH'$ can be counted per $x_1$-level-sets and hence
		\begin{equation*}
			\size{\sH'}=  \lr{2\cdot \left(\sum_{k=m+1}^{2m} k \right)+ (2m+1)} = 3m^2+3m+1 \;.  
		\end{equation*}
		The sweep shape $\sH'$ has hexagonal structure and it is therefore easy to cover the grid with the work bands.
		For $s = 3\cdot k$, $k \in \mathbb{N}$, the evaluation bands resulting from the sweep shape of size $m$ are identical to the work bands resulting from a sweep shape of size $m-2k$.
		Hence, in this case the grid can be covered with non-overlapping evaluation bands without adapting the sweep shape and we can choose
		\begin{equation}\label{eq:hexaSweepShapeS3}
			\sH = \sH' \qquad \text{for } s =3\;.
		\end{equation}

		For $s \notin 3\mathbb{N}$ the evaluation bands, however, do not cover the grid in a nice fashion (see Fig.~\ref{fig:gapForEvalBands} for an example).
		Consider the cases $s \in \{1,2\}$.
		In summary, when the evaluation bands are placed without overlap  there is one vertex missing per evaluation band and $H_h$.
		If the evaluation bands overlap such that the whole grid is covered the overlap would be of order $\mO{m}$ per $H_h$ and would affect the leading term of the non-compulsory I/Os.
		Therefore, the sweep shape $\sH'$ is enlarged by one vertex in a manner that the grid can be covered with non-overlapping evaluation bands.
		The enlarged sweep shapes\footnote{
		These are not the only possible choices to enlarge $\sH$ but there are up to six different choices for each $s$.
		}
		 $\sH$ are given by
		\begin{align}
				&s=1: \qquad \sH = \sH' \cup (-m,-1,(m+1))\qquad \qquad   \text{and} \label{eq:hexaSweepShapeS1}\\
				&s=2: \qquad \sH = \sH' \cup (-1,-m,(m+1)) \; . \label{eq:hexaSweepShapeS2}
		\end{align}
		In any case, $c=3$ holds.

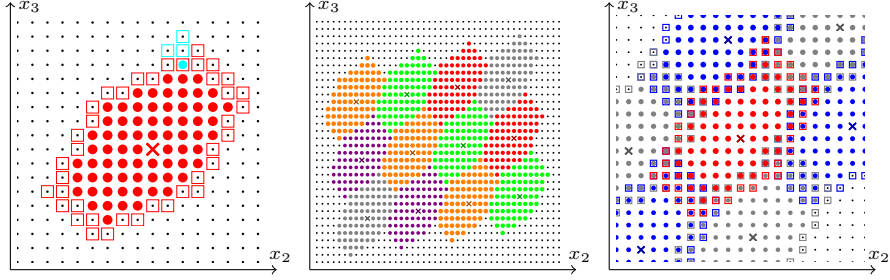
\begin{figure}[tbp]
			\centering
			\vspace{2ex}
			\setlength\fboxrule{0.0pt}
			\begin{tabular}{ccc}
				\includegraphics[clip=true,width =.27\columnwidth, bb=0.0cm 0.0cm 15.5cm 15.5cm,fbox]{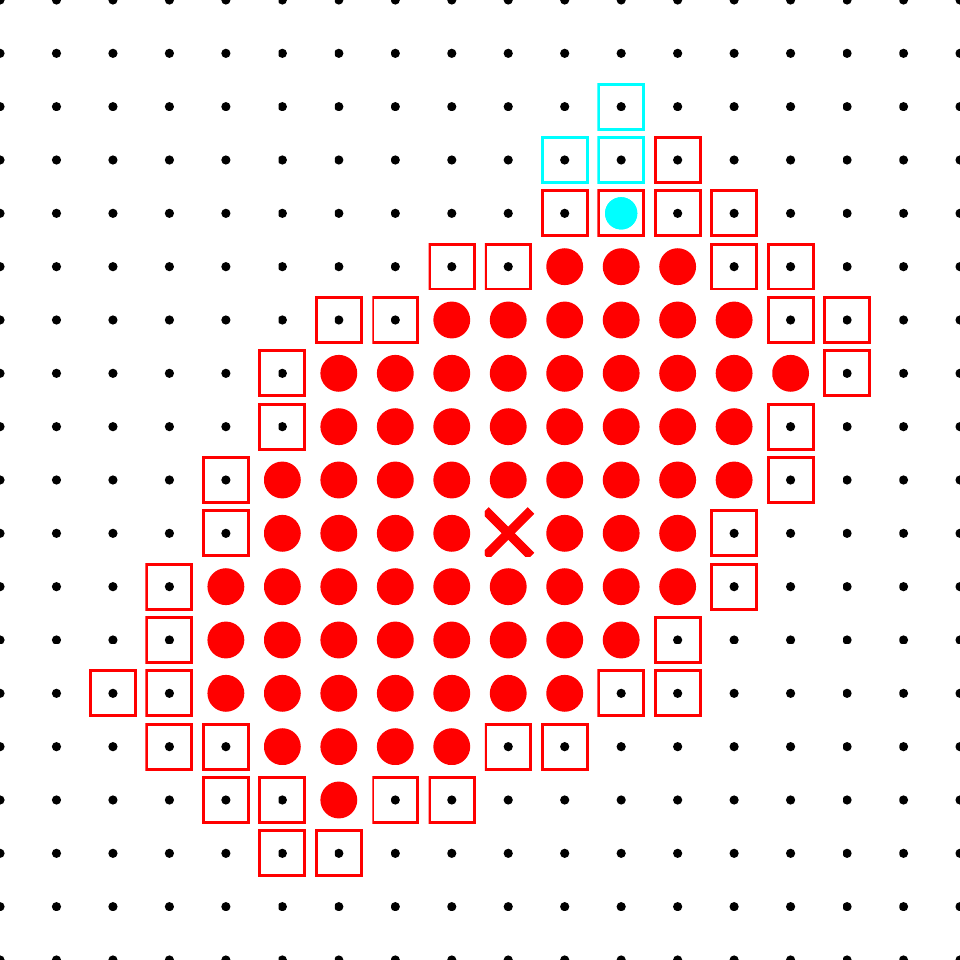}
				\begin{tikzpicture}[overlay]
					\draw[->] (-.295\columnwidth,-.1) -- (0,-.1);
					\draw (.05,-.1) node[above] {\scriptsize $x_2$};
					\draw[->] (-.295\columnwidth,-.1) -- (-.295\columnwidth,.29\columnwidth);
					\draw (-.295\columnwidth,.285\columnwidth) node[right] {\scriptsize $x_3$};
				\end{tikzpicture}
				&
				\includegraphics[width = 0.27\columnwidth, bb=0cm 0cm 9.2cm 9.2cm,fbox]{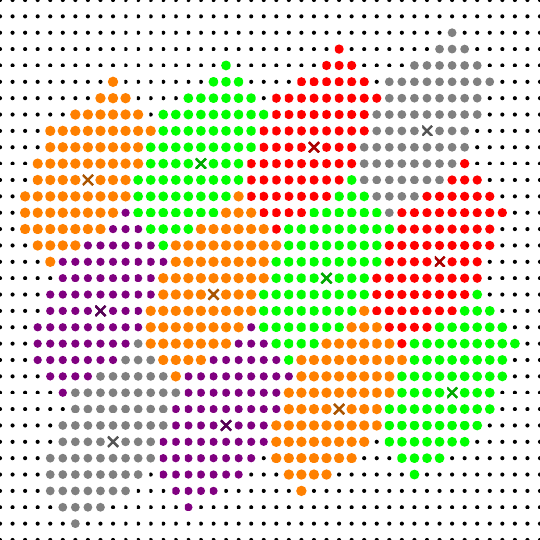}
				\begin{tikzpicture}[overlay]
					\draw[->] (-.295\columnwidth,-.1) -- (0,-.1);
					\draw (.05,-.1) node[above] {\scriptsize $x_2$};
					\draw[->] (-.295\columnwidth,-.1) -- (-.295\columnwidth,.29\columnwidth);
					\draw (-.295\columnwidth,.285\columnwidth) node[right] {\scriptsize $x_3$};
				\end{tikzpicture}
				&
				\includegraphics[width = 0.27\columnwidth, bb=0cm 0cm 10.9cm 10.9cm,fbox]{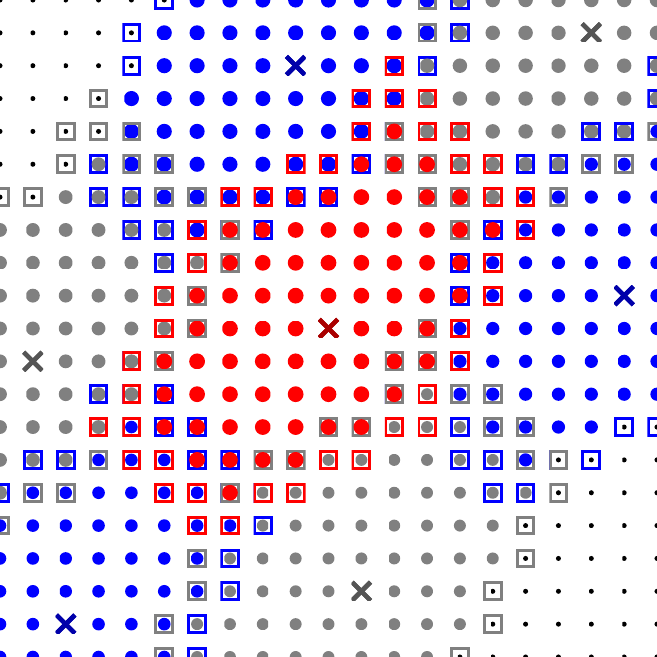} 
				\begin{tikzpicture}[overlay]
					\draw[->] (-.295\columnwidth,-.1) -- (0,-.1);
					\draw (.05,-.1) node[above] {\scriptsize $x_2$};
					\draw[->] (-.295\columnwidth,-.1) -- (-.295\columnwidth,.29\columnwidth);
					\draw (-.295\columnwidth,.285\columnwidth) node[right] {\scriptsize $x_3$};
				\end{tikzpicture}
			\end{tabular}
			\caption{
				$\boldsymbol{s=1}$ and $m=3$. The center of $W^\infty_h$ is marked with a cross. 
				\emph{Right:} $W^\infty_h$. Vertices $w\in W^\infty_h$ that can be evaluated because of $P_s(w)\subseteq W^\infty_h$ are depicted as circles, vertices for which $P_s(w)\nsubseteq W^\infty_h$ holds are squares. Vertices of the original sweep shape $\sH'$ in red, additional vertices of the adapted sweep shape $\sH$ in cyan.
				\emph{Middle:} a partial, non overlapping cover of the grid $[k_2]\times[k_3]$ with  the sets $\lrC{w \in W^\infty_h: P_s(w)\subseteq W^\infty_h }\subseteq E^\infty_h$.
				\emph{Left:} Overlapping work bands to estimate the vertices in the $k$-intersections.
				}
			\label{fig:3dHexEvalTilingS1}
		\end{figure}

\begin{figure}[tbp]
			\centering
			\vspace{2ex}
			\setlength\fboxrule{0.0pt}
			\begin{tabular}{ccc}
				\includegraphics[width =.27\columnwidth, bb=0cm 0cm 9cm 9cm,fbox]{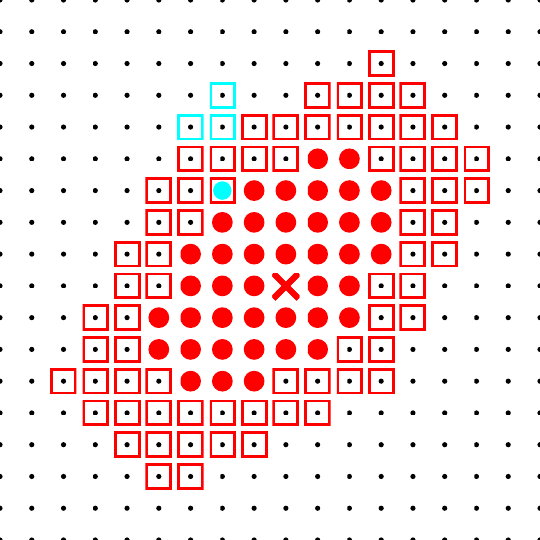}
				\begin{tikzpicture}[overlay]
					\draw[->] (-.295\columnwidth,-.1) -- (0,-.1);
					\draw (.05,-.1) node[above] {\scriptsize $x_2$};
					\draw[->] (-.295\columnwidth,-.1) -- (-.295\columnwidth,.29\columnwidth);
					\draw (-.295\columnwidth,.285\columnwidth) node[right] {\scriptsize $x_3$};
				\end{tikzpicture}
				&
				\includegraphics[width = 0.27\columnwidth, bb=0cm 0cm 9cm 9cm,fbox]{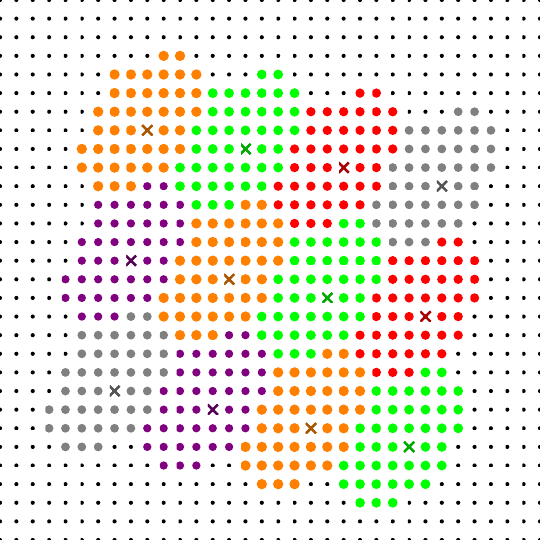}
				\begin{tikzpicture}[overlay]
					\draw[->] (-.295\columnwidth,-.1) -- (0,-.1);
					\draw (.05,-.1) node[above] {\scriptsize $x_2$};
					\draw[->] (-.295\columnwidth,-.1) -- (-.295\columnwidth,.29\columnwidth);
					\draw (-.295\columnwidth,.285\columnwidth) node[right] {\scriptsize $x_3$};
				\end{tikzpicture}
				&
				\includegraphics[width = 0.27\columnwidth, bb=0cm 0cm 11.2cm 11.2cm,fbox]{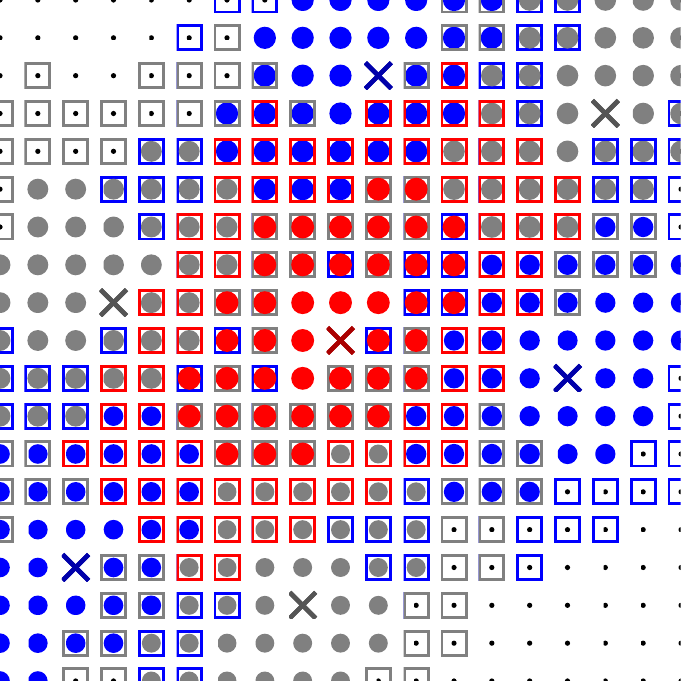} 
				\begin{tikzpicture}[overlay]
					\draw[->] (-.295\columnwidth,-.1) -- (0,-.1);
					\draw (.05,-.1) node[above] {\scriptsize $x_2$};
					\draw[->] (-.295\columnwidth,-.1) -- (-.295\columnwidth,.29\columnwidth);
					\draw (-.295\columnwidth,.285\columnwidth) node[right] {\scriptsize $x_3$};
				\end{tikzpicture}
			\end{tabular}
			\caption{$\boldsymbol{s=2}$ and $m=3$. The center of $W^\infty_h$ is marked with a cross. 
				\emph{Right:} $W^\infty_h$. Vertices $w\in W^\infty_h$ that can be evaluated because of $P_s(w)\subseteq W^\infty_h$ are depicted as circles, vertices for which $P_s(w)\nsubseteq W^\infty_h$ holds are squares. Vertices of the original sweep shape $\sH'$ in red, additional vertices of the adapted sweep shape $\sH$ in cyan.
				\emph{Middle:} a partial, non overlapping cover of the grid $[k_2]\times[k_3]$ with  the sets $\lrC{w \in W^\infty_h: P_s(w)\subseteq W^\infty_h }\subseteq E^\infty_h$.
				\emph{Left:} Overlapping work bands to estimate the vertices in the $k$-intersections.
				}
			\label{fig:3dHexEvalTilingS2}
		\end{figure}
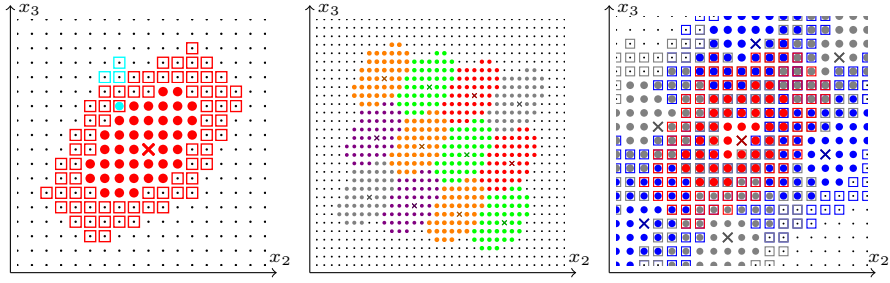

		
		Let us now discuss the evaluation bands and how they cover the grid for $s \in \lrC{1,2}$ in detail.
		Therefore, consider the intersection of a (infinite) work band with a $H_h$.
		This intersection is depicted in Fig.~\ref{fig:3dHexEvalTilingS1} for $s=1$ and in Fig.~\ref{fig:3dHexEvalTilingS2} for $s=2$. 
		The intersection consists of $\size{W^\infty_h} = 3\cdot\size{\sH'}= 3\cdot (3m^2+3m+2 )$ vertices, as we employ three different unit shifts and hence each subset of $\sH$ of a fixed $x_1$-coordinate appears three times in this intersection.

		To determine which vertices of the work band belong to the evaluation band and the $k$-intersections, regard how the shifts of the sweep shape affect the $s$-star stencil. 
		We define $P_s(\cdot)$ (see Fig. \ref{fig:stencilProjection}), the two dimensional projection of the $s$-star stencil $S_s(\cdot)$, as 
		\begin{align*}
			P_s(w) :=& \{ v\in \lr{ h\times  [k_2]\times[k_3]} : ||w-v||_1\leq s\} \;\cup  \\ 
				&\cup \{y: ||w-v||_{\infty}\leq s \wedge \lr{w_i\leq 0 \wedge  v_i \leq 0 \; \text{ for } \; i = 2,3} \;\}\; \cup \\
				&\cup\{y: ||w-v||_{\infty}\leq s \wedge \lr{ w_i \geq 0 \wedge  v_i \geq 0 \; \text{ for } \; i = 2,3 }\;\}\; ,
		\end{align*}
		We will argue next that a $w\in W$ can be evaluated, if the projection of the $s$-star stencil of $w$ is in $W \cap H_h$, i.e.
		\begin{equation}\label{eq:projStencilEval}
			P_s(w) \subset \lr{W \cap H_h} \quad \Rightarrow \quad w \in E_W \; .
		\end{equation}
		If $P_s(w) \subset \lr{W \cap H_h}$, all vertices of $P_s(w)$ belong to some sweep shape of $W$.
		In particular, every fixed vertex $v \in P_s(w)$  has a \emph{trace} in the 3-dimensional grid  depending on the sweep shape(s) it belongs to and the shifts that are applied to it and its sweep shape(s).
		It is important that the whole trace of a vertex belongs to $W$ if $v$ itself belongs to $W$.
		As the three unit shifts alternate as sweeps, the vertices $v\pm k \cdot (1,1,1)$ for $k\in \mathbb{Z}^d$ are in the trace of $v$ independently of which shift is next.
		As example consider, the vertex $v = (-1,-1,0)$ which has the vertex $(0,0,1)$ in its trace.
		Verifying the $\supseteq$ part of the equality
		\begin{equation*}
			P_s(w) = \bigcup_{k = -s}^s \Big(\lr{S_s(w)\cap H_{k}}-k \cdot (1,1,1)\Big)
		\end{equation*}
		yields that the traces of the vertices of the stencil projection $P_s(w)$ cover the stencil $S_s(w)$ itself.
		Hence $S_s(w)\subset W$ holds from which $w \in E_W$ follows.

		The structure of $W^\infty_h$ (see Figs.~\ref{fig:3dHexEvalTilingS1} and~\ref{fig:3dHexEvalTilingS2}) can be described as follows:
		the vertices of $W^\infty_h$ can be split into groups of three diagonals that all correspond to vertices of the same $x_1$-value of the sweep shape.
		The three diagonals within a group correspond to the three different unit shifts.
		Within such a group, the three diagonals have the same number of vertices.
		If the group contains vertices of $x_1$-value $u$, then there are $(2m+1)-|u|$ vertices in the group.
		The only exception is the group which contains the additional vertex, $(m,1,-(m+1))$ for $s=1$ and $(-1,-m,(m+1))$ for $s=2$. 
		This group contains one additional vertex per diagonal, or three additional vertices in total.

		To estimate $E^\infty_h$ apply the projected stencil $P_s$ to  $W^\infty_h$.
		By \eqref{eq:projStencilEval}: if the projected stencil of a vertex $w$ is in $W^\infty_h$, then the $w$ itself belongs to the evaluation band. 
		Hence
		\begin{equation*}
			E \supset \{w\in W: \;  P_s(w) \subseteq W \} \qquad \text{and}\qquad 	W\setminus E \subset \{w\in W: \;  P_s(w) \nsubseteq W \} \; .
		\end{equation*}

		The vertices in the set $A := \{w\in W: \;  P_s(w) \nsubseteq W \}$ can be counted per $H_h$.
		To treat all cases for different $s$ at once, we apply $P_s(\cdot)$ to the work band $(W')^\infty$ resulting from the original $\sH'$.
		If $P_s(\cdot)$ is within $(W')^\infty$ then it is for sure within $W^\infty$ and hence this underestimates the number of vertices that can be evaluated.
		Further, enlarging the sweep shape adds at  most three vertices $W^\infty_h$ and hence at most 3 vertices to $A$. 
		The first and the last $2s$ diagonals of $W^\infty_h$ ($m+1$ vertices each) are in the set $A$.
		For each group of three diagonals  with $x_1 \neq 0$, there are $2\cdot 2s$ vertices in $A$ at the ends of the diagonals.
		This holds also for the first and last group for which we have already excluded several diagonals completely.
		Again, this double accounting only decreases the vertices in the evaluation band and weakens the analysis.
		The middle group for $x_1 = 0$ contains $2s+3s$ vertices for which  $ P_s(w) \nsubseteq(W')^\infty $ holds. 
		In total (counting diagonal by diagonal from left to right), 
		\begin{align} \label{eq:countNotEval}
			\size{  \lr{W \setminus E}_h} \leq (m+1)\cdot 2s+ m\cdot4s+5s+m\cdot 4s+(m+1)\cdot 2s+3 = \nonumber\\
				= 12ms+\mO{1} \;.
		\end{align}
		Hence, a lower bound for the size of $E^\infty_h$ is given by
		\begin{equation}\label{eq:hexaEInfty}
			\size{E^\infty_h} \geq 3 \cdot (3m^2 +3m+2) - (12 ms+\mO{1}) = 9 m^2 - \mO{m}
		\end{equation}
		and $e = 9$ follows (Assumption~\ref{assum:evalSize}).

		Besides the sheer number of vertices in $E^\infty_h$, we also need to know the shape 
		of $E^\infty_h$ to cover the grid with few evaluation bands.
		Examine the sets $W^\infty_h$  and $A^\infty_h = \{w\in W: \;  P_s(w) \nsubseteq W \}\cap H_h$ depicted in Figs.~\ref{fig:3dHexEvalTilingS1} and~\ref{fig:3dHexEvalTilingS2} for $ m = 3$ and $s=1$ respectively $s=2$.\footnote{
		$E^\infty_h$ can also contain further vertices but a subset of $E^\infty_h$ is sufficient for the following analysis.
		}
		From the hexagonal structure of the sweep shapes follows that the two sets $W^\infty_h$ and $A_h$ also have a hexagonal like structure.
%
		Figs.~\ref{fig:3dHexEvalTilingS1} and~\ref{fig:3dHexEvalTilingS2}  
		show partial covers of the grid $[k_2]\times[k_3]$ with non-overlapping sets $A_h$ which can be extended to cover the whole grid  $[k_2]\times[k_3]$.
		Hence we can also cover the 3-dimensional grid $[k_1]\times[k_2]\times[k_3]$ with non-overlapping evaluation bands which gives rise to the list of corresponding work bands $\wL$
		and Assumption~\ref{assum:evalBandsCover} is fulfilled.

		As the sweep shape and also $E^\infty_h$ are hexagonal it follows that a work band overlaps with at most 6 other work bands and Assumption~\ref{assum:workbandConst} is satisfied.
		Further, Assumption~\ref{assum:evalWidth} holds by construction for the adapted $\sH$.
		Assumption~\ref{assum:workEvalDistance} holds for the unmodified sweep shape~$\sH'$. 
		This is sufficient as Assumption~\ref{assum:workEvalDistance} is only needed for Lem.~\ref{lem:nbrOfWorkBandsArb} estimating the total number of work bands which we apply to the unmodified sweep shape $\sH'$ in \eqref{eq:hexaNbrOfWorkBands}.

		Let us now check the $k$-intersections.
		The vertices  in the $k$-intersections
		\begin{equation}
			B := \bigcup_{k=2}^\infty \Phi_{(W,k,h)}= \{w \in W: w \in H_h \text{ and } \exists \, V \in \wL: W \neq V \text{ and } w \in V \}
		\end{equation}
		can be counted similarly to those in $W\setminus E$ (see \eqref{eq:countNotEval}).
		Again, let us first count the vertices in $B$ for the unmodified work band $(W')^\infty_h$.
		By construction, the $\{w\in W': \;  P_s(w) \subseteq W' \} $ (subsets of evaluation bands) do not overlap.
		Further, as the  $W'_h$ are similar to convex shapes, every vertex in $W'_h$ is in the projected stencil $P_s(\cdot)$ of some vertex $\{w\in W': \;  P_s(w) \nsubseteq W' \}  $.
		Hence, work bands neighboring $W'_h$ intrude into $\{w\in W': \;  P_s(w) \subseteq W' \}  $  by at most this projected stencil. 
		Therefore we can account for the vertices of $W'$ that are also part of other work bands similar to \eqref{eq:countNotEval}. 
		At most the first and last $2\cdot 2s$ diagonals of  $W'_h$  ($\leq m+1$ vertices each for the original $\sH'$) are shared with other work bands.
		For each group of three diagonals for $x_1 \neq 0$ there are $2\cdot 2\cdot 2s$ vertices at the ends of the diagonals that can also belong to other work bands, possibly being identical with the vertices of the first and last diagonals that have already been accounted for completely.
		The middle group for $x_1 = 0$ contains $2\cdot 5s$ vertices which are also part of other work bands. 
		When we now consider $\sH$ instead of $\sH'$ there are at most 3 more vertices in $W_h$ then in $W'_h$.
		Conservatively, we assume that these three vertices are in $k$-intersections for $k\geq 2$.
		Similarly, all 6 work bands adjacent to $W$ contain at most 3 more vertices than per $H_h$ than assumed for $\sH'$.
		Hence, at most another $3\cdot 6$ vertices of $W_h$ belong to $k$-intersections for $k \geq2$. 
		All in all, the number of vertices in $k$-intersections for $k\geq2$ is at most (counting diagonal by diagonal from left to right as in \eqref{eq:countNotEval})
		\begin{align} \label{eq:countKIntersects}
			\size{\bigcup_{k=2}^\infty \Phi_{(W,k,h)}} 
			\leq 2 \cdot \lr{12 ms+\mO{1}}+(3+6\cdot 3 )
				= 24 ms +\mO{1} \; .
		\end{align}
		This also gives the bound $\size{\Phi_{(W,2,h)}} \leq 24 ms+\mO{1}$ and therefore Assumption~\ref{assum:2intersects} is satisfied with $b= 24s$.
		Furthermore, looking at the effective placement of the work bands, only $k$-intersections for $k\leq 3$ are non-empty.
		Hence, the sets $\Phi_{(W,k)}$  are empty  for $k \geq 4$.
		If $k = 3$, every $k$-intersection is limited in all three unit directions to constant width  within a hyperplane $H_h$.
		Hence, any $k$-intersection for $k=3$ contains only a constant number of vertices per $H_h$.
		Using Lemma~\ref{lem:kIntersectsBounded} this yields $\size{\Phi_{(W,k,h)}}= \mO{1}$ for $k =3$, all work bands $W$ and all $h\in [k_1]$.
		Hence Assumption~\ref{assum:3intersects} holds.

		Assumption~\ref{assum:evalPoints} is proven by the following Lemma. 
		\begin{lem}
			Given the setup of Def.~\ref{def:memEffBandAlgo}, employ the sweep sequence $\sweepSeq = \{e_1, e_2, e_3 \}$, sweep shape $\sH$ as specified in one of \eqref{eq:hexaSweepShapeS1}, \eqref{eq:hexaSweepShapeS2} or \eqref{eq:hexaSweepShapeS3}  and list of work bands $\wL$ specified in this section.
			Then, for any $W \in \wL$ and any $w \in E_W$ the following equality  holds:
			\begin{equation*}
				I(w) = [o_W(x)-\delta, o_W(x)+\delta] \quad \text{ with } \quad \delta = s\cdot \size{\sH} +\mO{\sqrt{M}}\;.
			\end{equation*}
		\end{lem}
		\begin{proof}

			The interval $I(w)$ is described by~\eqref{eq:IwCharacterization} and hence it is left to determine $w_{\min}$ and $w_{\max}$ and their distance to $w$ in the work band order.
			By the definition of $I(w)$ we know that $w_{\min}, w_{\max}\in S_s(w)$.
			From the lexicographic order it then follows that
			\begin{equation*}
				w_{\min} = w +(-s,0,0) \qquad \text{and} \qquad w_{\max} = w + (s,0,0) \;.
			\end{equation*}
			The vertex $v$ to which $w$ is shifted in he next $s$ shifts has distance  $||v-w||_{W}=  s\cdot \size{\sH} $ from $w$ in the work band order.
			For $\sH_s$, the sweep shape proceeding $\sH$ by $s$ shifts, consider
				$\sH_s \cap S_s(w)$.
%
			Both $v$ and $w_{\max}$ are in $\sH_s \cap S_s(w)$.
			Consider the projection of the set $\sH_s \cap S_s(w)$ in the $x_1x_2$-plane.
			Up to translations, this projection is given by $\{x \in \mathbb{Z}^2: x_1 \geq 0 \wedge x_2 \geq 0 \wedge x_1+x_2 \leq s \}$.
			As all vertices of  $\sH_s \cap S_s(w)$ belong to the $s$-point stencil $S_s(w)$ they are distributed over at most $s$ adjacent rows and at most $s$ adjacent columns of the sweep shape $\sH_s $.
			By definition, a row of the projection of the sweep shape contains less than  $(2m+1)+1= \mO{m}$ vertices.
			Hence, the positions of the lexicographic minimum and maximum vertices of  $\sH_s \cap S_s(w)$ differ by at most $(2m+2)\cdot s+s= \mO{m}$ in the work band order.
			Hence, also $||w_{\max}- v||_W = \mO{m}$ and therefore
			\begin{align*}
				||w - w_{\max}||_W &\leq ||w-v||_W \!+\! ||v- w_{\max}||_W = \\
				& =  s\cdot \size{\sH}\!+\! \mO{m} \stackrel{\eqref{eq:sweepSize}}{=} s\cdot \size{\sH}\!+\!\mO{\sqrt{M}} .
			\end{align*}
			The same argument can be used to show that $||w - w_{\min}||_W \leq s\cdot \size{\sH}+\mO{\sqrt{M}}$.
			Hence, the claim follows.
		\end{proof}

		It is left to estimate the total number of work bands. 
		Identify a start and an end of each work band with respect to the order of the shifts.
		As all shifts are unit directions, we say that work bands start at the $2$ dimensional faces of the grid which contain the point $(0, 0,0)$.\footnote{
		Similarly, work bands end at the 3 faces of the grid which contain the vertex $(k_1-1,k_2-1, k_3-1)$.
		}
		These are the 3 sets 
		\begin{equation*}
			0 \times [k_2]\times [k_3]\;, \qquad  [k_1] \times 0\times [k_3]\qquad \text{and} \qquad [k_1] \times [k_2]\times 0 \; .
		\end{equation*}
		The face $ 0 \times [k_2] \times[k_3]$ of the grid can also be written as $H_0$.
		Hence, Lemma~\ref{lem:nbrOfWorkBandsArb} can be applied to give an upper bound for the number of work bands that start in this face of the grid.
		All other faces of the grid in which work bands start can be treated similarly.
		Denote by $H^i_h$ the hyperplane of normal $e_i$ and distance $h$ from the origin.
		Considering the unmodified sweep shape $\sH'$, the sweep shape and the work bands are symmetric with respect to the three coordinates $x_1$, $x_2$ and $x_3$, i.e. permuting the coordinates does not affect the the sweep shape.
		In particular, the intersections $(E')^\infty \cap H^i_0$ 
		are identical for all $i \in \{1,2,3\}$ up to translations and an isomorphism of the coordinates.
		As in \eqref{eq:hexaEInfty}, $(E')^\infty \cap H^i_0 
		= 9m^2-\mO{m}$.
		As the modified sweep shape $\sH$ enlarges $\sH'$ we get $\sH$, $\lr{(E')^\infty \cap H^i_0} \subset\lr{ E^\infty \cap H^i_0}$. 
		Hence, the technique of Lemma~\ref{lem:nbrOfWorkBandsArb} (first enlarging the sides of grid by $2l_i+4s$ for last $d-1$ directions and then dividing the number of vertices in one face of the grid by the number of vertices of $E^\infty_h$) can be applied to every face of the grid to bound the number of work bands which start at this face.
		All in all, the number of work bands needed to cover the grid with evaluation bands is at most (Assumption~\ref{assum:nbrOfWorkBands})
		\begin{align}\label{eq:hexaNbrOfWorkBands}
			\sum_{j=1}^3 \lr{ \frac{\prod_{\substack{i =1\\ i \neq j}}^3 \lr{k_i+\mO{m}}}{\size{(E')^\infty_h}-\mO{m}} } 
			\stackrel{\eqref{eq:hexaEInfty}}{=}3\cdot \frac{ \prod_{i=1}^{2} \lr{k_i+\mO{m}}}{\lr{9m^2-\mO{m}}-\mO{m}}
			\stackrel{\eqref{eq:sweepSize}}{=} \mO{\frac{k_1k_2}{M}} \;.
		\end{align}

		As all assumptions of Def.~\ref{def:memEffBandAlgo} are satisfied, Theorem~\ref{thm:algoFrameworkUpperBounds} gives an upper bound for the number of non-compulsory I/Os performed by the Hexagonal Band Algorithm ($c=3$, $e = 9$ and $b=24s$), 
		\begin{align*}
			\frac{8 \sqrt{2}s^{3/2}}{\sqrt{3}}\cdot \frac{k_1k_2k_3}{B\cdot \sqrt{M}}+\mO{\frac{k_1k_2k_3}{\sqrt{B \cdot M^2}}} \;.
		\end{align*}

	\section{Discussion and Future Work}
\label{sec:discussion}

	This section concludes the paper by first discussing variants of the theoretical model that are closer to real caches and the effects of the model upon the algorithms and lower bounds.
	We then summarize the results and discuss open problems.

	\subsection{Variants of the Theoretical Model}



			\label{sec:upperBoundsAndDiscussion}
			The main contribution of this paper are the lower bounds as they do not only provide part of the complexity result but actually guided the construction of the data layouts and algorithms improving the upper bounds.
			The theoretical model chosen explicitly manages the cache, assumes simple I/Os and counts read and write operations.
			Further, the given task assumes that we do not work in-place and considers stencil operations to be atomic.
			We now examine the consequences of dropping these assumptions to model scenarios closer to current hardware.
			In short, besides working not-in-place none of these assumptions are crucial and matching bounds in one model translate to matching bounds in the other models.
			

			Let us first drop the assumption that I/Os are simple and consider non-simple I/Os, i.e. copy a block of data from external to internal memory when it is accessed.
			For the moment still assume that the cache is managed explicitly, i.e. we can decide what happens if a block is evicted from internal memory (either store the block back to external memory or simply delete it) and we can write to blocks of the external memory before loading them first.
			For the upper bounds, non-simple I/Os would mean that the number of compulsory I/Os stays constant while the leading term of the non-compulsory I/Os halves.
			I/Os regarding the output are either part of the compulsory term or lower order terms.
			For the input, we never need to store the input values back to external memory as the original copy remains unchanged.
			As a non-compulsory read in the model using simple I/Os was always preceded by one non-compulsory write, the leading term of the non-compulsory I/Os in the non-simple model halves.
			The same argument holds for the lower bounds.
			Consider a vertex and the series of I/Os it causes.
			In the simple model, the first compulsory read of an input vertex is followed by a series of alternating non-compulsory writes and reads, ending with a read.
			Hence, also for the lower bounds the number of compulsory I/Os stays constant at $2\cdot \prod_{i=1}^d k_i$ while the number of non-compulsory I/Os halves when switching from simple to non-simple I/Os.
			Hence, dropping the assumption that I/Os are simple reduces the number of non-compulsory I/Os  by a factor of 2 for both the lower and the upper bounds.



			Switching from an explicit to an implicit cache affects the cache replacement strategy as well as how data is evicted from internal memory and the access to blocks we only want to store data in.
			Regarding the cache replacement strategy, assuming an implicit cache replacement strategy as LRU instead of managing the cache replacement strategy explicitly does neither affect the lower nor the upper bounds.
			An explicit cache replacement strategy can be seen as an optimal cache replacement strategy and hence lower bounds for an explicitly managed cache replacement strategy hold for any implicit strategy.
			For the upper bounds, any implicit cache replacement strategy can be simulated with dummy accesses to items that should be kept in cache.
			
			A block that is evicted from internal memory can either be stored back to external memory or deleted within internal memory and forgotten.
			In an implicit cache it is common to write the block to external memory, causing a write operation, only if it has been altered in internal memory.
			Hence, if the block has not been modified in internal memory, and is hence identical to the external memory copy, the block is deleted within internal memory saving the I/O.
			For the algorithm as well as the lower bound this is exactly the behavior we employed for non-simple I/Os in the explicit model and hence the bounds do not change.
			
			Finally in an implicit cache, a block has to be loaded to internal memory before results can be stored in it.
			This increases the number of compulsory I/Os to 3 times the number of grid points, as we need to read the input as well as all output blocks and store the output as well.
			It does, however, not change the leading term of the non-compulsory I/Os 
			as I/Os caused by the output either affect the compulsory I/Os or lower order terms of the non-compulsory I/Os.
			Hence, for the lower as well as the upper bounds in an implicit cache model using non-simple I/Os 
			the number of compulsory I/Os increases to $3\cdot \prod_{i=1}^d k_i$ while the leading term of the non-compulsory I/Os remains unchanged with respect to the explicit model using non-simple I/Os. 

			If we would just be interested in reads and disregard the write operations in an implicit cache using non-simple I/Os, the compulsory term goes back down to 2 I/Os per grid point for both the lower and the upper bounds.
			Also, the non-compulsory term would be unchanged, as the leading term of the non-compulsory I/Os results from reading the 2-intersections of the input grid.
			These vertices are never written and hence counting only reads does not change the term.
			The additional reads only amount to lower order terms as they are just caused by the first and the last block of each $k$-intersections of the output grid.

			As next step, let us consider to work in-place in an implicit model using non-simple I/Os and counting reads only.
			By working in-place, the compulsory read of the output grid can be avoided and the number of compulsory read operations drops to 1 per grid point.
			The presented algorithms, however, are not capable to work completely in-place.
			Some $k$-intersections for $k\geq 2$ need to be buffered in additional space in external memory as they are needed to evaluate vertices after the $k$-intersection itself has been evaluated.
			A na\"ive buffer for all vertices in the union of all $k$-intersections for $k\geq 2$  would require $\mO{\lr{\prod_{i=1}^{d}k_i}\big/\sqrt[d-1]{M}}$ additional space.
			This can be reduced to $\mO{\lr{\prod_{i=1}^{d-1}k_i}\big/\sqrt[d-1]{M}}$ by working through adjacent work bands first and reusing the buffer of one $k$-intersection when all the work bands it is part of have been evaluated.
			When the input values of a $k$-intersection for $k\geq 2$ are stored to the buffer when the $k$-intersection itself is evaluated, the number of non-compulsory reads a vertex of that $k$-intersection takes part in rises from $k-1$ to $k$.
			This additional non-compulsory read is caused by loading the block of the buffer into which we want to store the input data.
			In the worst case, for $k=2$, this means that the number of non-compulsory I/Os doubles and hence the leading term of the non-compulsory I/Os doubles.
			The lower bound transfers to the in-place setting, as working in-place is more restrictive.
			It is likely that the lower bound can be improved for the in-place setting such that previously matching lower and upper bounds match again. 

			Last but not least, let us address the assumption that evaluating a stencil is an atomic operation which cannot be split.
			Allowing partial evaluations of the stencil requires a more general lower bound while the upper bounds still apply.
			Further, although partial stencil evaluations are possible in practice, none of the implementations discussed in the related work takes advantage of partial evaluations of the stencil.
			Regarding the lower bounds, 
			we think the assumption that stencil operations are atomic can be dropped without weakening the lower bounds. 
			Given a set of vertices which we want to evaluate in one round of the algorithm, the isoperimetric inequalities yield how many grid points need to be transferred to (or have already been transferred from) other rounds. 
			This does not assume that the stencil is indivisible but only states that neighboring values are needed to evaluate the stencil.
			Reducing the number of vertices that need to be transferred from one round to another would  mean to compress the data which has to be disallowed for the I/O model to make sense.

		\subsection{Summary and Future Work}

			This work examined the constant of the leading term of the non-compulsory I/Os caused by one update of the grid according to the $s$-star stencil in the external memory model and the parallel external memory model.
			In two dimensions, matching lower and upper bounds were given closing a multiplicative gap of 4. 
			In three dimensions the provided bounds match up to a factor of $\sqrt{2}$ improving the known results by a factor of $2 \sqrt{3}\sqrt{B}$.
			For dimensions $d\geq 4$, the lower bound is improved between a factor of $4$ and $6$.
			For arbitrary dimension~$d$, 
			the first analysis of the constant of the leading term of the non-compulsory I/Os was presented. 
			For $d\geq 3$ the lower and upper bound match up to a factor of $\sqrt[d-1]{d!}$.


			The lower bound combines a round argument with an isoperimetric inequality to bound the progress that can be achieved in each round.
			The isoperimetric results needs to be deduced and adapted carefully such that no constant factors are lost in the analysis.
			To analyze the upper bounds in a uniform manner, the framework of (memory efficient) band algorithms was introduced.
			This framework also defines the data layout which stores vertices that are read, written or read and written in contiguous memory.

			An experimental consideration on how to turn the proposed algorithms into high performance code remains open.
			Although memory access is very important for high performance code, it is not the only factor influencing the runtime.
			It needs to be determined in a process of algorithms engineering if and to which extent the benefits from on optimized data layout and access lead to faster code.
			Other options that may influence runtime include the more complicated index computations, optimizing for several layers of the memory hierarchy, vectorization and loop unrolling enabling scalar replacement.

			Topics the paper does not address include a time step setup, matching bounds for non-trivial $B$ and standard layouts like a usual row or column layout, stencils different from star stencils and matching bounds for high dimensions. 
			We do not consider a time step as it introduces a directed dimension and hence changes the structure of the computation graph, the stencil defining the neighborhood of a set and hence the isoperimetric sets and inequalities. 
			However, in a setting with time step the number of computations is the product of the spatial and temporal dimensions whereas the number of compulsory I/Os solely depends on the spatial dimensions.
			This implies that an isoperimetric argument, as presented in this paper, would analyze the constant of the leading instead of the second order term.
			To transfer the lower bounds to a time step setting an isoperimetric inequality for the directed, multi-layer time step computation graph needs to be derived.
			With this new inequality, the rest of the argument can be applied as before.
%
			The best upper bounds in two (Diagonal Band Algorithm) and three dimensions (Hexagonal Band Algorithm) should easily be transferable to a time step setting as the structure of the two and three dimensional algorithms is compatible with the setting of one temporal and one respectively two spatial dimensions.
			Parallelizing the algorithms is, however, more difficult when there is a temporal dimension.
			Also, the lower bounds have not been tuned to account for different data layouts as this would also change the isoperimetric inequalities. 
			However, while accounting for a specific data layout further restricts the theoretical model it may be a key aspect to get matching lower and upper bounds for different layouts. 
			It would also be interesting to examine the I/O complexity of stencils different from the star stencils given by $\ell^1$ balls. 
			Canonical candidates are stencils described by $\ell^\infty$ balls and mixtures between $\ell^1$ and $\ell^\infty$ stencils appearing in finite element methods.
			Finally, it remains open if the complexity can be pinpointed for three and higher dimensions as both lower and upper bounds do not seem optimal.


	\section*{Acknowledgements}
		The authors like to thank Gero Greiner and Tobias Lieber for helpful discussions during the derivations of the lower and upper bounds.
		We also want to thank Marcel Sch\"ongens for lending us his unbiased mind to improve the presentation of the problem.
		Further, we would like to thank the anonymous referee.

	\bibliographystyle{plain}
	\bibliography{bibPhilipp}
\end{document}